\def\CLEANVERSION{}
\documentclass[journal]{IEEEtran}
\IEEEoverridecommandlockouts
\usepackage{setspace}
\usepackage{cite}
\usepackage{amsmath,amssymb,amsfonts,mathtools,bm}
\usepackage{amsthm}
\usepackage{algorithm}
\usepackage{algpseudocode}
\usepackage{graphicx}
\usepackage{textcomp}
\usepackage{xcolor}
\usepackage{tabularx}
\usepackage{booktabs}
\usepackage{makecell}
\usepackage{microtype}
\usepackage{multirow}
\usepackage{hyperref}
\usepackage{pifont}
\usepackage{array}
\usepackage{tikz}

\providecommand{\captionsetup}[1]{}
\usepackage{changes}
\ifdefined\CLEANVERSION
  \makeatletter
  \def\Changes@AuthorColor{black}
  \makeatother
\fi
\DeclareRobustCommand{\rev}[1]{\ifdefined\CLEANVERSION\textcolor{black}{#1}\else\textcolor{blue}{#1}\fi}

\makeatletter
\let\revision@bibitem\@bibitem
\def\@bibitem#1{\revision@bibitem{#1}\ifboolexpr{test {\ifstrequal{#1}{shen2026gaigs}} or test {\ifstrequal{#1}{cheng2026rmsampler}} or test {\ifstrequal{#1}{lu2024newrf}}}{\ifdefined\CLEANVERSION\color{black}\else\color{blue}\fi}{\color{black}}}
\makeatother

\definecolor{lime}{HTML}{A6CE39}
\DeclareRobustCommand{\orcidicon}{%
    \begin{tikzpicture}
    \draw[lime, fill=lime] (0,0) 
    circle [radius=0.16] 
    node[white] {{\fontfamily{qag}\selectfont \tiny ID}};    \draw[white, fill=white] (-0.0625,0.095) 
    circle [radius=0.007];    \end{tikzpicture}
    \hspace{-2mm}}
\foreach \x in {A, ..., Z}{%
    \expandafter\xdef\csname orcid\x\endcsname{\noexpand\href{https://orcid.org/\csname orcidauthor\x\endcsname}{\noexpand\orcidicon}}
    }

\allowdisplaybreaks
\renewcommand{\baselinestretch}{0.96}
\begin{document}

\title{A Tutorial on Learning-Based Radio Map Construction: Data, Paradigms, and Physics-Awareness}
\author{
Xiucheng~Wang\orcidA{},~\IEEEmembership{Graduate Student Member,~IEEE},
Yuhao~Pan\orcidB{},
Nan~Cheng\orcidL{},~\IEEEmembership{Senior Member,~IEEE},
\c{C}a\u{g}kan~Yapar\orcidC{},
Ruijin~Sun\orcidD{},~\IEEEmembership{Member,~IEEE},
Zhisheng~Yin\orcidE{},~\IEEEmembership{Member,~IEEE},
Conghao~Zhou\orcidF{},~\IEEEmembership{Member,~IEEE},
Wenchao~Xu\orcidG{},~\IEEEmembership{Member,~IEEE},
Yuxiang~Zhang\orcidH{},~\IEEEmembership{Member,~IEEE},
Jianhua~Zhang\orcidI{},~\IEEEmembership{Fellow,~IEEE},
Shuguang~Cui\orcidJ{},~\IEEEmembership{Fellow,~IEEE},
and Xuemin~(Sherman)~Shen\orcidK{},~\IEEEmembership{Fellow,~IEEE}

\thanks{
\par This work was supported by the National Key Research and Development Program of China under Grant 2024YFB907500.
\par Xiucheng Wang, Nan Cheng, Ruijin Sun, Zhisheng Yin, and Conghao Zhou are with the State Key Laboratory of Integrated Services Networks and School of Telecommunications Engineering, Xidian University, Xi'an 710071, China (e-mail: xcwang\_1@stu.xidian.edu.cn; dr.nan.cheng@ieee.org; \{sunruijin, zsyin, zhouconghao\}@xidian.edu.cn). (Corresponding author: Nan Cheng.)
\par Yuhao Pan and Wenchao Xu are with the Division of Integrative Systems and Design, The Hong Kong University of Science and Technology, Hong Kong, China (e-mail: ypanca@connect.ust.hk; wenchaoxu@ust.hk).
\par \c{C}a\u{g}kan Yapar is with the Institute of Telecommunication Systems, Technical University of Berlin, 10623 Berlin, Germany (e-mail: cagkan.yapar@tu-berlin.de).
\par Yuxiang Zhang and Jianhua Zhang are with the State Key Laboratory of Networking and Switching Technology, Beijing University of Posts and Telecommunications, Beijing, China (e-mail: zhangyx@bupt.edu.cn; jhzhang@bupt.edu.cn).
\par Shuguang Cui is with the School of Science and Engineering, the Shenzhen Future Network of Intelligence Institute (FNii-Shenzhen), and the Guangdong Provincial Key Laboratory of Future Networks of Intelligence, The Chinese University of Hong Kong (Shenzhen), Shenzhen, Guangdong 518172, China (e-mail: shuguangcui@cuhk.edu.cn).
\par Xuemin (Sherman) Shen is with the Department of Electrical and Computer Engineering, University of Waterloo, Waterloo,
N2L 3G1, Canada (e-mail: sshen@uwaterloo.ca).
}
}
    \maketitle

\IEEEdisplaynontitleabstractindextext

\IEEEpeerreviewmaketitle

\begin{abstract}
Radio maps (RMs) provide the digital representation of the wireless propagation environment, mapping complex geographical and topological boundary conditions to critical spatial-spectral metrics that range from received signal strength to full channel state information matrices. The integration of artificial intelligence into next generation wireless networks further necessitates the accurate construction of RMs as a foundational prerequisite for electromagnetic digital twins. This paper presents a comprehensive survey of learning-based RM construction, systematically addressing three intertwined dimensions: data, paradigms, and physics-awareness. From the data perspective, we review physical measurement campaigns, ray tracing simulation engines, and publicly available benchmark datasets, identifying their respective strengths and fundamental limitations. From the paradigm perspective, we establish a core taxonomy that categorizes RM construction into source-aware forward prediction and source-agnostic inverse reconstruction, and examine five principal neural architecture families spanning convolutional neural networks, vision transformers, graph neural networks, generative adversarial networks, and diffusion models. We further survey optics-inspired methods adapted from neural radiance fields and 3D Gaussian splatting for continuous wireless radiation field modeling. From the physics-awareness perspective, we introduce a three-level integration framework encompassing data-level feature engineering, loss-level partial differential equation regularization, and architecture-level structural isomorphism. Open challenges including foundation model development, physical hallucination detection, and amortized inference for real-time deployment are discussed to outline future research directions. The project page is at \url{https://github.com/UNIC-Lab/Awesome-Radio-Map-Categorized}.
\end{abstract}


\section{Introduction}\label{sec:introduction}
Next-generation wireless networks increasingly rely on artificial intelligence to manage complex, high-dimensional data distributions~\cite{dang2020should, 6g}. These distributions include massive multiple-input multiple-output (MIMO) channel matrices, heterogeneous traffic patterns, and dynamic network topologies~\cite{chen2017measurement, boban2023white}. To support the realization of electromagnetic digital twins, the accurate construction of radio maps (RMs) has become a fundamental prerequisite~\cite{zeng2024tutorial, jiang2024learnable}. An RM provides a digital representation of the wireless propagation environment over a geographic region of interest. It maps geographical and topological boundary conditions to spatial-spectral metrics. These metrics range from received signal strength (RSS) and time of arrival (ToA) to full complex channel state information (CSI) matrices~\cite{shen2023toward}. A high-fidelity RM therefore serves as the cornerstone for downstream tasks such as resource allocation, interference management, and predictive network planning~\cite{becvar2024machine, okawa2024optimal}, thereby supporting digital-twin applications and intelligent connectivity~\cite{liu2020vision6g}.

Traditionally, acquiring ground-truth radio environment data has relied on dedicated physical measurement campaigns~\cite{chen2017measurement}. Drive tests and walk tests capture the stochastic nature of real-world propagation channels~\cite{guo2013mobile}. However, these methods incur high operational costs and limited spatial coverage~\cite{sallouha2024rem}. Indoor environments and three-dimensional vertical domains remain particularly difficult to survey~\cite{li2024intelligent}. The industry has adopted crowdsourcing approaches such as the minimization of drive tests (MDT) to alleviate spatiotemporal sparsity~\cite{sallouha2024rem, guo2013mobile}. Nevertheless, commercial user equipment introduces significant measurement noise due to device heterogeneity and uncalibrated antenna configurations~\cite{10640063}.

To bypass the constraints of physical data collection, computational electromagnetic methods including ray tracing (RT) and the dominant path model (DPM) have been widely employed~\cite{dpm,irt}. These methods provide simulation tools for generating propagation data. The shooting and bouncing rays (SBR) algorithm nevertheless faces severe scalability bottlenecks because the number of traced rays and candidate interaction paths grows rapidly with scene complexity and maximum interaction order~\cite{dpm,deschamps1972ray,bakirtzis2024rigorous}. High-order city-scale simulation is therefore expensive, especially when a dynamic scene must be regenerated~\cite{10640063,mladenovic2022overview}. Differentiable RT frameworks such as Sionna support gradients with respect to scene and radio parameters, but differentiability does not remove the underlying path-tracing cost~\cite{hoydis2023sionna}. Pure simulation must consequently be benchmarked against the complete update interval before it is used for a real-time digital twin~\cite{11152929}. This constraint motivates the digital twin channel (DTC) paradigm, which represents physical channel characteristics in a synchronized digital model for proactive communication decisions~\cite{wang2025digitaltwinchannel,wang2025radioenvironmentknowledgepool,zhang2025wirelessenvironmental}.

To reduce this latency, recent studies investigate stepwise wireless environment information (WEI) compression, multi-task adaptive RT platforms, and pretrained channel models for environment-conditioned prediction~\cite{zhang2025foursteps,yu2025road,yu2025channelgpt}. These developments have accelerated the shift toward data-driven deep learning (DL) surrogates~\cite{10640063,mladenovic2022overview,sagir2024machine}. A learned surrogate approximates propagation patterns represented in its training data; it does not automatically encode or enforce a wave equation~\cite{levie2021radiounet,lee2023pmnet}. Its latency and fidelity must therefore be validated against a declared physical solver or measurement protocol in the intended deployment regime. Based on this distinction, this tutorial categorizes RM construction into source-aware forward prediction and source-agnostic inverse reconstruction.

In source-aware RM construction, the neural network receives a declared set of environment and transmitter parameters, such as geometry, location, frequency, power, and antenna settings~\cite{levie2021radiounet,lee2023pmnet,li2025rmtransformer}. The network acts as a fast surrogate within the variables and regimes represented by those inputs; ``source-aware'' does not imply that every material, boundary, or hardware parameter is known. Source-agnostic RM construction addresses scenarios where one or more transmitter configurations are latent or unavailable~\cite{teganya2021deep,chaves2023deeprem,fang2025radioformer}. Measurements are spatially sparse in these settings. From a signal-processing perspective, this formulation is generally an ill-posed nonlinear inverse problem~\cite{sallouha2024rem}. The network must infer unobserved field quantities and, when required, latent source variables from limited observations and stated priors.

Driven by these diverse problem formulations, the underlying neural architectures have evolved significantly. Early methods relied on convolutional neural networks (CNNs) to model propagation as a two-dimensional image-to-image mapping~\cite{levie2021radiounet, lee2023pmnet, ansari2021prediction}. However, fixed and localized receptive fields limit CNNs' ability to capture long-range spatial dependencies~\cite{10640063}. Macroscopic shadowing and diffraction around distant obstacles require global spatial reasoning.

Vision transformers (ViTs) address this limitation through self-attention mechanisms~\cite{li2025rmtransformer, fang2025radioformer, liaq2025visual, liu2025paying}. They process environmental features globally and establish correlations between distant geographic structures and local signal attenuation. This capability is valuable for inverse problems where sparse observations must be contextualized within the full environment~\cite{fang2025radioformer, hehn2023transformer}. Token-based variants can accept variable-size sparse point sets when their positional encoding, masks, and batching strategy are designed for set inputs. This avoids dense-grid zero padding, but it is not an automatic property of every vision Transformer~\cite{fang2025radioformer,hehn2023transformer}.

Graph neural networks (GNNs) have also been adopted when the data are represented as an irregular graph of sensors, links, scatterers, or network entities~\cite{li2024radiogat,chen2023graph,perdomo2025wirelessnet,shibli2024data}. Radio propagation still occurs in physical space; the non-Euclidean object is the chosen interaction graph. Nodes represent spatial or network elements, while edges encode assumed visibility, communication, interference, or neighborhood relations. This formulation can translate selected propagation relations into learnable graph priors~\cite{li2024radiogat}, but graph construction remains a model assumption that must be validated.

Under severe spatial sparsity, deterministic point estimators can average over materially different solutions. Generative adversarial networks (GANs) instead learn a conditional generator~\cite{zhang2023rme,zhou2025tire,chen2024act,pan2025sc,sarkar2024recugan,ma2025radio}, while diffusion models learn repeated denoising or score evaluations~\cite{wang2024radiodiff,11278649,11083758,luo2024rm,luo2025denoising,liu2025wifi,fu2025generative,zhao20253d,dai2025bs}. Their reported stability and sample fidelity are objective-, data-, and implementation-dependent rather than automatic properties of every RF task. Recent token-prediction models replace iterative diffusion with coarse-to-fine autoregressive decoding~\cite{zhang2026radiovar}. This is sampling-free only in the source's sense of avoiding iterative diffusion sampling; sequential token decoding and its associated latency remain. Diffusion sample dispersion can provide an empirical uncertainty proxy, but it requires calibration on held-out environments before operational use~\cite{wang2024radiodiff,ha2025bayesian}.

To overcome resolution bottlenecks inherent in discrete grid representations, researchers have adapted optical neural rendering techniques to the radio-frequency domain. Neural radiance fields (NeRF) and 3D Gaussian splatting (3DGS) provide continuous scene representations that support complex-valued channel reconstruction in the cited studies~\cite{jiang2024learnable,zhao2023nerf2,wen2025wrf,zhang2026rf}. Their fidelity is specific to the target modality, measurements, scene-fitting protocol, and RF rendering rule; it is not a universal advantage over a grid baseline. Per-scene optimization also limits rapid reuse when the environment or transmitter configuration changes~\cite{jiang2024learnable}. Section~\ref{subsec:learning_primer} translates these model names into elementary training and inference operations before the detailed comparisons below.

Purely data-driven models have no built-in guarantee of electromagnetic consistency outside the conditions represented in training, so a plausible output can violate an applicable physical constraint~\cite{11278649,karniadakis2021physics}. Physics-informed methods address this risk by introducing electromagnetic information into the data representation, loss, or architecture~\cite{karniadakis2021physics,jiang2024physics,liu2025pinn,jia2025rmdm,feng2025physics}. Maxwell- or Helmholtz-derived quantities can guide learning, but their validity still depends on the predicted modality, discretization, material model, and boundary data~\cite{11278649,jia2025rmdm,chen2026radiodun}. This tutorial organizes these choices into data-level feature engineering, loss-level partial differential equation (PDE) regularization, and architecture-level structural isomorphism~\cite{11278649,jia2025rmdm,chen2026radiodun,wang2025iradiodiff}.

\subsection{Related Work}\label{subsec:related_works}
Although RM construction and wireless deep learning have attracted significant attention, existing surveys focus on different aspects. Doha and Abdelhadi~\cite{doha2025deep} provide a comprehensive roadmap for integrating deep learning into wireless receivers. Their scope is confined to temporal and symbol-level processing at the receiver. This focus is fundamentally distinct from the spatial, environment-scale field reconstruction addressed in this tutorial. Vasudevan and Yuksel~\cite{10640063} offer a broad overview of machine learning for radio propagation. Their coverage emphasizes traditional regression-based algorithms such as random forests and support vector machines. Zeng et al.~\cite{zeng2024tutorial} present a visionary tutorial on the channel knowledge map concept and its network-level applications. Regarding map construction, their discussion relies on classical spatial interpolation techniques like Kriging and tensor completion. Our tutorial, in contrast, treats RM construction as a complex inverse problem and explores advanced generative paradigms. Feng et al.~\cite{feng2025recent} classify RM estimation methods into model-driven, data-driven, and hybrid approaches. Their architectural coverage extends to ViTs and conditional GANs. However, the existing literature largely views RM construction as a deterministic interpolation or image-to-image translation task. Our tutorial advances beyond this view in three respects. First, we review probabilistic generative paradigms for extreme spatial sparsity and treat their sample dispersion as an empirical uncertainty proxy that requires calibration on held-out environments~\cite{wang2024radiodiff,ho2020denoising}. Second, we systematically review and mathematically adapt optics-inspired neural rendering techniques to the complex-valued RF domain~\cite{jiang2024learnable, zhang2026rf}. Third, we propose a rigorous three-level taxonomy for physics-informed integration and formally define the concept of physical hallucinations~\cite{11278649}. Table~\ref{tab:survey_comparison} makes this boundary explicit by comparing the prior surveys along scope, problem formulation, physical representation, and evidence type rather than by claiming superiority from a longer paper list.

The distinction is not merely broader architectural coverage. This tutorial develops four design principles that connect a deployment question to a modeling choice. First, identifiability is considered before architecture selection: known transmitter parameters define a forward surrogate problem, whereas unknown parameters and sparse observations define an inverse problem that requires an explicit prior. Second, the generalization boundary is determined by the physical variables represented at the input, including frequency, material, antenna, and geometry. Third, the appropriate depth of physics integration is selected jointly with data availability, model mismatch, and the computation budget. Fourth, numerical results from different studies are treated as source-reported evidence rather than a common leaderboard unless their datasets, units, masks, and training budgets match. These principles turn the taxonomy into a decision process instead of a catalogue of additional models.
\begin{table*}[t]
\centering
\caption{Scope and organizing principles of this tutorial compared with representative related surveys. The entries describe coverage and evidence boundaries; they are not an accuracy or citation-count ranking.}
\label{tab:survey_comparison}
\footnotesize
\renewcommand{\arraystretch}{1.0}
\vspace{-4pt}
\resizebox{0.99\linewidth}{!}{%
\begin{tabular}{@{}l|l|l|l|l|l@{}}
\toprule
Dimension & Doha and Abdelhadi~\cite{doha2025deep} & Zeng et al.~\cite{zeng2024tutorial} & Vasudevan and Yuksel~\cite{10640063} & Feng et al.~\cite{feng2025recent} & This tutorial \\
\midrule
Primary scope & \shortstack[l]{DL for physical-layer\\receiver modules} & \shortstack[l]{CKM concepts and\\network use} & \shortstack[l]{ML for propagation\\and path loss} & \shortstack[l]{Data/model-driven\\RM estimation} & \shortstack[l]{End-to-end learning-based\\RM construction} \\
Problem view & \shortstack[l]{Detection, decoding,\\and equalization} & \shortstack[l]{Map interpolation\\and querying} & \shortstack[l]{Propagation regression\\and classification} & \shortstack[l]{Interpolation and\\image mapping} & \shortstack[l]{Identifiability-first\\forward/inverse split} \\
Architecture and representation & \shortstack[l]{MLP/CNN/RNN\\receiver features} & \shortstack[l]{Map abstraction\\and interpolation} & \shortstack[l]{Regressors and\\propagation features} & \shortstack[l]{CNN/ViT/GAN\\estimators} & \shortstack[l]{CNN, ViT, GNN, generative,\\and neural-field models} \\
Optics-inspired RF fields & Not covered & Not covered & Not covered & Not covered & \shortstack[l]{RF-NeRF/3DGS with\\continuous complex fields} \\
Extreme sparsity & Outside main scope & \shortstack[l]{Kriging and\\matrix completion} & \shortstack[l]{Interpolation and\\drive-test reduction} & CNN/GAN inpainting & \shortstack[l]{Generative and physical\\inverse recovery} \\
Physics integration & Receiver features & \shortstack[l]{Map abstraction; no\\unified PDE taxonomy} & \shortstack[l]{Propagation features\\and measurements} & \shortstack[l]{Model knowledge as\\estimation dependency} & \shortstack[l]{Three-level integration and\\hallucination tests} \\
Practical guidance & Receiver selection & Network-level map use & Learner selection & Method categorization & \shortstack[l]{Dataset, cost, code, and\\deployment protocols} \\
Evidence and decision rule & Receiver-task fit & Map-decision fit & Propagation-task fit & Dependency-task fit & \shortstack[l]{Match identifiability, inputs,\\physics, cost, and protocol} \\
\bottomrule
\end{tabular}}
\end{table*}

\subsection{Contributions}
\label{subsec:intro_contributions}

The main contributions of this tutorial are summarized as follows.
\begin{enumerate}
    \item  We present a comprehensive data ecosystem for RM construction. We systematically review physical measurement techniques~\cite{guo2013mobile, sallouha2024rem}, RT simulation engines~\cite{dpm, irt, hoydis2023sionna}, and publicly available benchmark datasets~\cite{levie2021radiounet, lee2023pmnet, 11083758}. We identify their respective strengths and limitations while providing practitioner-oriented guidance for dataset selection.
    \item We establish a unified architectural taxonomy organized by the forward-inverse problem dichotomy. This taxonomy spans CNNs~\cite{levie2021radiounet, lee2023pmnet}, ViTs~\cite{li2025rmtransformer, fang2025radioformer, liaq2025visual}, GNNs~\cite{li2024radiogat, chen2023graph, perdomo2025wirelessnet}, GANs~\cite{zhang2023rme, zhou2025tire, chen2024act}, diffusion models~\cite{wang2024radiodiff, 11278649, 11282987, 11083758}, and optics-inspired neural rendering methods~\cite{jiang2024learnable, zhang2026rf, zhao2023nerf2, wen2025wrf, umer2025neural, zhou20256d}. Cross-architecture comparisons clarify the trade-offs among inference latency, data efficiency, and spatial modeling capability. The taxonomy is accompanied by an identifiability-first rule and four deployment principles that specify when an architecture, physical input, or uncertainty mechanism is justified.
    \item  We introduce a physics-informed integration framework that categorizes embedding strategies by their depth of physical coupling~\cite{11278649, jiang2024physics, liu2025pinn, chen2026radiodun, wang2025iradiodiff}. This framework provides actionable recipes for practitioners to incorporate electromagnetic knowledge incrementally. Open challenges including foundation model development, physical hallucination detection, and amortized inference for real-time deployment are discussed throughout.
\end{enumerate}
\section{Preliminaries}
\label{sec:preliminaries}
This section introduces only the mathematical foundations needed to understand the later design choices. It first explains the learning pipeline in terms familiar to a communications reader, then defines the radio map and its physical properties. The section concludes with concise introductions to diffusion models, neural radiance fields, and 3D Gaussian splatting. Detailed derivations that do not change the model-selection decision have been omitted to preserve a tutorial path from intuition to equations.

\subsection{Deep Learning Primer}
\label{subsec:learning_primer}

A learning-based RM constructor is a parameterized mapping from available evidence to a spatial field. During training, labeled examples are used to adjust parameters $\boldsymbol{\theta}$ so that a prediction $\hat{\mathbf{M}}=f_{\boldsymbol{\theta}}(\mathbf{X})$ approaches a reference map $\mathbf{M}$. The input $\mathbf{X}$ can contain a building map, transmitter configuration, carrier frequency, and sparse measurements. A loss function converts the disagreement between $\hat{\mathbf{M}}$ and $\mathbf{M}$ into a scalar training signal. During inference, the learned parameters are fixed and only the forward evaluation of $f_{\boldsymbol{\theta}}$ is required. This distinction explains why a method may be expensive to train but still suitable for real-time deployment.

\begin{align}
&\boldsymbol{\theta}^{\star}=\arg\min_{\boldsymbol{\theta}}\frac{1}{N_{\mathrm{tr}}}\sum_{n=1}^{N_{\mathrm{tr}}}\mathcal{L}\!\left(f_{\boldsymbol{\theta}}(\mathbf{X}_n),\mathbf{M}_n\right).
\label{eq:generic_learning}
\end{align}

Equation~\eqref{eq:generic_learning} states the common learning objective in plain terms: select the parameter set that gives the smallest average training error. The word ``architecture'' describes how a model transforms the input tensor rather than a particular software package. A convolutional neural network slides a small kernel over neighboring cells, aggregates local features, and normally reduces the spatial resolution through pooling or strided convolution. An encoder--decoder network then restores the grid resolution, while skip connections copy fine boundary features from the encoder to the decoder. This local-to-global path explains why CNNs are efficient on regular maps but can blur a shadow boundary that depends on a distant obstacle. A Transformer first divides the map or measurement set into tokens and computes attention weights between token pairs. The query, key, and value projections determine which remote tokens contribute to each output token; this attention operation is unrelated to a physical antenna beam, even though both use the word ``attention'' or ``beam'' in different contexts.

A graph neural network replaces the raster neighborhood with nodes and edges. Each node stores a sensor, transmitter, or spatial-cell feature, and a message-passing layer transforms neighboring node features before an aggregation step updates the node state. The edge definition therefore becomes a modeling assumption about which propagation interactions are visible to the network. A generative adversarial network uses a generator to propose a map and a discriminator to distinguish generated maps from reference maps, so the discriminator is needed during training but not during deployment. A diffusion model instead adds a known noise schedule during training and learns a denoising direction; inference repeats the denoiser for several steps and can return multiple conditional samples. The extra sampling steps are the price paid for representing non-unique inverse solutions.

Neural rendering uses a different computational object. A neural radiance field evaluates a coordinate-conditioned function along many sampled rays and integrates the returned field values, whereas 3D Gaussian splatting stores explicit spatial primitives and projects visible primitives into a query view. Both representations can answer arbitrary-coordinate queries after a scene has been fitted, but the scene-fitting cost is paid again when the environment changes unless an amortized initializer is available. These distinctions give a practical reading rule for the later sections. Use convolution when the grid and latency dominate, attention or graphs when non-local or irregular interactions dominate, a generative model when observations admit multiple maps, and a neural field when continuous within-scene queries justify per-scene construction.

\rev{Table~\ref{tab:learning_translation} links familiar communications concepts to the learning inputs, parameters, and outputs used throughout this tutorial.}

\begin{table}[!t]
\centering
\captionsetup{font={small}, skip=8pt}
\caption{Translation from communications objects to learning objects in an RM constructor.}
\vspace{-4pt}
\footnotesize
\setlength{\tabcolsep}{3pt}
\begin{tabularx}{0.99\linewidth}{@{}>{\raggedright\arraybackslash}p{0.25\linewidth}|>{\raggedright\arraybackslash}X|>{\raggedright\arraybackslash}p{0.32\linewidth}@{}}
\toprule
Communications object & Learning interpretation and RM example & Practical check \\
\midrule
Known geometry and source settings & Condition/input tensor: building height, transmitter, frequency, and beam & Which physical variables can change at test time? \\
Reference field or solver output & Label/target: power, path loss, CSI, or angular map & Which quantity is predicted, and in what domain? \\
Sparse drive-test reports & Observation channel and mask: RSS values with coordinates and validity flags & Are complete regions, rather than only random pixels, held out? \\
Trainable filters or weights & Parameters $\boldsymbol{\theta}$: CNN kernels, attention projections, or a denoiser & What is learned once, and what must be recomputed? \\
Engineering error or constraint & Loss or residual: masked error, PDE residual, or link-budget check & Does the metric test the intended task and physical consistency? \\
New streets, frequencies, or beams & Held-out test environment or condition used to measure transfer & Does the claimed generalization cover the omitted variable? \\
\bottomrule
\end{tabularx}
\label{tab:learning_translation}
\end{table}

A running example clarifies the two problem types used throughout this tutorial. Consider an urban block with a building-height map and one base station. If the base-station location, transmit power, frequency, and antenna pattern are known, the input completely specifies the excitation and the task is source-aware forward prediction. If these transmitter parameters are unavailable and only a set of drive-test received-signal-strength samples is observed, the same task becomes source-agnostic inverse reconstruction. The first task learns a fast substitute for a propagation solver. The second must infer unobserved causes and therefore needs stronger structural, generative, or physical priors.

The learning loop is easier to interpret when it is related to a familiar link-budget study. The dataset is divided into training, validation, and test environments rather than only random pixels. A mini-batch contains an input tensor, a target field, and any validity mask, and gradient descent updates the model parameters using the loss evaluated on that batch. The validation split is used to select the checkpoint, noise schedule, or early-stopping point without exposing the final test environments. If the training error decreases while the validation error rises, the model has memorized the observed scenes; additional layers or diffusion steps do not solve this problem. Weight decay, data augmentation, conditioning on the omitted physical variables, and scene-disjoint sampling are alternative ways to control overfitting. These choices are part of the experiment protocol, not implementation details that can be omitted from a reproducibility report.

Several training terms used throughout the tutorial have specific operational meanings. A mini-batch is the subset of training examples processed before one parameter update, and an epoch is one pass through the training set. The optimizer computes the update direction from the loss gradient, while the learning rate controls the update magnitude. A hyperparameter is selected outside the gradient update, such as the batch size, regularization weight, or number of diffusion steps. A checkpoint stores the model parameters at a particular training stage so that validation can select a deployable version. Overfitting occurs when training error improves without a corresponding improvement on held-out environments, which is why test scenes must remain untouched until the final evaluation.

A prediction also differs from a measurement in its uncertainty. A deterministic CNN returns one field, even when several fields are consistent with the sparse observations. A diffusion model can generate several conditional samples, and their mean and spread provide an empirical uncertainty estimate, but the spread is useful only when it is calibrated on held-out environments. Calibration asks whether regions assigned a nominal coverage probability contain the true field at the stated frequency. In practice, the report should include both a point error and a coverage or interval score, together with the sampling cost required to obtain the interval. This distinction prevents a visually smooth map from being mistaken for a reliable map.

\subsection{Definition and Characterization of Radio Map}
\label{subsec:rm_definition}

A radio map is a spatially indexed, queryable representation of one or more radio quantities over a geographic region of interest~\cite{zeng2024tutorial,shen2023toward}. It may be stored on a discrete grid, at irregular points, or as a continuous function, and it need not contain the complete electromagnetic field. Unlike isolated measurements, an RM organizes spatial variation for inference, planning, and optimization tasks~\cite{becvar2024machine,sallouha2024rem}.

\subsubsection{Electromagnetic Foundation}
\label{subsubsec:em_foundation}

The electromagnetic field is governed by Maxwell's vector equations~\cite{taflove2005computational,balanis2016antenna}. For time-harmonic operation in a linear isotropic medium, a decoupled polarization or scalar field component $u(\mathbf r)$ can be modeled by the inhomogeneous Helmholtz equation
\begin{equation}
\nabla^2 u(\mathbf{r}) + k^2(\mathbf{r})\, u(\mathbf{r}) = -f(\mathbf{r}),
\label{eq:helmholtz}
\end{equation}
where $f(\mathbf r)$ is a source term and $k^2(\mathbf r)$ is generally complex in a lossy medium. Reflection, diffraction, and penetration follow only after the material interfaces, sources, and boundary or radiation conditions are specified. The scalar reduction is not exact for anisotropic, strongly vector-coupled, or polarization-resolved problems, which require Maxwell's vector equations. Accordingly, a complex-field RM may sample a solution $u$, whereas a power-, angle-, or delay-domain RM contains quantities derived from the field or channel measurements rather than being itself a scalar Helmholtz solution.

\begin{table*}[!t]
\centering
\caption{Systematic comparison of Radio Map, Channel Knowledge Map, and Radio Environment Map~\cite{zeng2024tutorial, sallouha2024rem, shen2023toward}.}
\label{tab:rm_ckm_rem}
\footnotesize
\renewcommand{\arraystretch}{1.0}
\setlength{\tabcolsep}{3pt}
\begin{tabularx}{0.99\linewidth}{@{}>{\raggedright\arraybackslash}p{2.1cm}|>{\raggedright\arraybackslash}X|>{\raggedright\arraybackslash}X|>{\raggedright\arraybackslash}X@{}}
\toprule
Attribute & Radio Map & Channel Knowledge Map & Radio Environment Map \\
\midrule
Core quantity & EM field state: power, angle, delay & Transmitter--receiver impulse response & Multi-layer knowledge base: spectrum, interference, policy \\
\midrule
Physical basis & Helmholtz/Maxwell solution; field-theoretic~\cite{karniadakis2021physics} & System-theoretic: link input--output response~\cite{zeng2024tutorial} & Network-level abstraction; not tied to a single PDE \\
\midrule
Representation & $P(\mathbf{r})\propto|u(\mathbf{r})|^2$ under a fixed impedance/normalization, or a multi-channel field tensor & $h(\mathbf{r};\tau)$ for a declared Tx--Rx pair & Aggregated records without one universal form \\
\midrule
Source dependency & May be source-aware or source-agnostic & Usually tied to a specific Tx--Rx pair~\cite{zeng2024tutorial} & May aggregate multiple links \\
\midrule
Data acquisition & RSS, sounding, ray tracing, or field inversion~\cite{guo2013mobile} & Channel sounding and impulse-response extraction~\cite{chen2017measurement} & Sensors, databases, and operator records~\cite{sallouha2024rem} \\
\midrule
Typical application & Coverage, beam management, localization~\cite{becvar2024machine, levie2021radiounet} & Channel database and precoder design~\cite{zeng2024tutorial} & Cognitive radio and dynamic spectrum access~\cite{sallouha2024rem} \\
\bottomrule
\end{tabularx}
\end{table*}

The PDE-rooted view carries an important implication: spatial RM values are coupled rather than statistically independent. Physics-informed neural networks (PINNs) can exploit this coupling by penalizing a residual of~\eqref{eq:helmholtz} during training~\cite{karniadakis2021physics,liu2025pinn,jiang2024physics}. A finite penalty encourages consistency with the selected wave model in unmeasured regions; it does not guarantee Maxwell-consistent output when the field modality, discretization, materials, sources, or boundary conditions are incomplete.

\subsubsection{Multi-Dimensional Representation}
\label{subsubsec:rm_multidim}

An RM need not contain the complete electromagnetic state. A full time-harmonic vector-field description includes complex amplitude, polarization, and frequency, whereas a channel description can additionally resolve delay and arrival or departure direction~\cite{zeng2024tutorial}. Application-specific RMs retain only the quantities needed by the downstream task and can be organized into a hierarchy of increasing dimensionality.

At the most compact level, the power-domain RM captures the spatially distributed RSS at each grid point. It is expressible as a single-channel image $\mathbf{P} \in \mathbb{R}^{H \times W}$~\cite{levie2021radiounet, lee2023pmnet}. This form aligns naturally with standard CNN input conventions.

An angular- or delay-domain RM can store an angular power spectrum (APS) or power-delay profile (PDP) as multi-channel images~\cite{11083758}. Each channel corresponds to a declared angular bin or delay tap. These are power distributions over resolved coordinates, not the complete complex multipath channel unless phase and the relevant antenna dimensions are also retained.

One higher-dimensional example is a delay--angle--power tensor $\mathbf T\in\mathbb R^{H\times W\times N_\tau\times N_\theta}$~\cite{shen2023toward,11083758}. Here, $\theta$ must be declared as an arrival angle, departure angle, or a composite angular index. Representing DoA and DoD jointly requires separate angular dimensions (and potentially polarization, frequency, transmitter, receiver, or array indices). Thus the displayed four-dimensional tensor is an application-specific representation rather than the highest possible RM dimensionality.

\subsubsection{Distinction from Related Concepts}
\label{subsubsec:rm_distinction}

Two concepts frequently appear alongside the RM: the channel knowledge map (CKM)~\cite{zeng2024tutorial} and the radio environment map (REM)~\cite{sallouha2024rem}. Each rests on a different abstraction. Table~\ref{tab:rm_ckm_rem} compares them along seven key attributes.

The CKM is a spatial distribution of impulse responses $h(\mathbf{r};\tau)$~\cite{zeng2024tutorial}. Each response characterizes the input--output behavior between a declared transmitter and a receiver at $\mathbf r$. Under a fixed impedance and field normalization, a power-domain RM is proportional to $|u(\mathbf r)|^2$; it is not numerically equal to that quantity under every field convention. For cochannel fields $u_m$ observed over the same averaging interval, the received power contains both self and cross terms:
\begin{equation}
\begin{aligned}
P(\mathbf r)&\propto\left|\sum_{m=1}^{M}u_m(\mathbf r)\right|^2\\
&=\sum_{m=1}^{M}|u_m(\mathbf r)|^2
+2\operatorname{Re}\!\sum_{m<n}u_m(\mathbf r)u_n^*(\mathbf r).
\end{aligned}
\label{eq:superposition}
\end{equation}
The time- or ensemble-averaged cross terms vanish only under an applicable incoherence assumption, such as uncorrelated phases, asynchronous signaling with sufficient averaging, or separation by the receiver into orthogonal resources~\cite{levie2021radiounet}. Signals on distinct carriers should be summed after band-selective power measurement rather than inserted as same-frequency phasors in~\eqref{eq:superposition}. Coherent joint transmission retains the cross terms, so a multi-source power map is not generally the sum of per-source maps.

The REM is a system-level knowledge repository~\cite{sallouha2024rem}. It encompasses spectrum occupancy, regulatory constraints, interference metrics, and user-behavior data. The RM defined here is a proper subset of the REM, constituting its physical field layer.

\subsubsection{Intrinsic Properties}
\label{subsubsec:rm_properties}

Several field properties guide the later taxonomy. In a linear medium, complex fields obey superposition; received powers add only after cross terms vanish through distinct carriers, uncorrelated phases, or suitable time averaging, as stated in~\eqref{eq:superposition}~\cite{levie2021radiounet}. Within each homogeneous region, the regularity of $u(\mathbf r)$ follows from the governing PDE and its source and boundary data, while material interfaces can introduce derivative discontinuities~\cite{karniadakis2021physics}. Environment coupling through $k^2(\mathbf r)$ makes geometry and material maps informative in data-sparse regions~\cite{levie2021radiounet,lee2023pmnet}. Power-domain RMs can be estimated from calibrated RSS without full channel sounding~\cite{guo2013mobile,sallouha2024rem}, whereas delay-, angle-, and phase-resolved maps require richer measurements.

\subsection{Diffusion Model}
\label{subsec:diffusion_model}

RM construction under severe spatial sparsity can require a conditional distribution when the observations admit materially different fields~\cite{wang2024radiodiff,luo2025denoising}. Diffusion models provide a likelihood- or score-based alternative to generative adversarial networks and variational autoencoders~\cite{ho2020denoising,kingma2019introduction}. They decompose generation into repeated denoising evaluations and can represent several outputs compatible with the learned conditional distribution. Statistical plausibility under that distribution is not equivalent to electromagnetic validity, and sample quality remains dependent on the training distribution, conditioning information, solver, and evaluation protocol.

\subsubsection{Discrete-Time Formulation}
\label{subsubsec:ddpm}

The denoising diffusion probabilistic model (DDPM) creates a sequence of progressively noisier maps and learns to reverse that corruption~\cite{ho2020denoising}. Let $\beta_t$ control the noise added at step $t$, define $\alpha_t=1-\beta_t$, and let $\bar{\alpha}_t=\prod_{i=1}^{t}\alpha_i$. Any noisy training state can then be sampled directly:
\begin{equation}
\mathbf{x}_t = \sqrt{\bar{\alpha}_t}\,\mathbf{x}_0 + \sqrt{1 - \bar{\alpha}_t}\,\boldsymbol{\epsilon}, \quad \boldsymbol{\epsilon} \sim \mathcal{N}(\mathbf{0}, \mathbf{I}).
\label{eq:ddpm_marginal}
\end{equation}

The first term retains part of the clean RM and the second injects Gaussian noise. A neural network receives $\mathbf{x}_t$ and $t$ and predicts the injected noise. This gives the practical training objective
\begin{equation}
\mathcal{L}_{\text{DDPM}}(\theta) = \mathbb{E}_{t, \mathbf{x}_0, \boldsymbol{\epsilon}} \left[ \left\| \boldsymbol{\epsilon} - \boldsymbol{\epsilon}_\theta(\mathbf{x}_t, t) \right\|^2 \right].
\label{eq:ddpm_loss}
\end{equation}
At inference, repeated noise removal converts an initial random field into one plausible RM. Repeating the procedure with different initial noise produces alternative maps, which is useful when sparse observations do not determine a unique solution. The cost is proportional to the number of denoising evaluations.

\subsubsection{Continuous-Time SDE Generalization}
\label{subsubsec:sde}

The discrete denoising chain also has a continuous-time interpretation through a stochastic differential equation (SDE)~\cite{song2020score}. The forward dynamics gradually erase structure by adding noise, while the reverse dynamics follow a learned score that points toward regions of higher data probability. For model selection, the important consequence is not the full SDE derivation but the solver freedom it provides: the same learned dynamics can be evaluated with different numerical step sizes. Fewer steps reduce latency, whereas more steps usually improve numerical fidelity.

\subsubsection{Relevance to Wireless Channel Modeling}
\label{subsubsec:diffusion_wireless}

For wireless applications, the continuous-time view supports adaptive solvers and physics-guided drift terms~\cite{song2020score, 11278649}. The discrete denoising diffusion probabilistic model remains easier to implement and train, so it is still common in RM construction~\cite{wang2024radiodiff, 11083758, luo2024rm, zhao20253d}. A practitioner should therefore choose the formulation according to the latency budget and the need to modify the generation dynamics, rather than treating the SDE as a separate model family.

\subsection{Neural Radiance Fields}
\label{subsec:nerf}

NeRF represents a 3D scene as a continuous volumetric function parameterized by a neural network~\cite{mildenhall2021nerf}. The optical formulation maps a spatial location $\mathbf{x}=(x,y,z)$ and viewing direction $\mathbf{d}=(\theta,\phi)$ to volume density $\sigma$ and emitted radiance $\mathbf{c}=(r,g,b)$ through a multilayer perceptron (MLP), written as $F_\Theta:(\mathbf{x},\mathbf{d})\mapsto(\mathbf{c},\sigma)$.

\subsubsection{Volume Rendering}
\label{subsubsec:nerf_rendering}

NeRF answers a coordinate query by sampling points along a ray and accumulating their contributions. For ray $\mathbf{r}(t)=\mathbf{o}+t\mathbf{d}$, the continuous rendering rule is
\begin{equation}
C(\mathbf{r}) = \int_{t_n}^{t_f} T(t)\, \sigma(\mathbf{r}(t))\, \mathbf{c}(\mathbf{r}(t), \mathbf{d}) \, \mathrm{d}t,
\label{eq:nerf_rendering1}
\end{equation}
where $T(t)$ is accumulated transmittance, $\sigma$ controls the local contribution, and $\mathbf c$ is directional emission. Numerical rendering evaluates a quadrature approximation at finitely many ray samples. Additional samples can reduce integration error or resolve variation already supported by the learned field, but they do not create spatial information absent from the measurements or representation; they also increase query cost and, during training, stored activations. For coherent or phase-sensitive RF targets, the optical real-valued accumulation rule must be replaced or augmented by a complex-field rendering rule; power-only targets may instead use a declared incoherent energy-accumulation approximation~\cite{jiang2024learnable,jiang2024physics}.

\subsubsection{Positional Encoding}
\label{subsubsec:nerf_pe}

A plain multilayer perceptron tends to learn smooth coordinate functions before fine variations. Positional encoding supplies sine and cosine basis features at several frequencies and can mitigate this spectral bias. The largest encoded frequency limits the finest directly supplied basis scale, but the recoverable variation also depends on sampling, network capacity, optimization, and noise. In RF applications, the encoding bandwidth should therefore be related to wavelength and the target phase resolution rather than copied from an optical implementation~\cite{zhao2023nerf2}. Excessively high frequencies can increase optimization difficulty, whereas an insufficient basis can smooth interference fringes.

\begin{table*}[!t]
\centering
\captionsetup{font={small}, skip=10pt}
\caption{Taxonomy of input representations, condition embedding strategies, and generalization boundaries for radio map construction. The rightmost column states the missing transfer mechanism or the validation needed before a transfer claim is made.}
\vspace{-6pt}
\resizebox{0.98\linewidth}{!}{
\begin{tabular}{@{}l|l|l|l|l|l@{}}
\toprule
\textbf{Deployment Scenario} & \textbf{Env. Representation ($C_{geo}$)} & \textbf{Tx Position Encoding} & \textbf{Condition Embedding} & \textbf{Representative Methods} & \textbf{Generalization Boundary or Evidence} \\
\midrule
\multirow{2}{*}{\shortstack[l]{Single-floor indoor \\ (offices, corridors)}}
& 2D binary occupancy map & Binary mask channel & \multirow{2}{*}{\shortstack[l]{Input-level channel \\ concatenation}} & RadioUNet \cite{levie2021radiounet}, & No explicit height or \\
& $\mathbf{M}_{occ}\in\{0,1\}^{H\times W}$ & $\mathbf{S}_{bin}\in\{0,1\}^{H\times W}$ & & PMNet \cite{lee2023pmnet} & 3D transfer mechanism \\
\midrule
\multirow{2}{*}{\shortstack[l]{Sub-6\,GHz urban \\ micro/macro-cell}}
& 2.5D normalized height map & Binary mask or & \multirow{2}{*}{\shortstack[l]{Input-level channel \\ concatenation}} & \cite{jiang2024physics}, \cite{li2024intelligent}, & Material transfer requires \\
& $\mathbf{M}_{height}\in [0,1]^{H\times W}$ & Gaussian heatmap & & \cite{jiang2024learnable} & explicit $\epsilon_r,\sigma$ inputs \\
\midrule
\multirow{2}{*}{\shortstack[l]{Indoor with materials \\ and directional antennas}}
& 2D multi-channel tensor & Binary mask + & \multirow{2}{*}{\shortstack[l]{Input-level channel \\ concatenation}} & SIP2Net \cite{lu2025sip2net}, & Cross-band transfer requires \\
& (reflectance, transmittance, free-space path loss) & antenna pattern channel & & \cite{ansari2021prediction} & held-out-band validation \\
\midrule
\multirow{2}{*}{\shortstack[l]{3D indoor \\ (multi-floor, furniture)}}
& 2D floor-plan slice & Binary mask channel & Input-level concatenation & RadioResUNet \cite{pyo2022radioresunet} & \shortstack[l]{Fixed Rx height; \\ no explicit height output} \\
& 2D height-encoded pixel values & Tx height $\to$ pixel value & \shortstack[l]{Input concat.; output \\ channels as Rx heights} & R$^2$Net \cite{rao2026r2net} & \shortstack[l]{Discrete Rx heights; \\ no arbitrary 3D queries} \\
\midrule
\multirow{2}{*}{\shortstack[l]{UAV / air-to-ground \\ (3D volumetric)}}
& 3D voxel grid or & Coordinate vector & \multirow{2}{*}{\shortstack[l]{MLP embedding or \\ Transformer cross-attention}} & CAED \cite{ivanov2024deep}, & Fixed voxel resolution; \\
& point cloud $\in\mathbb{R}^{H\times W\times D}$ & $(x,y,z,f,P_t)$ & & \cite{11083758} & high memory for large scenes \\
\midrule
\multirow{2}{*}{\shortstack[l]{Dynamic environments \\ (moving vehicles)}}
& Decoupled static + dynamic & Binary mask + & \multirow{2}{*}{\shortstack[l]{Cross-attention \\ (separate prompts)}} & RadioDiff \cite{wang2024radiodiff}, & Heuristic decoupling; \\
& obstacle matrices $(\mathbf{H}_s, \mathbf{H}_d)$ & Gaussian heatmap & & RadioDiff-Flux \cite{11282987} & no scatterer-level detail \\
\midrule
\multirow{2}{*}{\shortstack[l]{Beam-aware CKM \\ (continuous beamforming)}}
& 2.5D building map + & Binary mask channel & \multirow{2}{*}{\shortstack[l]{Adaptive layer norm. \\ (adaLN) for beam $\mathbf{w}$}} & \cite{jin2025channel}, & Requires beam-disjoint and \\
& Tx position map $\mathbf{T}$ & for Tx location & & \cite{zhao2026beamckmdiff} & scene-disjoint validation \\
\midrule
\multirow{2}{*}{\shortstack[l]{Multi-band / cross-freq. \\ radio mapping}}
& Height map + ITU material & Binary mask + & \multirow{2}{*}{\shortstack[l]{Physics-derived feature \\ tensors (FSPL, gain)}} & \cite{feng2025physics}, & Transfer limited to represented \\
& tensors ($\epsilon_r(f), \sigma(f)$) & freq.\ as scalar input & & \cite{chen2024diffraction} & material and frequency regimes \\
\midrule
\multirow{2}{*}{\shortstack[l]{Universal / foundation \\ model paradigm}}
& 3D building density + & Sparse measurements & \multirow{2}{*}{\shortstack[l]{Memory-based prompt \\ learning (key-value pool)}} & \cite{jiang2025unirm}, & Depends on prompt-bank coverage \\
& height topology & + frequency + Rx height & & \cite{jaiswal2025data} & and adaptation protocol \\
\bottomrule
\end{tabular}
}
\label{tab:input_taxonomy}
\end{table*}

\subsubsection{Limitations for RF Adaptation}
\label{subsubsec:nerf_rf_limit}

The optical NeRF formulation operates on real-valued RGB intensities under ray-optics assumptions. Wireless RF signals are inherently complex-valued, involving both amplitude and phase. RF propagation at centimeter and millimeter wavelengths is dominated by multipath interference and diffraction~\cite{jiang2024physics, balanis2016antenna}. These physical disparities require substantial mathematical redesign, which is detailed in Section~\ref{subsec:nerf_rm}.

\subsection{3D Gaussian Splatting}
\label{subsec:3dgs}

While NeRF evaluates an implicit network repeatedly along every ray, 3D Gaussian splatting (3DGS) stores the scene explicitly as anisotropic Gaussian primitives~\cite{kerbl20233d}. Each primitive has a center, orientation, spatial extent, and appearance coefficients. Rendering visits only the primitives that influence a query, which reduces repeated neural-network evaluation. The number of retained primitives becomes the main storage and rendering cost.

\subsubsection{Directional Appearance}
\label{subsubsec:3dgs_sh}

View-dependent appearance in optical 3DGS is commonly parameterized by a low-order spherical-harmonic (SH) expansion~\cite{kerbl20233d}, $\mathbf c(\mathbf d)=\sum_{l=0}^{L_{\max}}\sum_{m=-l}^{l}\mathbf k_l^mY_l^m(\mathbf d)$. SH is not intrinsically unsuitable for RF; with sufficient order and complex coefficients it forms a general angular basis. The limitation is that the low-order real-valued parameterization used for optical intensity can smooth sharp multipath structure and cannot directly represent coherent phase~\cite{zhang2026rf,wen2025wrf}. RF adaptations therefore use complex coefficients or alternative angular bases, including Fourier--Legendre expansions, as discussed in Section~\ref{subsec:3dgs_rm}.

\subsubsection{Rendering}
\label{subsubsec:3dgs_render}

The covariance is projected onto the image plane as $\boldsymbol{\Sigma}' = \mathbf{J} \mathbf{W} \boldsymbol{\Sigma} \mathbf{W}^T \mathbf{J}^T$, where $\mathbf{W}$ is the world-to-camera transform and $\mathbf{J}$ is the projection Jacobian. Pixel color is computed via front-to-back alpha compositing:
\begin{equation}
C(\mathbf{x}') = \sum_{i \in \mathcal{N}} \mathbf{c}_i\, \alpha_i \prod_{j=1}^{i-1} (1 - \alpha_j),
\label{eq:3dgs_compositing}
\end{equation}
where $\alpha_i$ derives from the Gaussian footprint and an optimizable opacity. Adaptive density control clones or prunes Gaussians during training~\cite{kerbl20233d}.

\section{Problem Formulation and Taxonomy}
\label{sec:problem_formulation}

\subsection{Mathematical Abstractions and Modeling}
\label{subsec:math_abstraction}

The propagation environment occupies a continuous region of interest $\mathcal D\subset\mathbb R^3$. A grid $\mathcal G=\{g_1,\ldots,g_N\}$ contains $N$ spatial samples; its physical resolution is instead set by the grid spacing and extent~\cite{zeng2024tutorial,levie2021radiounet}. A two-dimensional raster representation can then be written as $\mathbf E\in\mathbb R^{H\times W\times C_E}$, while a volumetric representation requires an additional depth dimension.

The channel dimension $C_E$ concatenates distinct physical descriptors:
\begin{align}
C_E = C_{geo} + C_{mat} + C_{sem}.
\label{eq:channel_dim}
\end{align}
Here, $C_{geo}$ encodes spatial geometry such as obstacle locations and heights. $C_{mat}$ defines electromagnetic material properties, including relative permittivity $\epsilon_r$ and conductivity $\sigma$. $C_{sem}$ provides semantic indices that distinguish structures such as concrete walls, glass windows, and foliage \cite{chen2024diffraction, lu2025sip2net}. For a typical urban setup, one may set $C_{geo}=1$, $C_{mat}=2$, and $C_{sem}=1$, yielding a 4-channel input tensor.

The geometric dimensionality within $C_{geo}$ should follow the target quantity and deployment geometry~\cite{wang2024radiodiff,zeng2024tutorial}. A 2D binary occupancy map $\mathbf{M}_{occ}\in\{0,1\}^{H\times W}$ is a compact choice for single-floor or approximately flat scenes~\cite{levie2021radiounet,lee2023pmnet}. A 2.5D height map $\mathbf{M}_{height}\in[0,1]^{H\times W}$ retains one vertical value per horizontal cell and can represent rooftop blockage, but it cannot represent vertically overlapping structures~\cite{qiu2023deep,jiang2024physics,jiang2024learnable}. Multi-floor, urban-canyon, and unmanned aerial vehicle (UAV)-to-ground tasks may require a voxel grid, point cloud, mesh, or continuous 3D field when vertical interactions affect the declared output~\cite{teganya2021deep,11083758}. Geometry and materials provide inputs to a propagation model; they become complete electromagnetic boundary data only when interface conditions, sources, and material parameters are also specified~\cite{feng2025physics,chen2026radiodun}.

Table~\ref{tab:input_taxonomy} provides a taxonomy that maps deployment scenarios to environment representations, transmitter encoding strategies, condition embedding mechanisms, and generalization boundaries. This table serves as a practitioner-oriented reference and is discussed in the following subsection.

\subsection{Input Representation and Condition Embedding}
\label{subsec:input_representation}

Input representation and condition embedding determine which physical variations a model can address explicitly~\cite{zeng2024tutorial,10640063}. They do not by themselves establish generalization; that claim requires a split that holds out the relevant scene or physical variable. This subsection separates representational support from empirical transfer evidence.

\subsubsection{Environment Representation}
\label{subsubsec:env_representation}

The environment tensor $\mathbf{E}$ can be encoded at varying levels of physical fidelity. Each level presents a trade-off between representational richness and generalization scope.

The simplest encoding is a 2D binary occupancy map $\mathbf{M}_{occ}\in\{0,1\}^{H\times W}$, where each pixel marks obstacle presence. RadioUNet and PMNet use this representation with building-layout and transmitter-location channels~\cite{levie2021radiounet,lee2023pmnet}. It is lightweight and can be adequate for a declared single-floor or approximately flat target, but adequacy must be validated rather than inferred from the scene label~\cite{ansari2021prediction}. Because the representation contains no height or material value, any transfer across those variables is implicit and must be tested on held-out buildings.

A 2.5D height map retains one scalar height per horizontal cell at lower cost than a voxel grid~\cite{jiang2024physics,jiang2024learnable}. R$^2$Net reports the study-specific encoding $v_{env}=(h_{env}+0.1)/3.1$ after the study's specified height preprocessing~\cite{rao2026r2net}; these constants are not portable to heights in arbitrary physical units. The offset separates zero-height structures from the study's open-space code. Such a map can expose rooftop blockage but cannot represent vertically overlapping structures such as bridges or multiple floors.

A 3D voxel tensor $\mathbf{E}\in\mathbb{R}^{H\times W\times D\times C_E}$ preserves volumetric occupancy at a fixed resolution and is useful when vertical interactions matter~\cite{teganya2021deep,11083758}. It is not the only 3D representation: point clouds, meshes, and continuous fields trade regular memory access for adaptive spatial resolution. All volumetric choices incur greater storage or query cost than a 2D raster. RadioResUNet~\cite{pyo2022radioresunet} provides a fixed-height indoor baseline, whereas R$^2$Net encodes environment and transmitter heights in 2D and maps discrete receiver heights to output channels~\cite{rao2026r2net}. The latter predicts a set of height slices rather than an arbitrary continuous 3D field.

Beyond geometry, frequency-dependent permittivity $\epsilon_r(f)$ and conductivity $\sigma(f)$ provide a mechanism for material-aware conditioning~\cite{feng2025physics,lu2025sip2net,chen2024diffraction}. Their presence does not establish transfer unless training or the embedded physics covers the target material and band. A model that omits these variables has no explicit material mechanism, although correlations in the training distribution can still yield incidental transfer.

\subsubsection{Transmitter Position Encoding}
\label{subsubsec:tx_encoding}

The transmitter location $\mathbf{S}$ can be represented through several strategies. The most common approach uses a binary spatial mask $\mathbf{S}_{bin} \in \{0,1\}^{H\times W}$ \cite{levie2021radiounet, lee2023pmnet}. A single activated pixel marks the transmitter position. This mask is concatenated with $\mathbf{E}$ along the channel dimension to form $[\mathbf{E}; \mathbf{S}_{bin}]$. While simple, binary masks quantize the transmitter to discrete grid locations.

To enable continuous positioning, several works parameterize the transmitter as a 2D Gaussian heatmap \cite{wang2024radiodiff}:
\begin{align}
\mathbf{P}(j,k) = \exp\!\left(-\frac{(j-a)^2 + (k-b)^2}{2\sigma_s^2}\right),
\label{eq:gaussian_heatmap}
\end{align}
Here, $(a,b)$ are continuous transmitter coordinates and $\sigma_s$ is a chosen or learned spread in grid units. The representation is differentiable with respect to $(a,b)$ and is therefore one option for joint RM generation and transmitter localization~\cite{wang2025radiodiff-loc,zhen2025radiation}. A fixed $\sigma_s$ is only an encoding bandwidth: it is not a calibrated location uncertainty and does not sharpen merely because model training converges. An uncertainty interpretation requires $\sigma_s$ or a separate posterior to be learned and calibrated against held-out localization errors.

For additional transmitter parameters such as transmit power $P_t$, carrier frequency $f_c$, or antenna downtilt $\theta_{tilt}$, these scalars can be embedded via multi-layer perceptron (MLP) mappings \cite{zheng2024transformer, li2025rmtransformer}. Including such parameters as explicit inputs provides a mechanism for cross-configuration transfer. A model with frequency as an input can in principle generalize to new bands, provided the training data cover the relevant frequency-dependent material and propagation regimes; a model trained only at a fixed frequency has no explicit basis for such transfer.

\subsubsection{Condition Embedding Mechanisms}
\label{subsubsec:condition_embedding}

The mechanism by which conditions are injected into the architecture determines how effectively the model uses them. Three strategies are employed in the literature.

\textbf{Input-level channel concatenation} is the simplest approach. All condition tensors are stacked along the channel dimension and fed as a unified input \cite{levie2021radiounet, lee2023pmnet}. For example, RadioUNet concatenates the building map and transmitter mask into a 2-channel tensor. SIP2Net \cite{lu2025sip2net} extends this to a 5-channel tensor with reflectance, transmittance, distance, free-space path loss (FSPL), and antenna pattern maps. This strategy is simple and requires no additional components. However, it treats all channels homogeneously, without selective modulation based on specific physical parameters.

\textbf{Cross-attention conditioning} provides a more expressive alternative \cite{fang2025radioformer, liu2025paying}. The environment or transmitter features are encoded separately and injected via cross-attention layers. The network's intermediate features serve as queries, and the condition embeddings serve as keys and values. RadioDiff \cite{wang2024radiodiff} uses cross-attention to inject static and dynamic obstacle features as separate prompts. This mechanism accommodates conditions of different resolutions or modalities. It also allows spatially varying attention that aligns structures with local signal behavior. The computational cost scales with the number of attention layers and sequence length.

\begin{figure*}[!t]
    \centering
    \captionsetup{font={small}, skip=10pt}
    \includegraphics[width=0.85\linewidth]{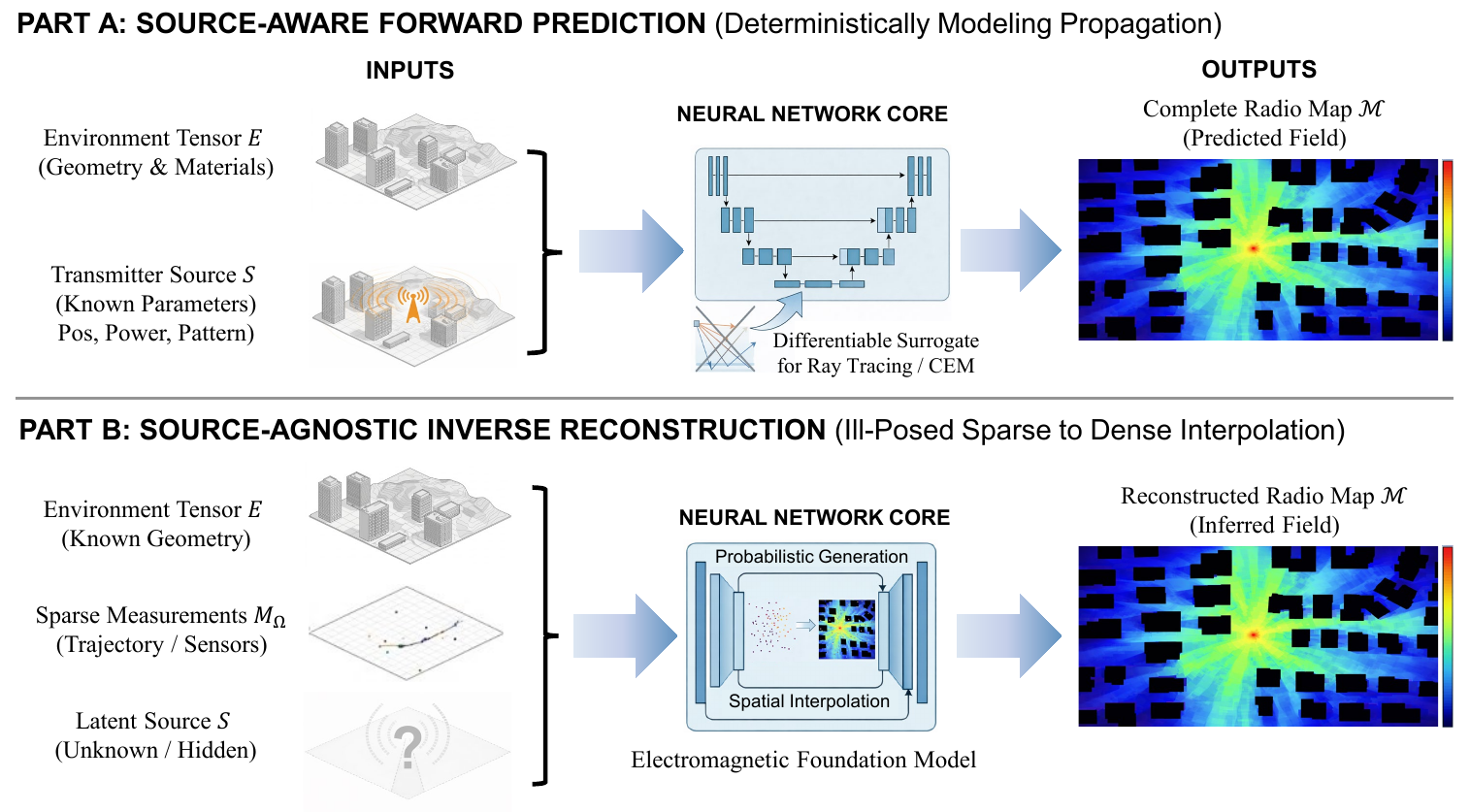}
    \caption{Two primary paradigms of neural RM construction. Part~A illustrates source-aware modeling, where a neural network acts as a surrogate under declared environment and transmitter inputs. Part~B illustrates source-agnostic reconstruction, an ill-posed inverse problem that can use deterministic regularization or probabilistic generation to infer unobserved field quantities from sparse measurements. Neither paradigm is physically valid outside the input, modality, and evaluation assumptions stated by the model.}
    \label{fig-problem}
\end{figure*}

\textbf{Adaptive layer normalization} (adaLN) offers a parameter-efficient alternative for scalar or low-dimensional conditions \cite{11278649}. It dynamically generates scale and shift parameters from the condition embedding $\mathbf{c}_{emb}$:
\begin{align}
\text{adaLN}(\mathbf{f}, \mathbf{c}_{emb}) &= (1 + \gamma) \odot \mathcal{N}(\mathbf{f}) + \beta,
\label{eq:adaln_forward} \\
[\gamma, \beta] &= \text{MLP}(\mathbf{c}_{emb}).
\label{eq:adaln_mlp}
\end{align}
The scale factor $\gamma$ adjusts feature amplitudes and the shift $\beta$ modulates the baseline level. This mechanism suits globally applied conditions such as carrier frequency or transmit power~\cite{jin2025channel,qiu2024channel}. Spatially varying geometry normally requires a spatial conditioning path or another mechanism that preserves location; a global adaLN vector alone does not encode where an obstacle occurs.

In practice, state-of-the-art frameworks combine multiple strategies \cite{jiang2025unirm, jaiswal2025data}. For instance, one may use input concatenation for the spatial environment map, a convolutional encoder for spatial features, and adaLN for continuous beam vectors.

\subsubsection{Generalization Boundaries Determined by Input Design}
\label{subsubsec:generalization_boundary}

A key insight from the above discussion is that generalization is bounded by what is explicitly represented in the input \cite{zeng2024tutorial, 10640063}. A model has no explicit mechanism or supervised signal for reliable generalization over a physical dimension absent from its input.

If the environment is represented only by a binary occupancy map, the model has no explicit height information with which to distinguish buildings of different heights. A hard-coded transmitter location supplies no direct mechanism for transfer to unseen positions, and an implicit single-band training set supplies no direct frequency signal for held-out-band transfer~\cite{feng2025physics,chen2024diffraction}. Correlated variables may still yield incidental transfer, but such behavior must be demonstrated on a split that holds out the claimed physical variable.

This observation leads to a practical design rule. Practitioners should identify the geometry, material, transmitter, frequency, antenna, and beam variables that change at deployment. Each varying quantity should be represented explicitly or handled by a stated invariant or physical model, and the claimed transfer dimension should be held out during evaluation. Table~\ref{tab:input_taxonomy} distinguishes this representational mechanism from the evidence required to support generalization.

\subsection{The Forward Problem: Source-Aware Radio Mapping}
\label{subsec:forward_problem}

Fig.~\ref{fig-problem} contrasts the two primary paradigms of 
neural RM construction: source-aware forward prediction and 
source-agnostic inverse reconstruction. 
Source-aware radio mapping is a forward prediction problem. The objective is to predict the declared RM quantity---for example path loss, received power, or a complex channel field---from an environment representation $\mathbf E$ and known input subset $\mathbf S$~\cite{levie2021radiounet,wang2024radiodiff}.

A network trained against ray-tracing, full-wave, or measured labels approximates the corresponding input--output map, not the solver's unobserved internal physics~\cite{feng2025physics,chen2026radiodun}. Accuracy can be sensitive to geometry and material perturbations because path visibility, phase, and multipath interference can change nonlinearly~\cite{chen2024diffraction}. The sensitivity is target-dependent: carrier-resolved complex fields vary at a finer spatial scale than locally averaged power or path loss.

With multiple transmitters, aggregation depends on waveform separation, receiver filtering and averaging, synchronization, and phase coherence. The following equations are two limiting regimes rather than an exhaustive dichotomy. For mutually incoherent contributions measured in the same declared output band, power addition gives
\begin{align}
\mathbf{M}(\mathbf{r}) = \sum_{k=1}^K P_k |H_k(\mathbf{r})|^2,
\label{eq:incoherent}
\end{align}
where $P_k$ is transmit power and $H_k$ is the complex channel response under a consistent normalization. For phase-synchronized cochannel contributions, coherent superposition gives
\begin{align}
\mathbf{M}(\mathbf{r}) = \left| \sum_{k=1}^K \sqrt{P_k} H_k(\mathbf{r}) e^{j\phi_k} \right|^2,
\label{eq:coherent}
\end{align}
which retains the initial phases $\phi_k$ and the resulting cross terms. Partially coherent, time-varying, or cyclostationary signals require the applicable cross-correlation or cross-spectral-density model rather than either limiting expression alone.

The choice of regime depends on the physical coherence conditions \cite{wang2022joint, oh2004mimo}. Incoherent addition applies when transmitters operate independently, as in conventional orthogonal frequency division multiple access (OFDMA) inter-cell interference \cite{lim2023interference, zhao2025imnet}. Each base station transmits asynchronously, and cross-cell signals combine on a power basis. Coherent superposition applies when transmitters share a common phase reference \cite{you2020wireless}. Examples include distributed MIMO with joint transmission and reconfigurable intelligent surface (RIS)-assisted links. Ignoring phase coherence in such scenarios underestimates the signal-to-noise ratio in constructive regions and overestimates it in destructive zones. This distinction requires architectures that handle global context and complex signal interactions across the spatial domain \cite{li2025rmtransformer, fang2025radioformer}.

\subsection{The Inverse Problem: Source-Agnostic Radio Mapping}
\label{subsec:inverse_problem}

Source-agnostic radio mapping is an inverse reconstruction problem. The transmitter parameters $\mathbf{S}$ are latent or unavailable \cite{chaves2023deeprem, teganya2021deep}. The goal is to reconstruct the complete RM $\mathbf{M}$ from the environment tensor $\mathbf{E}$ and sparse measurements $\mathbf{M}_{\Omega} = \{ (x_i, y_i, v_i) \}_{i=1}^{m}$, where $m \ll N$.

This task is an ill-posed non-linear inverse problem \cite{ha2025bayesian, wang2025radiodiff-inv}. The observed measurements $\mathbf{y}$ can be formulated as:
\begin{align}
\mathbf{y} = \mathcal{A}(\mathbf{P}_{\Omega} \odot \mathbf{M}) + \mathbf{n},
\label{eq:inverse_observation}
\end{align}
where $\mathbf{n}$ is additive measurement noise. Here, $\mathbf{P}_{\Omega}$ is a linear binary sampling mask applied via the Hadamard product $\odot$ to select $m$ locations. $\mathcal{A}(\cdot)$ is the non-linear degradation function of receiver hardware, including sensitivity thresholds, dynamic range saturation, and quantization \cite{wang2025radiodiff-inv}.

Reconstruction is generally ill-posed because the masked observation operator is non-injective: distinct fields, and potentially distinct source--environment configurations, can agree on the sampled values~\cite{teganya2021deep,ha2025bayesian}. A prior regularizes this ambiguity but does not establish uniqueness~\cite{chen2026radiodun,feng2025physics}. Angular sparsity is physically motivated when only a limited set of resolvable scattering clusters carries most energy~\cite{chen2017measurement}; it should not be assumed for rich diffuse scattering without measurement evidence.
The second commonly used prior is low rank, but its physical scope must be stated carefully~\cite{sayeed2002deconstructing}. A single-frequency power map from one transmitter, including a single-antenna received-signal-strength map, is not inherently low rank. Diffraction edges, local shadowing, and spatially varying materials can produce a slowly decaying singular-value spectrum even when the map is physically valid. Low rank has a physical basis for narrowband massive MIMO channel matrices with limited angular spread relative to the array aperture. It can also emerge after several frequency slices are stacked into a spatial-frequency tensor whose slices share propagation paths~\cite{xie2025cf}. Outside these settings, nuclear-norm or low-rank regularization is an algebraic modeling choice rather than a universal electromagnetic law. Practitioners should inspect singular-value decay on representative maps and compare against non-low-rank regularizers before adopting this assumption. The learning model must combine only validated priors with $\mathbf{E}$ to infer the latent source $\mathbf{S}$ and reconstruct unobserved shadow regions~\cite{wang2025radiodiff-loc, luo2025denoising, fang2025radioformer}.

\subsection{Environmental Constraints and Domains}
\label{subsec:env_constraints}

The propagation domain constrains the useful representation, but it does not prescribe one architecture~\cite{zeng2024tutorial,10640063}. Indoor scenes often involve penetration, reflection, and vertical interactions~\cite{lu2025sip2net,ansari2021prediction}. A floor plan can be sufficient for a fixed-height power map, whereas multi-floor or material-sensitive outputs may require height, material, point-cloud, mesh, or building-information-model inputs. Graph neural networks and 3D CNNs are candidate encoders for irregular adjacency and volumetric structure, respectively; neither is mandatory for every indoor task~\cite{chen2023graph,li2024radiogat}.

Outdoor scenes span larger areas and are shaped by line-of-sight (LoS) blockage, shadowing, and rooftop diffraction~\cite{levie2021radiounet,wang2024radiodiff}. They are commonly represented by 2.5D digital surface models and often have sparse measurements~\cite{lee2024scalable,zheng2023cell}. Global attention, multiscale convolution, and graph propagation are alternative mechanisms for modeling long-range blockage; their relative value must be tested under the same scene split and latency budget~\cite{li2025rmtransformer,fang2025radioformer,liu2025paying}.

\subsection{Target RM and Mapping Function}
\label{subsec:target_rm}

The target RM is a tensor $\mathbf{M} \in \mathbb{R}^{H \times W \times C_M}$. The channel dimension $C_M$ encodes specific metrics of the wireless channel state \cite{zeng2024tutorial, you2020wireless}. For example, $C_M = 3$ may comprise an RSS channel in dBm, a time of arrival (ToA) channel in nanoseconds, and a signal-to-interference-plus-noise ratio (SINR) channel in dB.

The mapping function parameterized by weights $\boldsymbol{\theta}$ is defined differently for each problem type. For the source-aware forward task, the network acts as a differentiable surrogate solver \cite{levie2021radiounet, wang2024radiodiff}:
\begin{align}
\hat{\mathbf M}=\mathcal F_{\boldsymbol\theta}(\mathbf E,\mathbf S_{\mathrm{known}}).
\label{eq:forward_mapping}
\end{align}
For the source-agnostic inverse task, the network acts as a spatial interpolator and latent source estimator \cite{chaves2023deeprem, teganya2021deep}:
\begin{align}
\hat{\mathbf M}=\mathcal R_{\boldsymbol\theta}(\mathbf E,\mathbf M_{\Omega}).
\label{eq:inverse_mapping}
\end{align}

\subsection{Deployment-Driven Model Selection}
\label{subsec:deployment_model_selection}

Architecture selection should follow a fixed decision order rather than the popularity of a model family. First identify the information regime: known transmitter and environment variables define a source-aware forward surrogate, whereas latent sources and sparse observations define a source-agnostic inverse problem. Next match the output topology to the representation: use a regular grid for dense raster outputs, a graph for irregular sensors or entities, and a continuous field only when arbitrary-coordinate queries or three-dimensional interactions are required. Then ask whether the observations determine one plausible map. A deterministic predictor is adequate for a nearly unique solution; a generative model is appropriate when materially different maps remain consistent with the evidence~\cite{wang2025radiodiff-inv}. Finally select the depth of physics integration from the reliability of the metadata and the available compute budget, and reject any design whose feature-generation, iterative-solver, or verification latency exceeds the deployment interval. A claimed transfer dimension, such as frequency, material, antenna, geometry, or beam, should have an explicit input, invariant, or physical mechanism and must be evaluated on a split that withholds that dimension.

\subsection{Discussion on Sampling Paradox}
\label{subsec:sampling_paradox}

A sampling paradox arises in source-agnostic inversion. Under idealized free-space propagation, three non-collinear line-of-sight (LoS) range measurements can identify a two-dimensional source location only when anchor coordinates, a valid range or path-loss relation, transmit power for RSS ranging, noise statistics, and LoS status are known~\cite{balanis2016antenna}. This localization result neither identifies a full radio field nor establishes a sufficient sample count for RM reconstruction. In practice, multipath fading and non-line-of-sight (NLoS) blockage violate the ideal ranging model~\cite{goldsmith2005wireless,chen2017measurement,wang2024radiodiff}. Additional LoS samples can become redundant under a calibrated geometric model, while samples near shadow and diffraction regions can be more informative for field recovery. Uniform or random sampling may miss these regions~\cite{chaves2023deeprem,teganya2021deep}. Physics-aware adaptive sampling should therefore be evaluated against uniform and random masks at the same budget, using diffraction corners, shadow boundaries, or calibrated predictive uncertainty as candidate acquisition scores~\cite{feng2025ipp,gao2025time,li2026uavadaptive3d,chen2024diffraction,ha2025bayesian}.

\subsection{Two Deployment Workflows}
\label{subsec:deployment_workflows}

The source-aware workflow begins with a deployment question such as evaluating candidate base-station configurations in a known district. The practitioner rasterizes the building heights and available material attributes, represents the transmitter location, power, frequency, and antenna pattern explicitly, and generates reference maps with measurements or a calibrated propagation solver. A CNN is a suitable first baseline when the grid is regular and latency requirements are strict. A Transformer becomes useful when remote obstacles create long-range dependencies, while a 3D model is justified only when vertical interactions affect the target metric. Training and validation environments must be geographically disjoint to test cross-scene generalization. Deployment then consists of one forward pass per candidate configuration, followed by a physical consistency check before the RM is used for coverage or site-selection decisions.

The source-agnostic workflow begins with sparse field measurements such as crowdsourced or drive-test RSS. Each sample must retain its coordinates, device metadata, time stamp, and calibration status. A measurement mask and the environment tensor form the model input. For moderate regular sampling, masked CNN completion provides an interpretable baseline; irregular sparse sets favor attention or graph processing; severe non-uniqueness favors a generative prior that can expose alternative plausible maps. Validation must hide complete spatial regions rather than random pixels when the intended use is extrapolation into unsurveyed streets. The final report should state the error domain, evaluation mask, inference latency, uncertainty calibration, and failure rate under missing or corrupted sensors. These steps convert architecture selection into a reproducible deployment procedure.

\subsubsection{Worked Beam-Aware Urban Example}
\label{subsubsec:worked_beam_example}

Consider a city operator that must select a beam for a newly installed massive-MIMO base station before a complete drive test is available. The operator has a digital surface model, a building-material database, the array geometry, a codebook of candidate beam vectors, and a small set of calibrated reference measurements. The target is not only a scalar received-power image. It is a beam-conditioned angular power spectrum or channel-knowledge map, because the downstream decision is beam selection at each receiver location. Recent BeamCKMDiff and RadioDiff-v2 studies illustrate this shift from map appearance to a task-oriented angular output~\cite{zhao2026beamckmdiff,wang2026radiodiffv2,huang2026map2aps}.

The first step is to define the information boundary. Building geometry, frequency, antenna pattern, and beam vector are explicit inputs, while the held-out street and the final receiver coordinates are excluded from training. If the target output is a regular raster, a conditional diffusion model is a candidate when the conditional solution is demonstrably non-unique; if arbitrary receiver coordinates or a continuous angular query are required, an RF radiance-field representation can be evaluated as a second family~\cite{zhang2026rf,yang2025graf,yang2025gsrf}. The operator should retain a lightweight CNN baseline because it exposes whether the additional representation improves the decision metric. Simulated maps can support pretraining, but an operational fidelity claim requires validation against reference measurements. The calibration residual---the declared error metric between calibrated simulator outputs and reference measurements on a calibration-only set---must be logged separately from the learned reconstruction error.

The evaluation then follows the operational sequence rather than a single image score. First, the model is tested on a scene-disjoint set of streets and on beam vectors absent from the training set. Second, angular-distribution error, top-$K$ beam-selection regret, and localization error are reported together with RMSE or NMSE. Here, top-$K$ beam-selection regret is the downstream utility loss relative to the best beam in the declared codebook after the predictor is allowed to return $K$ candidates. Third, repeated conditional samples are used to determine whether the selected beam remains stable under posterior uncertainty, i.e., variation across repeated samples from the learned conditional distribution; this variation is treated as operational confidence only after calibration on held-out scenes. Finally, the predicted map is passed through the same link-budget or ray-tracing check used for the reference data, and a coverage decision is withheld when the physical residual exceeds a declared threshold. This example shows how a generalist can move from a communications question to an input representation, a model family, and an auditable deployment test without treating a reported PSNR as the sole criterion.

A deployment protocol also needs a recovery path when one of these checks fails. A large calibration residual indicates a mismatch between the propagation solver and the reference measurements, whereas a large reconstruction residual with a small calibration residual points to the learned model or its conditioning. An unstable top-$K$ beam across posterior samples should trigger a conservative fallback, such as retaining the current beam, requesting an additional measurement, or selecting the beam with the smallest worst-case loss. A failure concentrated near a newly built structure should invalidate the cached geometry features before the network is retrained. These actions separate data, model, and environment failures instead of hiding them in one aggregate score. The log should retain the input version, beam vector, calibration residual, physical residual vector, uncertainty interval, and final decision so that a later map update is auditable.

The recovery logic also clarifies the role of recent 2026 methods. RadioDiff-v2 and Map2APS make beam selection and localization explicit downstream tasks, while RMPrior preserves a coarse propagation structure during reverse diffusion~\cite{wang2026radiodiffv2,huang2026map2aps,guo2026rmprior}. These methods should therefore be compared by decision stability and calibrated abstention as well as by image error. A model that slightly improves RMSE but changes the selected beam under small perturbations may be unsuitable for an operational controller. Conversely, a model with a modest point error can be useful if its uncertainty identifies the streets where another measurement is most valuable. This decision-oriented view is the practical bridge between a generative field model and a communications deployment.

\section{Data Acquisition and Open Datasets}
\label{sec:data_acquisition}

\begin{figure*}[!t]
    \centering
    \captionsetup{font={small}, skip=10pt}
    \includegraphics[width=0.85\linewidth]{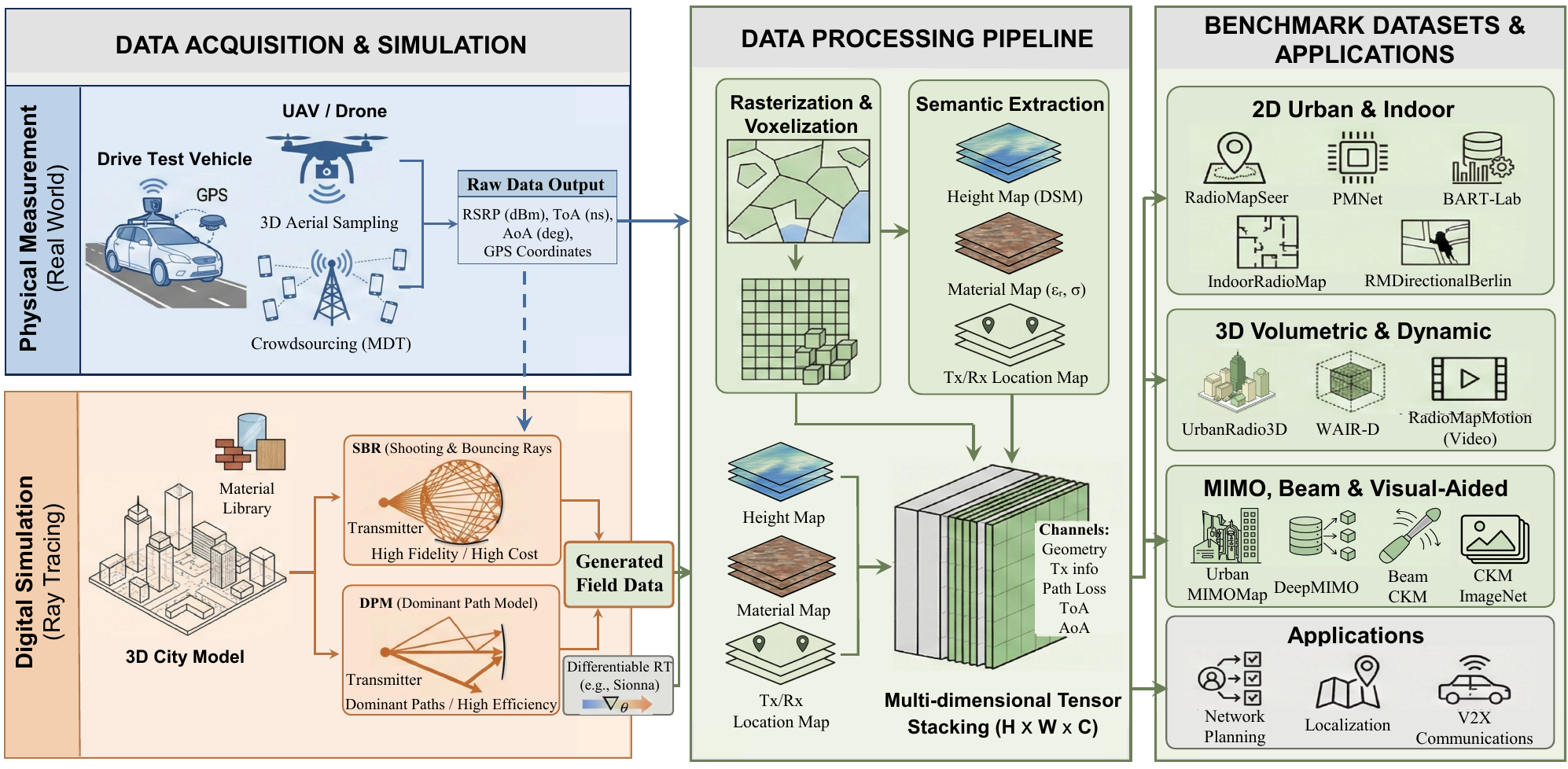}
    \caption{Illustration of the RM data collection workflow. A comprehensive workflow bridging physical measurements and ray tracing simulations to generate standardized multi-dimensional benchmark datasets.}
    \label{fig:data_collection}
\end{figure*}

As outlined in Section~\ref{sec:introduction}, this tutorial addresses 
RM construction along three intertwined dimensions: data, paradigms, 
and physics-awareness. Beginning with the data dimension, high-quality 
datasets are essential for data-driven RM construction~\cite{zeng2024tutorial, 10640063}. 
This section reviews three data acquisition paradigms: physical measurement, 
ray tracing simulation, and differentiable simulation. It then discusses 
preprocessing strategies and catalogs publicly available benchmarks. The 
section concludes with dataset limitations and selection guidance. 
Fig.~\ref{fig:data_collection} illustrates the complete workflow from 
physical measurements and ray tracing simulations to standardized 
benchmark datasets.


\subsection{Real-World Measurement Techniques}
\label{subsec:real_measurement}

Physical measurement campaigns provide reference observations for validating RM algorithms~\cite{sallouha2024rem,christopher2021characterizing}. They are not error-free ground truth: calibration, positioning, antenna response, temporal variation, noise floor, and sampling support must be reported. Compared with simulation, measurements expose stochastic channel variation and hardware impairments that a ray-optical model may omit or approximate~\cite{boban2023white}.

Traditional drive tests (DTs) and walk tests (WTs) collect geolocated signal indicators~\cite{guo2013mobile}. The classical Lee procedure estimates a local mean from multiple samples over a total route length commonly reported in the range $20\lambda$--$40\lambda$, where $\lambda$ is the carrier wavelength~\cite{lee2006estimate}. Let $W$ denote the total averaging-window length. A continuous one-dimensional idealization is
\begin{align}
P_{\mathrm{local}}(x) = \frac{1}{W} \int_{x-W/2}^{x+W/2} P_{\mathrm{inst}}(y)\, \mathrm{d}y,
\label{eq:lee_averaging}
\end{align}
where $P_{\mathrm{inst}}$ and $P_{\mathrm{local}}$ are linear-power quantities; averaging dBm values would instead compute a geometric mean in linear power. Here $W$, not $W/2$, is the total route length. The often quoted $40\lambda$ value is the conservative end of the classical range, not a universal decorrelation theorem: estimator variance also depends on sample count, spacing, fading statistics, and antenna motion. For representative 3--6~GHz carriers, $\lambda\approx5$--$10$~cm, so the $40\lambda$ upper guideline gives $W\approx2$--$4$~m; lower sub-6~GHz bands require a longer physical route~\cite{lee2012digital}. Samples within the window remain correlated unless their spacing and channel statistics justify an independence approximation.

To address sparse dedicated campaigns, commercial user equipment (UE) can contribute minimization-of-drive-tests (MDT) reports~\cite{sallouha2024rem,becvar2024machine}. Such data can improve temporal coverage but introduce device and campaign heterogeneity. In a dB-domain schematic, $P_{\mathrm{meas,dB}}=P_{\mathrm{true,dB}}+\eta_{\mathrm{sf}}+\eta_{\mathrm{device}}$, where the terms can include fading, calibration, antenna, and hardware effects~\cite{cisse2024fine}; the additive form should not be applied unchanged to linear power. Calibration should use reference devices or nodes and report residual bias rather than assuming that per-device normalization removes it. UAV sampling extends measurements into three dimensions~\cite{zeng2022uav,feng2025ipp}. For an existing noisy observation set $\mathbf Y_{\Omega}$ and candidate noisy measurements $\mathbf Y_{\mathcal T}$, target-directed GP information gain can be written as the reduction in entropy of unobserved latent field values:
\begin{align}
\mathcal I_{\bar\Omega}(\mathcal T)=H[\mathbf M_{\bar\Omega}\mid\mathbf Y_\Omega]-H[\mathbf M_{\bar\Omega}\mid\mathbf Y_\Omega,\mathbf Y_{\mathcal T}].
\label{eq:info_gain}
\end{align}
For i.i.d. Gaussian measurement noise of variance $\sigma_n^2$, a common sampled-location objective is $I(\mathbf M_{\mathcal T};\mathbf Y_{\mathcal T}\mid\mathbf Y_\Omega)=\tfrac12\log\det(\mathbf I+\sigma_n^{-2}\mathbf K_{\mathcal T,\mathcal T\mid\Omega})$~\cite{ha2025bayesian}. It is not automatically identical to~\eqref{eq:info_gain} for a finite target set $\bar\Omega$; exact target-directed selection also depends on the cross-covariance between candidate and target locations. Either objective balances conditional variance against redundancy through covariance, rather than merely selecting the largest marginal variances. A selected trajectory concentrates in shadowing or NLoS regions only when the posterior and motion constraints assign information there~\cite{gao2025time}. Li et al.~\cite{li2026uavadaptive3d} couple a 3D estimator with a diffusion-based trajectory planner, but its acquisition policy remains model- and protocol-dependent.

\subsection{Ray Tracing for Data Generation}
\label{subsec:ray_tracing}

Synthetic data generation relies on computational electromagnetics (CEM). Three algorithmic families are predominant: the shooting and bouncing rays (SBR) algorithm, the dominant path model (DPM), and differentiable ray tracing \cite{dpm, irt, hoydis2023sionna}.

\subsubsection{Classical Forward Ray Tracing}
\label{subsubsec:sbr}

SBR launches millions of ray tubes from the transmitter \cite{deschamps1972ray}. The total received field $\mathbf{E}_{total}$ is calculated by coherent superposition of all ray tubes intersecting the receiver volume:
\begin{align}
\mathbf{E}_{total} = \sum_{n=1}^{N_{ray}} \mathbf{E}_n \cdot \prod_{i=1}^{K_n} \Gamma_i(\theta_i) \cdot \prod_{j=1}^{M_n} T_j(\psi_j) \cdot \frac{e^{-j k d_n}}{d_n},
\label{eq:sbr}
\end{align}
Equation~\eqref{eq:sbr} is a schematic ray sum: $\mathbf E_n$ is a launch coefficient, $k=2\pi/\lambda$, $d_n$ is path length, and $\Gamma_i$ and $T_j$ are Fresnel coefficients. Antenna factors, diffraction coefficients, polarization transformations, and segment-dependent spreading must be included in $\mathbf E_n$ or as additional path factors when the solver models them. A naive worst-case enumeration can grow as $\mathcal O(N_{\mathrm{rays}}N_{\mathrm{obj}}^{K_{\max}})$, but visibility tests, acceleration structures, path pruning, and power thresholds make measured cost implementation- and scene-dependent~\cite{bakirtzis2024rigorous}. Reproducible dataset generation should report launched rays, interaction order, retained paths, modeled mechanisms, scene geometry, hardware, and wall-clock time.

\subsubsection{Dominant Path Model}
\label{subsubsec:dpm}

The dominant path model (DPM) builds on intelligent ray tracing with precomputed visibility relations~\cite{irt}. Its cited motivation is that a small path set accounts for most received energy in the evaluated scenarios~\cite{wolfle1998dominant}; the reported percentage is not universal across frequency, materials, antenna pattern, or environment. Retained paths can include line of sight, selected reflections, diffractions, and penetrations. With nonnegative loss terms under a declared dB sign convention, a schematic total path loss is
\begin{align}
&PL_{\mathrm{total}}=20\log_{10}\!\left(\frac{4\pi d}{\lambda}\right)
+\sum_{i=1}^{N_{\mathrm{ref}}}L_{\mathrm{ref},i}\notag\\
&\hspace{2.0em}+\sum_{j=1}^{N_{\mathrm{diff}}}L_{\mathrm{diff},j}
+\sum_{k=1}^{N_{\mathrm{pen}}}L_{\mathrm{pen},k}.
\label{eq:dpm}
\end{align}
The cited DPM implementation reports an approximately $\mathcal O(N_{\mathrm{Rx}}N_{\mathrm{obj}})$ query-stage dependence after visibility preprocessing~\cite{dpm}. This expression omits visibility-tree construction, path-order constants, memory, and hardware effects. DPM reduces cost by retaining selected interactions, whereas an SBR configuration can preserve a richer phase-resolved path set at higher cost~\cite{levie2021radiounet}; neither family has one transferable runtime or fidelity ranking without matched settings.

\subsubsection{Differentiable Ray Tracing}
\label{subsubsec:diff_rt}

Differentiable ray tracing frameworks, exemplified by Sionna \cite{hoydis2023sionna}, implement the entire propagation computation graph using automatic differentiation. This enables end-to-end gradient computation with respect to scene parameters such as material properties and transmitter configurations \cite{bakirtzis2024solving}. The significance for RM construction is twofold. First, it provides a physics-grounded data generation engine that integrates into gradient-based training loops. Second, it supports inverse problems where environmental parameters are recovered from observations \cite{hoydis2023sionna}. Several datasets in Table~\ref{tab:datasets}, notably BeamCKM \cite{wang2025beamckm}, employ Sionna with $10^9$ rays following ITU-R standards. Practitioners are encouraged to evaluate such frameworks when constructing custom datasets, especially for domain adaptation from simulation to measurements.

\begin{table*}[!t]
\centering
\captionsetup{font={small}, skip=10pt}
\caption{Comparative summary of public datasets for RM construction. ``Source'' indicates ray tracing simulation (Sim.) or physical measurement (Meas.). $\dagger$ denotes parametric generation scripts rather than fixed maps.}
\vspace{-6pt}
\resizebox{0.98\linewidth}{!}{
\begin{tabular}{@{}cl|c|l|c|c|c|l@{}}
\toprule
& \textbf{Dataset} & \textbf{Dim.} & \textbf{Modality} & \textbf{Freq.} & \textbf{Source} & \textbf{Size} & \textbf{Key Features} \\
\midrule
\multicolumn{8}{l}{\textbf{2D Urban \& Indoor Path Loss}} \\
\midrule
1 & RadioMapSeer \cite{levie2021radiounet} & 2D & Path loss  & 5.9 GHz & Sim. (DPM/IRT) & 56,080 maps & Dual-engine; 701 maps $\times$ 80 Tx; 1 m res.; foundational benchmark \\
2 & PMNet \cite{lee2023pmnet} & 2D & PL + height & Multi & Sim. (WI) & $\sim$2,000 maps & Cross-city transfer learning \\
3 & BART-Lab \cite{li2024radiogat} & 2D & PL (multi-band) & 1.75--5.75 GHz & Sim. (Feko/WP) & 2,100 maps & 5 frequency bands; coarse and fine subsets \\
4 & IndoorRadioMap \cite{indoordataset2024} & 2D & PL & Multi & Sim. (Ranplan) & 25 envs. & 0.25 m res.; material-encoded RGB; directional antennas \\
5 & DeepREM \cite{chaves2023deeprem} & 2D & PL & 2.4/LTE GHz & Sim. (WI) & 8,400 maps & Rosslyn micro-scale; inverse benchmark \\
6 & RMDirectionalBerlin \cite{jaensch2025radio} & 2D & PL + RGBIR & Sub-6 GHz & Sim. (RT) & 74,000 maps & 9.3 TB; sector antennas; vegetation separation \\
\midrule
\multicolumn{8}{l}{\textbf{3D Volumetric \& Continuous}} \\
\midrule
7 & UrbanRadio3D \cite{11083758} & 3D & PL, ToA, DoA & 5.9 GHz & Sim. (WP) & 11.2M pts & 1 m vertical res.; 20 height layers; UAV support \\
8 & RadioMap3DSeer \cite{yapar2024overview} & 3D & Path loss & 3.5 GHz & Sim. (IRT) & 56,080 maps & 3D extension of RadioMapSeer \\
9 & NeRF$^2$ Benchmarks \cite{zhao2023nerf2} & 3D & Implicit field & Multi & Meas. & Multi-scene & BLE, MIMO CSI, RFID; real measurements \\
10 & WAIR-D \cite{huangfu2022wair} & 3D & Ray params & Multi & Sim. (RT) & 10,000 envs & 40 cities; sub-6 GHz to THz; foundation model \\
\midrule
\multicolumn{8}{l}{\textbf{MIMO, Beam-Aware \& Precoding}} \\
\midrule
11 & UrbanMIMOMap \cite{qiu2024channel} & 2D & $4\!\times\!4$ CSI & 3.5 GHz & Sim. (RT) & 3,000 maps & Full complex $\mathbf{H}$; 0.5 m res. \\
12 & BeamCKM \cite{wang2025beamckm} & 2D & Beam RSRP & 3 GHz & Sim. (Sionna) & 10,000 maps & $10^9$ rays; DFT codebook; 256$\times$256 pixel \\
13 & DeepMIMO$^\dagger$ \cite{alkhateeb2019deepmimo} & Config. & Channel matrix & mmWave & Sim. (WI) & Parametric & Arbitrary arrays; 28/60 GHz; script-based \\
\bottomrule
\end{tabular}
}
\label{tab:datasets}
\end{table*}

\subsection{Data Preprocessing and Representation}
\label{subsec:preprocessing}

Raw spatial data must be transformed into a representation compatible with the selected model~\cite{zeng2024tutorial}. A grid model rasterizes vector CAD or GIS geometry, whereas a graph, point-set model, mesh, or neural field can retain an irregular or continuous interface. Beyond occupancy, feature engineering can expose selected physical variables to the learning model~\cite{lu2025sip2net,chen2024diffraction}.

Edges and wedges can generate diffracted fields when geometrical-optics rays alone do not reach the receiver~\cite{kouyoumjian1974uniform}. Sobel or Laplacian filters can expose rasterized wall boundaries through an edge map $\mathbf F_{\mathrm{edge}}$, while Harris or Shi--Tomasi operators provide a corner-candidate map $\mathbf F_{\mathrm{corner}}$~\cite{chen2024diffraction,chen2026radiodun}. These image features identify geometry; they do not compute a UTD diffraction coefficient. The coefficient also depends on wedge angle, incidence and observation directions, polarization, material/boundary model, frequency, and distance. Such maps can reduce the burden of discovering candidate interaction sites from data, but their benefit is source- and resolution-dependent.

These geometric features can be combined with material maps encoding relative permittivity $\epsilon_r$ and conductivity $\sigma$~\cite{feng2025physics}. Values assigned from ITU-R P.2040 or another catalogue remain uncertain relative to the installed material, contributing to the simulation-to-measurement gap~\cite{bakirtzis2024rigorous}. The concatenated tensor $\mathbf E\in\mathbb R^{H\times W\times C_{\mathrm{in}}}$ conditions the learning model; it is not, by itself, a complete electromagnetic boundary-value specification, which additionally requires sources, interfaces, and boundary or radiation conditions.

The number of input channels $C_{in}$ represents a design trade-off. Including more physics-derived features, such as free-space path loss (FSPL) maps, distance fields, and antenna gain projections, reduces the network's representational burden \cite{lu2025sip2net}. However, each additional channel increases preprocessing cost and may introduce approximation errors. Section~\ref{subsec:input_representation} provides detailed guidance on how input design determines generalization boundaries.

\subsection{Dataset Analysis and Benchmarks}
\label{subsec:datasets}

Numerous publicly available datasets address specific challenges in 6G environment-aware communications \cite{zeng2024tutorial, boban2023white}. We organize them by primary data modality. Table~\ref{tab:datasets} provides a comparative summary.

\subsubsection{2D Urban and Indoor Path Loss Datasets}
\label{subsubsec:2d_datasets}


\textbf{RadioMapSeer} \cite{levie2021radiounet} is the foundational benchmark for 2D path loss prediction. It comprises 701 city maps with 80 transmitter positions per map, yielding a total of 56,080 radio maps simulated using both DPM and IRT engines at 5.9~GHz with 1~m resolution. As the most widely used benchmark, most methods in Sections~\ref{sec:forward_methods} and~\ref{sec:inverse_methods} report results on this dataset. Its main limitation is the absence of vertical information and multipath parameters.

\textbf{PMNet Dataset} \cite{lee2023pmnet} features RMs from real-world urban environments including USC and downtown Boston. Generated via Wireless InSite, it explicitly includes subsets with varying building densities for cross-city transfer learning evaluation.

\textbf{BART-Lab Dataset} \cite{li2024radiogat} employs Feko and WinProp to simulate five frequency bands from 1.75 to 5.75 GHz within the same environment. A coarse subset with 2,000 maps enables rapid prototyping, while a fine subset with 100 maps supports high-fidelity reconstruction. This structure enables cross-band prediction research.

\textbf{IndoorRadioMap} \cite{indoordataset2024} was released for the First Indoor Pathloss RM Prediction Challenge \cite{bakirtzis2025first}. It includes 25 indoor environments with material properties encoded into RGB tensors at 0.25 m resolution. The simulation considers up to 8 reflections and 10 transmissions with directional antenna patterns.

\textbf{DeepREM} \cite{chaves2023deeprem} focuses on the Rosslyn urban scenario at 2.4 GHz. It serves as a standard benchmark for sparse measurement interpolation algorithms including tensor completion and kriging \cite{teganya2021deep}.

\textbf{RMDirectionalBerlin} \cite{jaensch2025radio} bridges academic assumptions and industrial reality by modeling sectorized base stations. This 9.3 TB dataset includes 74,000 RMs with normalized digital surface models separating buildings from vegetation, alongside RGBIR aerial imagery.

\subsubsection{3D Volumetric and Continuous Datasets}
\label{subsubsec:3d_datasets}

\textbf{UrbanRadio3D} \cite{11083758} contains 11.2 million labeled points across 20 height layers from 1 m to 20 m. Generated at 5.9 GHz using WinProp, it provides time of arrival (ToA) and direction of arrival (DoA) for every voxel. This makes it suitable for 3D network planning and UAV trajectory optimization \cite{zeng2022uav}.

\textbf{3DiRM3200} \cite{rao2026r2net} provides 3,200 indoor 3D RMs with paired building layouts, furniture layouts, and transmitter location maps. It is useful for evaluating height-embedded 2D architectures that estimate 3D path loss without full volumetric convolutions.

\textbf{Radio3DMix} \cite{chen2026radiogen3d} provides 50,000 synthetic 3D RMs at 2.4 GHz over 20 height layers. Unlike pure ray-tracing datasets, it uses a parametric target model calibrated by sparse UAV measurements, making it a useful benchmark for low-altitude 3D RM generation.

\textbf{NeRF$^2$ Benchmarks} \cite{zhao2023nerf2} represent a shift from discrete sampling to implicit continuous representation. Spanning BLE RSSI, MIMO CSI, and RFID spectra, they enable neural networks that encode the electromagnetic field as a continuous function. Notably, this is one of the few benchmarks grounded in physical measurements rather than simulation \cite{krijestorac2023deep}.

\textbf{WAIR-D} \cite{huangfu2022wair} provides 10,000 environments from 40 global cities, covering sub-6 GHz to terahertz frequencies. This breadth of environmental and spectral diversity supports foundation model training \cite{jaiswal2025data}.

\subsubsection{MIMO, Beam-Aware, and Precoding Datasets}
\label{subsubsec:mimo_datasets}

\textbf{UrbanMIMOMap} \cite{qiu2024channel} provides the full $4\times 4$ complex channel state information (CSI) matrix $\mathbf{H}$ at 0.5 m resolution. This enables direct computation of ergodic capacity and beamforming optimization \cite{you2020wireless}.

\textbf{BeamCKM} \cite{wang2025beamckm} covers 100 urban maps with uniform linear array antennas at 3 GHz. Sionna generates data with $10^9$ rays and ITU-R electromagnetic parameters. All DFT codebook codewords with $N_{BS}=8$ are simulated, producing beam-specific channel knowledge maps at $256\times 256$ pixel resolution with 40 dB dynamic range.

\textbf{Map2APS} provides geometry-conditioned angular power spectra for 99 urban environments and supports scene-disjoint testing of direct angular-map prediction~\cite{huang2026map2aps}. Its path-level output enables evaluation through distributional fidelity, beam-selection loss, and localization rather than only scalar path-loss error.

\textbf{DeepMIMO} is a parametric generation framework rather than a fixed dataset~\cite{alkhateeb2019deepmimo}. It generates channel matrices for user-specified array and OFDM configurations from released Wireless InSite scenario data, including commonly used 28 and 60~GHz settings. This flexibility supports custom-array studies within the available scenario, frequency, and ray-data interface; it does not by itself validate transfer to a new environment, carrier, antenna implementation, or hardware chain.

\subsubsection{Temporal and Dynamic Datasets}
\label{subsubsec:temporal_datasets}

\textbf{RadioMapMotion} \cite{zhao2023temporal} incorporates real-world vehicle trajectories to generate sequential RM frames. By formulating estimation as a video prediction task, it enables spatio-temporal architectures for predictive resource allocation in high-mobility scenarios \cite{zhao2025temporal, gao2025time}.

\subsubsection{Cross-Modal and Vision-Aided Datasets}
\label{subsubsec:crossmodal_datasets}

\textbf{CKMImageNet} \cite{zeng2024tutorial} pairs multi-resolution satellite and aerial imagery with channel parameters including angle of arrival (AoA), angle of departure (AoD), and delay spread. This facilitates vision-aided beam prediction with zero or minimal pilot overhead \cite{mkrtchyan2025vision, liaq2025visual}.

\subsubsection{Practitioner Guidance on Dataset Selection}
\label{subsubsec:dataset_guidance}

Dataset selection should follow the target modality rather than a permanent ranking. RadioMapSeer~\cite{levie2021radiounet} is useful for initial two-dimensional path-loss development because it has widely used baselines. PMNet~\cite{lee2023pmnet} and WAIR-D~\cite{huangfu2022wair} support cross-layout studies when their scene splits match the intended claim. UrbanRadio3D~\cite{11083758} supports volumetric or UAV-oriented outputs, while 3DiRM3200~\cite{rao2026r2net} represents receiver-height slices through a height-encoded two-dimensional interface. BeamCKM~\cite{wang2025beamckm} and UrbanMIMOMap~\cite{qiu2024channel} expose beam- or MIMO-related quantities, respectively. Measurement-grounded NeRF$^2$ data~\cite{zhao2023nerf2} are relevant to simulation-to-measurement studies, but no single measured dataset covers every geometry, frequency, array, or hardware condition. A valid choice must match output domain, dimensionality, carrier, antenna configuration, scene split, licensing, and calibration evidence.

\begin{figure*}[!t]
    \centering
    \captionsetup{font={small}, skip=10pt}
    \includegraphics[width=0.85\linewidth]{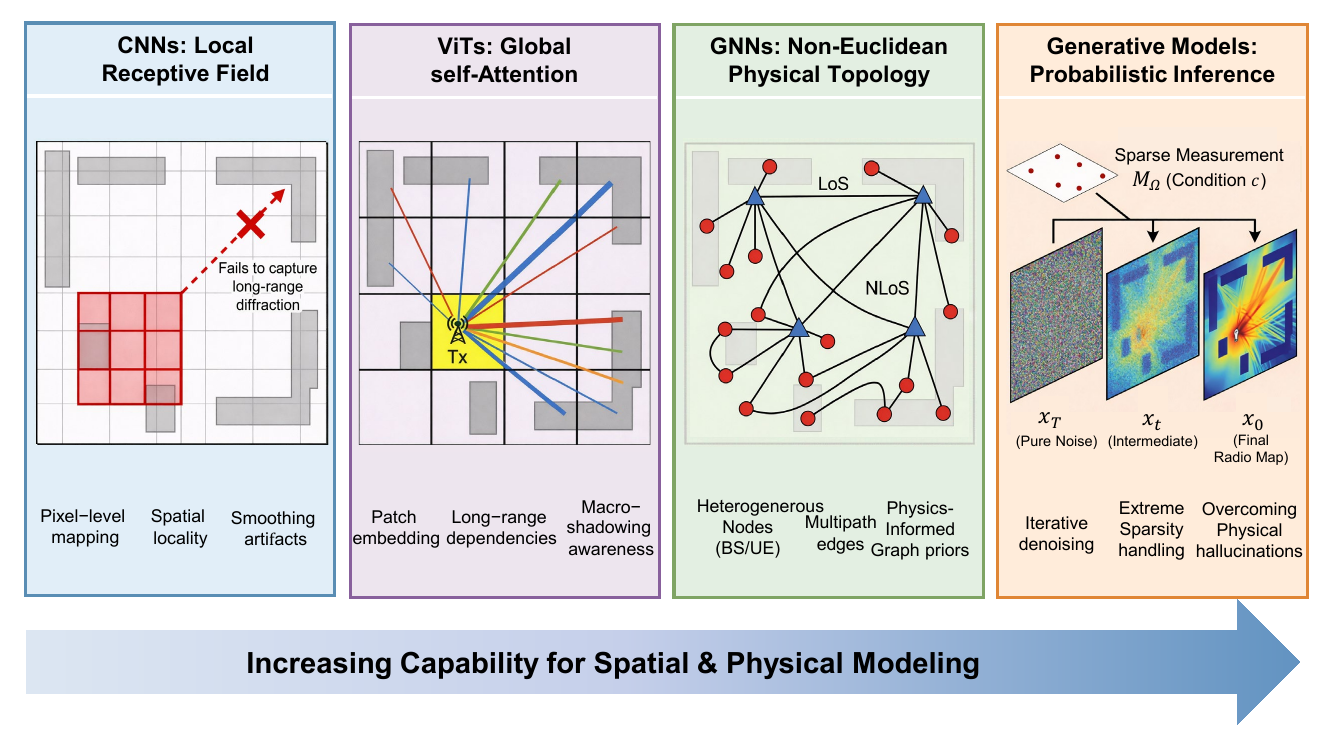}
    \caption{Conceptual progression of learning interfaces for RM construction. CNNs efficiently process regular local neighborhoods but require sufficient depth, dilation, or other mechanisms for remote dependencies. ViTs provide token-level long-range interactions at an attention cost that depends on token count. GNNs operate on an irregular interaction graph whose edges encode assumed propagation or network relations. Generative models represent a learned conditional distribution when sparse evidence admits multiple maps, but neither generation nor a finite physics penalty guarantees electromagnetic validity. The diagram summarizes modeling capabilities rather than a chronological dominance or universal accuracy ranking.}
    \label{fig-network-elvolution}
\end{figure*}

\subsection{Discussion on Dataset Limitations}
\label{subsec:dataset_limitations}

Open-source datasets have accelerated RM research, but three limitations require explicit reporting~\cite{boban2023white,10640063}. First, synthetic solvers introduce a simulation--measurement gap, and finite ray launching can add numerical sampling noise~\cite{bakirtzis2024rigorous,cisse2024fine,levie2022fast}. Second, an 8-bit export quantizes the declared clipping interval into 255 steps; a 100~dB linear mapping, for example, has a nominal step of about 0.39~dB before clipping and invalid-pixel handling. The error therefore depends on the mapping rather than on ``8 bit'' alone. A 1~m grid also cannot resolve carrier phase at centimeter wavelengths. Sub-wavelength measurements or a validated envelope model are required for carrier-resolved claims; a continuous neural field does not recover phase information absent from its data or physics~\cite{zhao2023nerf2}. Third, geometrical optics loses validity when relevant features are comparable with the wavelength~\cite{deschamps1972ray,taflove2005computational}. Statistical expansion of a single-input single-output result into a multiple-input multiple-output matrix requires a validated array and correlation model~\cite{chen2017measurement}, while homogeneous materials omit dispersive $\epsilon_r(f)$ and $\sigma(f)$~\cite{bakirtzis2024rigorous}. These limits motivate calibrated measurements, disclosed numerical precision, resolution appropriate to the target modality, and frequency-dependent material models~\cite{boban2023white,hoydis2023sionna}.

\subsection{Evaluation Metrics and Reproducible Comparison}
\label{subsec:evaluation_protocol}

A numerical error is meaningful only after the prediction domain and evaluation support are fixed. Let $m_i$ and $\hat m_i$ denote the reference and predicted values on an evaluation set $\Omega_e$, and let $n_e=|\Omega_e|$. The most common pointwise and image-structure metrics are defined as follows:
\begin{align}
&\begin{aligned}
\mathrm{RMSE}_{\Omega_e}&=\sqrt{\frac{1}{n_e}\sum_{i\in\Omega_e}(\hat m_i-m_i)^2},\\
\mathrm{NMSE}_{\Omega_e}&=\frac{\sum_{i\in\Omega_e}|\hat p_i-p_i|^2}{\sum_{i\in\Omega_e}|p_i|^2+\epsilon}.
\end{aligned}
\label{eq:rm_metrics_error}\\
&\begin{aligned}
\mathrm{PSNR}&=10\log_{10}\!\left(\frac{L^2}{\mathrm{MSE}_{\Omega_e}}\right),\\
\mathrm{SSIM}&=\frac{(2\mu_{\hat m}\mu_m+C_1)(2\sigma_{\hat m m}+C_2)}{(\mu_{\hat m}^2+\mu_m^2+C_1)(\sigma_{\hat m}^2+\sigma_m^2+C_2)}.
\end{aligned}
\label{eq:rm_metrics_structure}
\end{align}

RMSE preserves the unit of its input. RMSE on dBm or path loss therefore measures logarithmic-domain error and is reported in dB. Power ratios use $10\log_{10}$, whereas voltage, field-amplitude, or complex-channel magnitudes use $20\log_{10}$; using the wrong convention changes the dB value by a factor of two. The absolute squares in NMSE make the definition valid for either a real linear-power target or complex channel coefficients, and $\epsilon$ must have the same squared units as its denominator. A normalized-image NMSE must instead disclose its normalization and be labeled as such. PSNR depends on the declared range $L$, so different clipping intervals are not comparable. SSIM measures local contrast and structure but can remain high under an absolute power bias; it should accompany an absolute error metric.

The evaluation support is equally important. Full-grid evaluation measures overall reconstruction, a masked protocol measures only unobserved pixels, and a region-held-out protocol tests extrapolation into an unsurveyed area. Reporting an error on observed input pixels can conceal poor reconstruction elsewhere. Each result must therefore specify whether buildings and invalid pixels are excluded, whether the mask is shared across methods, and whether averaging occurs per map or over all pixels. The same train, validation, and test environments must be used for a controlled comparison.
For SSIM, $\mu_{\hat m}$ and $\mu_m$ are local means, $\sigma_{\hat m}^{2}$ and $\sigma_m^{2}$ are local variances, and $\sigma_{\hat m m}$ is the local covariance over a disclosed sliding window; $C_1$ and $C_2$ stabilize low-contrast patches. The window size, stride, and boundary handling must be reported.

Tables in the following sections summarize values reported by the original studies. They are literature-index tables, not leaderboards, because the underlying datasets, propagation solvers, normalizations, masks, hardware, and training budgets differ. A controlled benchmark must additionally report the software framework and version, numerical precision, accelerator, batch size, number of training updates, total accelerator-hours, peak allocated memory, preprocessing time, parameter count, and inference protocol. The inference protocol should state batch size, warm-up runs, timing repetitions, and whether data transfer and physical-feature generation are included. Checkpoints, configuration files, evaluation masks, and metric code are needed to reproduce the final number. A missing item should be labeled ``not reported'' rather than inferred from another implementation.
The companion Awesome Radio Map repository at \url{https://github.com/UNIC-Lab/Awesome-Radio-Map-Categorized} maintains links to public implementations for the surveyed papers and classifies the literature according to whether code is openly available. For auditability, the repository state checked for this revision is pinned to commit \href{https://github.com/UNIC-Lab/Awesome-Radio-Map-Categorized/tree/02aeeba8d35fe716732f95305059eed23f7d71c2}{\texttt{02aeeba8d35f}}; the main branch remains a living supplement, and no archival release or DOI is claimed. Fully reproducing every method without released source code, checkpoints, preprocessing pipelines, or simulator settings would require reconstructing undocumented implementation choices and exceeds the scope and page budget of this tutorial. We therefore distinguish public artifacts from unavailable ones and do not present unreleased implementations as independently reproduced evidence.

\section{Learning-Based Methods for Source-Aware Radio Mapping}
\label{sec:forward_methods}
This section reviews the principal neural architecture families for RM construction. Table~\ref{tab:arch_roadmap} compares them under one set of criteria: typical strength and limitation, compute driver, cross-scene scalability, accuracy-evidence class, uncertainty support, latency evidence, and recommended deployment. Accuracy is represented by whether evidence is matched within a study, source-reported under an unmatched protocol, or not reported; incompatible numerical scores are not merged. No family dominates all criteria~\cite{zeng2024tutorial,10640063}. The correct choice follows from the observation pattern and deployment budget, rather than from a cross-paper error number. Fig.~\ref{fig-network-elvolution} illustrates the progression from local convolutional processing to global and generative inference.


\begin{table*}[!t]
\centering
\captionsetup{font={small}, skip=8pt}
\caption{Compact cross-architecture guide for RM construction. Accuracy is recorded by evidence class rather than by incomparable cross-paper scores. \rev{Dominant costs are analytic estimates. Table~\ref{tab:within_study_resources} gives the numerical resource evidence.}}
\vspace{-4pt}
\footnotesize
\setlength{\tabcolsep}{3pt}
\begin{tabularx}{0.99\linewidth}{@{}l|>{\raggedright\arraybackslash}X|>{\raggedright\arraybackslash}p{3.0cm}|>{\hsize=1.15\hsize\linewidth=\hsize\raggedright\arraybackslash}X|>{\hsize=0.85\hsize\linewidth=\hsize\raggedright\arraybackslash}X@{}}
\toprule
\textbf{Family} & \textbf{Strength and limitation} & \textbf{Dominant cost} & \textbf{Generalization and accuracy evidence} & \textbf{Recommended use} \\
\midrule
CNN & Local grid mapping; may blur remote shadows & \rev{$HWk^2C_iC_o$ per layer. $HWC$ activation storage} & Paired scenes; mostly source-reported, unmatched metrics & Regular grids; low latency \\
Transformer & Long-range blockage; token cost grows & $T^2d$ attention and activations & Scene/token diversity; mostly source-reported, unmatched metrics & Remote geometry; sparse tokens \\
GNN & Irregular topology; edges encode assumptions & $|\mathcal{E}|d$ plus graph construction & Topology transfer; study-specific robustness evidence & Irregular sensors or links \\
GAN & One-pass conditional samples; mode coverage can be limited & Generator inference; discriminator training & Paired/cycle evidence; source-reported, unmatched metrics & Low-latency conditional generation \\
DDPM & Multiple conditional samples; iterative and calibration-dependent & $K$ denoiser passes & Sample diversity; within-study ablations; calibrate distributions & Ambiguous inverse problems \\
Turbo/flow & Generative output; fewer solver steps & $S$ evaluations, $S\ll K$ & Teacher trajectories; study-specific timing/fidelity evidence & Generative output with latency limit \\
NeRF & Continuous per-scene field; fitting limits reuse & \rev{$RQ$ field-network evaluations. $R_bQd$ feature storage} & Within-scene fit; study-specific reconstruction metrics & Fixed scene; arbitrary queries \\
3DGS & Fast post-fit rendering; RF compositing needed & \rev{Projection and sorting. Storage linear in primitive count} & Per-scene fit; study-specific rendering/channel metrics & Fixed scene; fast rendering \\
\bottomrule
\multicolumn{5}{l}{\footnotesize $H,W$: grid size; $k$: kernel width; $C$: channels; $T$: tokens; $d$: feature width; $K,S$: solver steps; \rev{$R,R_b,Q$: rays, ray batch size, and samples.}} \\
\end{tabularx}
\label{tab:arch_roadmap}
\end{table*}

\begin{table}[!t]
\centering
\captionsetup{font={small}, skip=8pt}
\caption{\rev{Published resource evidence and common-GPU profiles at the declared workloads.}}
\footnotesize
\begin{tabularx}{\columnwidth}{@{}Xcccc@{}}
\toprule
& \multicolumn{2}{c}{\textbf{Reported}} & \multicolumn{2}{c}{\rev{\textbf{Profiled}}} \\
\textbf{Method} & \rev{\textbf{ms}} & \textbf{GB} & \rev{\textbf{ms}} & \rev{\textbf{Peak GiB}} \\
\midrule
\multicolumn{5}{@{}l}{\rev{\textbf{Radio-map prediction}}} \\
RME-GAN & \rev{5.7} & 0.819 & \rev{--} & \rev{--} \\
RadioUNet & \rev{5.6} & 0.865 & \rev{$3.52\pm0.13$} & \rev{0.133} \\
\rev{PMNet} & \rev{--} & \rev{--} & \rev{$12.91\pm0.17$} & \rev{0.300} \\
\rev{SIP2Net} & \rev{--} & \rev{--} & \rev{$24.41\pm0.75$} & \rev{0.562} \\
RadioDiff & \rev{240} & 1.520 & \rev{$260.34\pm40.74$} & \rev{1.587} \\
RMDM & \rev{19,300} & 1.653 & \rev{--} & \rev{--} \\
RadioDiff-$k^2$ & \rev{760} & 3.020 & \rev{$278.46\pm45.03$} & \rev{1.587} \\
\rev{iRadioDiff} & \rev{--} & \rev{--} & \rev{$579.51\pm6.17$} & \rev{1.593} \\
\rev{PhyRMDM} & \rev{NR} & \rev{NR} & \rev{$985.52\pm4.23$} & \rev{0.829} \\
\rev{TransPathNet} & \rev{43.8} & \rev{NR} & \rev{--} & \rev{--} \\
\midrule
\multicolumn{5}{@{}l}{\rev{\textbf{Spatial-spectrum synthesis}}} \\
\rev{DCGAN} & \rev{90} & \rev{NR} & \rev{--} & \rev{--} \\
\rev{NeRF$^2$} & \rev{200} & \rev{NR} & \rev{$78.63\pm0.47$} & \rev{2.011} \\
\rev{WRF-GS} & \rev{5} & \rev{NR} & \rev{$3.68\pm0.14$} & \rev{0.065} \\
\rev{WRF-GS+} & \rev{8} & \rev{NR} & \rev{$7.36\pm0.24$} & \rev{0.657} \\
\rev{GAI-GS} & \rev{17.37} & \rev{NR} & \rev{$9.94\pm0.24$} & \rev{0.107} \\
\rev{GSRF} & \rev{4.18} & \rev{NR} & \rev{$12.54\pm1.28$} & \rev{0.747} \\
\bottomrule
\end{tabularx}
\label{tab:within_study_resources}
\label{tab:repro_resources}
\label{tab:optics_cost_axis}
\end{table}

\rev{Table~\ref{tab:within_study_resources} consolidates selected published experiments and implementation profiles by output task. The five published RM baselines follow one RTX Pro 6000 experiment~\cite{11278649}, including its source-aware RME-GAN adaptation and its RMDM baseline. PhyRMDM denotes Jia et al.'s distinct physics-aligned model~\cite{jia2025rmdm}. DCGAN, NeRF$^2$, WRF-GS, and WRF-GS+ report spectrum synthesis on an RTX 3090~\cite{wen2025neural}. GAI-GS reports model inference on an A100~\cite{shen2026gaigs}, GSRF on an RTX 3080 Ti~\cite{yang2025gsrf}, and TransPathNet reports $384\times384$ prediction on an RTX 4090~\cite{li2025transpathnet}. NR denotes a quantity unreported in the cited experiment. A dash in the published columns denotes a model outside those selected resource comparisons. In the profiled columns, it denotes an implementation not measured here. Published batch sizes, repetitions, and allocator conventions remain source-defined or unspecified.}

\rev{Training and scene construction are separate from repeated queries. RadioDiff-$k^2$ reports 30~h and 24~h for its two training stages on four GPUs, corresponding to 216 accelerator-hours~\cite{11278649}. In the A100 comparison, WRF-GS, WRF-GS+, and GAI-GS report fitting times of 312.38, 101.06, and 203.00~min, respectively. Their separately measured rendering times are 39.29, 1.43, and 0.91~ms~\cite{shen2026gaigs}. The WRF-GS inference time in that comparison is 434.21~ms. Its quality evaluation and timing subset have different sample scopes, and the combined inference/rendering boundary is unspecified.}

\rev{The profiles use PyTorch 2.10 and CUDA 12.8 on one RTX PRO 6000 Blackwell GPU with batch size 1. Computation uses BF16 with FP32 FFT, covariance decomposition, complex accumulation, and rasterization where required. Timing covers GPU-resident inference. Memory is peak allocated PyTorch memory including resident parameters, excluding driver context and allocations outside its allocator. Models use 20 warm-up calls and 64 timed calls, except PhyRMDM and NeRF$^2$ with five warm-up calls and 16 calls. RadioDiff and RadioDiff-$k^2$ each pool 128 calls across the pretrained DPM and IRT4 models. Latency is reported as the mean $\pm$ standard deviation. Other architectures are initialized for profiling. Reconstruction accuracy remains as reported in the original studies.}

\rev{Profiled RM queries use $256\times256$ inputs from eight RadioMapSeer environments or eight indoor layouts. PMNet uses its released stride-eight architecture~\cite{lee2023pmnet}, and SIP2Net uses five physical-feature channels~\cite{lu2025sip2net}. RadioDiff and RadioDiff-$k^2$ execute four denoiser evaluations, iRadioDiff ten, and PhyRMDM 50-step DDIM. Field methods synthesize $360\times90$ spectra at measured RFID poses. NeRF$^2$ uses 64 samples per ray in 8,192-ray batches. WRF-GS and GAI-GS use 20,000 initial primitives, WRF-GS+ 200,000, and GSRF 2,900 at FLE degree nine. Mapping or basis evaluation and RF rasterization are included. Physical-feature construction, input transfer, disk I/O, fitting, and verification are excluded. Post-fit primitive count and visibility affect cost. These are execution profiles at declared configurations. A quality--cost ranking additionally requires matched outputs, observations, and reconstruction error.}

\begin{table*}[!t]
\centering
\captionsetup{font={small}, skip=10pt}
\caption{Summary of representative CNN-based models for forward RM construction. Heterogeneous numerical scores are omitted because their datasets, targets, masks, and normalizations differ; the table records what each source evaluated.}
\vspace{-6pt}
\resizebox{0.98\linewidth}{!}{
\begin{tabular}{@{}l|l|l|l|l|l@{}}
\toprule
\textbf{Model} & \textbf{Core Architecture} & \textbf{Input Modality \& Dim.} & \textbf{Output Dim.} & \textbf{Test Dataset} & \textbf{Evaluation Target} \\
\midrule
RadioUNet \cite{levie2021radiounet} & U-Net & $256\!\times\!256\!\times\!2$ (Map, Tx) & $256\!\times\!256\!\times\!1$ & RadioMapSeer (DPM) & Normalized-map RMSE \\
PPNet \cite{qiu2023deep} & SegNet (regression) & $256\!\times\!256\!\times\!2$ (Height map, Tx) & $256\!\times\!256\!\times\!1$ & RadioMapSeer (5.9 GHz) & Normalized-map RMSE \\
UNetDCN \cite{jiang2024learnable} & UNet + deformable conv. & $H\!\times\!W\!\times\!C$ (RGB, Dist, Angle) & $H\!\times\!W\!\times\!1$ & Aerial + nDSM & Path-loss RMSE in dB \\
PMNet \cite{lee2023pmnet} & ResNet + atrous conv. + SPP & $256\!\times\!256\!\times\!2$ (Bldg, Tx) & $256\!\times\!256\!\times\!1$ & USC/Boston RT & Gray-scale normalized RMSE \\
SIP2Net \cite{lu2025sip2net} & U-Net + asym. conv. + ASPP & $256\!\times\!256\!\times\!5$ (multi-prior) & $256\!\times\!256\!\times\!1$ & SPGC 2025 indoor & Weighted RMSE in dB \\
R$^2$Net \cite{rao2026r2net} & 2D residual network & $256\!\times\!256\!\times\!3$ (height enc.) & $256\!\times\!256\!\times\!16$ & 3DiRM3200 (3D) & Normalized 3D RMSE \\
RadioResUNet \cite{pyo2022radioresunet} & Residual U-Net & $256\!\times\!256\!\times\!2$ (Layout, Tx) & $256\!\times\!256\!\times\!1$ & Indoor RT; later 2.4 GHz study \cite{pyo2023deep} & Absolute error in dB \\
Sat-Image PL \cite{wang2024channel} & Dual-branch (CNN+MLP)+SPP & $175\!\times\!p\!\times\!5$ (RGB, FSPL, Dist) & Scalar PL & 5.9 GHz V2I (real) & Path-loss RMSE and correlation \\
Geo2ComMap \cite{lin2025geo2commap} & AG U-Net-TP & $H\!\times\!W\!\times\!5$ (multi-metric) & $H\!\times\!W\!\times\!1$ & OSM+Sionna RT & Median absolute rate error \\
LPCGMN \cite{jin2024i2i} & Laplacian Pyramid + LRDC & High-res env.\ image & $H\!\times\!W\!\times\!1$ & DPM / DPMcars & Normalized MSE \\
UAV Multimodal \cite{mkrtchyan2025fusion} & CNN + Transformer (parallel) & Image, Tx/Rx pos, freq. & Vectors (PL, DS, AoA) & AirSim + WinSite & PL, delay-spread, and angle errors \\
\bottomrule
\end{tabular}
}
\label{tab:cnn_summary}
\end{table*}

\subsection{CNN-Based Methods}
\label{subsec:cnn_forward}

CNN-based forward RM construction establishes an image-to-image translation paradigm \cite{levie2021radiounet, lee2023pmnet}. Complex propagation phenomena are represented as mappings between 2D spatial grids. Table~\ref{tab:cnn_summary} summarizes representative CNN-based models. This subsection organizes the discussion around four recurring architectural challenges rather than individual models.

\subsubsection{Encoder-Decoder Paradigm and Resolution Bottlenecks}
\label{subsubsec:encoder_decoder}

RadioUNet \cite{levie2021radiounet} introduced the U-Net encoder-decoder architecture to RM construction. It takes a two-channel input of the building map and transmitter position to predict path loss maps. RadioUNet also proposed transfer learning from coarse simulation to sparse measurements, reducing the need for large empirical datasets \cite{levie2022fast}.

A key tension in this paradigm is spatial resolution versus receptive field and feature abstraction. Pooling or strided convolution reduces the resolution of the main encoder path~\cite{unet}. If high-frequency information is not retained in a skip path or another representation, it cannot be reconstructed uniquely from the downsampled activations. Skip connections therefore preserve selected pre-downsampling features, but boundary fidelity still depends on the stored features, alias control, decoder, loss, and target resolution.


PPNet~\cite{qiu2023deep} addresses this bottleneck through an architectural design inherited from SegNet. Pooling layers store maximum-activation indices for reuse during decoding; the indices preserve pooling locations but not discarded activation magnitudes. A two-phase strategy first learns binary building layouts and then fine-tunes with height-encoded inputs at a reduced learning rate. The cited experiments report improved boundary accuracy, while the mechanism does not guarantee recovery of all information removed by downsampling.

\subsubsection{Receptive Field Expansion}
\label{subsubsec:receptive_field}

Standard $3\!\times\!3$ convolutions have a fixed receptive field that grows only linearly with depth. This locality is mismatched with wave propagation physics \cite{levie2021radiounet, li2025rmtransformer}. A distant building can create a deep shadow zone hundreds of meters away. Capturing such long-range dependencies has motivated several innovations.

UNetDCN \cite{jiang2024learnable} integrates deformable convolutions that learn positional offsets for each sampling point. The receptive field expands non-uniformly to selectively attend to distant diffraction edges. A secondary capability is that the trained network can serve as a differentiable surrogate for antenna parameter optimization via backpropagation. However, gradient-based inversion through deep networks is susceptible to exploding gradients and local optima.

\begin{table*}[!t]
\centering
\captionsetup{font={small}, skip=10pt}
\caption{Summary of ViT-based methods for source-aware RM construction. Heterogeneous numerical scores are omitted; the final column records each source's evaluation target.}
\vspace{-6pt}
\resizebox{0.98\linewidth}{!}{
\begin{tabular}{@{}l|l|l|l|l|l@{}}
\toprule
\textbf{Model} & \textbf{Core Architecture} & \textbf{Input Dimensions} & \textbf{Output Dimensions} & \textbf{Test Dataset(s)} & \textbf{Evaluation Target} \\
\midrule
RMTransformer \cite{li2025rmtransformer} & MaxViT encoder + CNN decoder & $2 \times 256 \times 256$ & $1 \times 256 \times 256$ & USC dataset & Normalized-map RMSE \\
DAT-Unet \cite{liu2025paying} & Deformable attention + U-Net & $2 \times 256 \times 256$ (sparse pts) & $1 \times 256 \times 256$ & RadioMapSeer, SpectrumNet & Masked normalized RMSE \\
TransPathNet \cite{li2025transpathnet} & TransNeXt + EMCAD decoder & $C \times 384 \times 384$ (multi-prior) & $1 \times 384 \times 384$ & ICASSP 2025 challenge & Path-loss RMSE in dB \\
DINOv2-ViT \cite{mkrtchyan2025vision} & Pre-trained DINOv2 + UPerNet & $C \times 518 \times 518$ (multi-prior) & $1 \times 518 \times 518$ & ICASSP 2025 challenge & Weighted RMSE in dB \\
OSSN \cite{zheng2024transformer} & ViT + pyramid decoder & Building map + prompt & $1 \times H \times W$ & RadioMapSeer & Site-selection cost and map error \\
TxSTrans \cite{liaq2025visual} & ViT + ResNet + IFPN & $256 \times 256 \times 1$ + prompt & $256 \times 256 \times 1$ & RadioMapSeer/3DSeer & Sparse-region RMSE \\
UniRM \cite{jiang2025unirm} & UNet + self-attention + prompts & Sparse obs + 3D map + $f$ + $h$ & $W \times L$ (dense map) & Multi-scenario custom & Scenario-specific RMSE \\
\bottomrule
\end{tabular}
}
\label{tab:vit_methods}
\end{table*}

PMNet \cite{lee2023pmnet, lee2024scalable} uses atrous convolutions and spatial pyramid pooling (SPP). Atrous convolutions increase the effective receptive field by inserting gaps between kernel elements \cite{lee2024scalable}. The SPP module aggregates multi-scale features at the bottleneck. This captures building occlusion patterns at multiple spatial extents simultaneously. A transfer learning framework further enables generalization across city layouts.

SIP2Net \cite{lu2025sip2net} targets indoor scenarios with directional antennas. It combines asymmetric convolutions with atrous spatial pyramid pooling (ASPP) to reinforce linear wall edges \cite{ansari2021prediction}. Its composite objective combines numerical and adversarial terms to encourage both pointwise fidelity and realistic spatial structure; neither property is guaranteed outside the reported evaluation protocol.

\subsubsection{Dimensionality Reduction: Encoding 3D Physics in 2D Networks}
\label{subsubsec:dim_reduction}

Full 3D volumetric convolutions incur substantial computational overhead \cite{11083758}. Several methods encode 3D physics within 2D architectures through suitable tensor representations.

R$^2$Net \cite{rao2026r2net} maps height information into 2D inputs and predicts height-indexed 3D path-loss maps through residual learning. It further separates indoor and outdoor variants to capture penetration loss and diffraction loss more effectively. Discrete receiver heights are mapped to output channels. For 3DiRM3200, the result is a $256\!\times\!256\!\times\!3$ input and a $256\!\times\!256\!\times\!16$ output, avoiding 3D convolutions entirely. RadioResUNet \cite{pyo2022radioresunet} is an earlier indoor residual U-Net baseline. A later study from the same group reports 2.4~GHz ray-tracing-driven indoor radio estimation with practical validation \cite{pyo2023deep}. RadioResUNet is best treated as an indoor ray-tracing-driven CNN baseline, whereas R$^2$Net explicitly targets 3D RM estimation through height embedding.

For wide-area outdoor prediction with variable distances, the satellite-image model in~\cite{wang2024channel} uses spatial pyramid pooling to accept variable-sized inputs. Tying the crop extent to transmitter--receiver separation can avoid one fixed-size rescaling step, but it does not make the complete representation lossless: satellite resolution, cropping, registration, height omission, and material omission still bound the available information.

\subsubsection{Multi-Modal Fusion and Physical Prior Integration}
\label{subsubsec:multimodal_cnn}

Multi-modal fusion can expose geometry, material, distance, or image variables as separate inputs~\cite{lu2025sip2net,lin2025geo2commap}. This improves traceability of the information interface but does not by itself make the learned predictor explainable or causal.

Geo2ComMap \cite{lin2025geo2commap} extends CNN capabilities to multi-metric throughput evaluation. It combines attention gates with geographic databases, reducing anomalous errors at building edges. In UAV air-to-ground scenarios, a multimodal fusion network \cite{mkrtchyan2025fusion} processes visual images through a residual CNN in parallel with a Transformer branch. This captures global dependencies in time-series channel data and addresses channel uncertainty from rapid 3D movement. LPCGMN \cite{jin2024i2i} adopts a Laplacian pyramid reconstruction network. It retains residual information during downsampling and reconstructs the channel knowledge map as an image inpainting task.

\subsection{ViT-Based Methods}
\label{subsec:vit_forward}

To model long-range token interactions without relying only on convolutional depth, RM studies have adopted ViT architectures~\cite{li2025rmtransformer,liu2025paying,fang2025radioformer}. Their receptive-field advantage is accompanied by tokenization, positional-encoding, and attention-memory costs. Table~\ref{tab:vit_methods} summarizes representative methods.

RMTransformer \cite{li2025rmtransformer} integrates a multi-scale MaxViT encoder with a convolutional decoder. Block attention captures localized boundary details, while grid attention models global structural dependencies. Multi-scale features are fused into the decoder via skip connections. On the USC dataset, RMTransformer achieves an RMSE of 0.0071, approximately 62\% lower than RadioUNet \cite{levie2021radiounet} and 32\% lower than PMNet \cite{lee2023pmnet}.

When spatial observations are extremely sparse, the challenge intensifies. DAT-Unet \cite{liu2025paying} embeds deformable attention within a U-Net framework. It dynamically adjusts attention receptive fields by learning positional offsets. These offsets focus on structurally relevant regions such as building corners and known observation points. With only 50 sampling points, DAT-Unet reduces the RMSE by approximately 12\% compared to RadioUNet \cite{levie2021radiounet}.

Indoor propagation introduces further complexity. TransPathNet \cite{li2025transpathnet} uses a two-stage coarse-to-fine residual learning framework. A TransNeXt encoder predicts a coarse global map, which is then refined with physical priors in the second stage. \rev{Its ablation illustrates why added stages require independent testing. The full pipeline improves Kaggle RMSE from 9.93 to 9.73~dB, while full-test RMSE changes from 10.327 to 10.397~dB~\cite{li2025transpathnet}. The coarse-only model still achieves the lower error on the full test set.}

DINOv2-ViT \cite{mkrtchyan2025vision} uses large-scale foundation-model pre-training for cross-condition generalization. It employs a DINOv2 vision transformer as the encoder and a UPerNet convolutional decoder for reconstruction. The large-scale self-supervised pre-training of DINOv2 provides rich embeddings learned from diverse visual structures. Cross-condition evaluation across unseen buildings, frequencies, and antenna configurations demonstrates the robustness of this transfer learning approach.

\begin{table*}[!t]
\centering
\captionsetup{font={small}, skip=10pt}
\caption{Comparison of GAN-based methods for RM construction. ``Training Mode'' indicates the data pairing requirement. Source-reported results are descriptive because the evaluation protocols differ.}
\vspace{-6pt}
\resizebox{0.98\linewidth}{!}{
\begin{tabular}{@{}l|c|l|l|c|l@{}}
\toprule
\textbf{Method} & \textbf{Training Mode} & \textbf{Generator / Discriminator} & \textbf{Key Innovation} & \textbf{Phys. Prior} & \textbf{Best Suited Scenario} \\
\midrule
RME-GAN \cite{zhang2023rme} & Paired & 18-layer U-Net / 4-layer CNN & Physical template + micro-correction & \checkmark & Non-uniform sampling \\
FPTC-GANs \cite{wang2025two} & Paired & RMP-GAN + RMC-GAN (sequential) & First-predict-then-correct & \checkmark & Dynamic vehicular interference \\
RadioGen3D \cite{chen2026radiogen3d} & Paired synthetic & 3D U-Net / 3D PatchGAN & Radio3DMix + 3D cGAN & \ding{55} & Low-altitude 3D RM \\
CycleGAN \cite{roger2023deep} & Unpaired & Dual ResNet-9 / PatchGAN & Cycle-consistency; no paired GT & \ding{55} & No paired GT available \\
TiRE-GAN \cite{zhou2025tire} & Paired + task-driven & ResNet gen. / PatchGAN disc. & Task-incentivized outage loss & \checkmark & Spectrum management \\
\bottomrule
\end{tabular}
}
\label{tab:gan_methods}
\end{table*}

RM construction ultimately serves downstream deployment tasks. OSSN~\cite{zheng2024transformer} couples RM estimation and base-station site selection through cross-attention between building tokens and candidate-location embeddings. The source reports that its distilled recommendation stage changes its stated candidate-scoring dependence from $\mathcal O(NM)$ to $\mathcal O(M)$; feature extraction, training, constants, and memory are outside that expression. UniRM~\cite{jiang2025unirm} uses memory-based prompt learning for the multi-band and three-dimensional regimes evaluated by its source~\cite{jaiswal2025data}; ``universal'' is the framework's stated scope, not proof of transfer to an arbitrary environment or modality.

\subsection{GNN-Based Methods}
\label{subsec:gnn_forward}

GNNs are useful when sensors, links, or propagation hypotheses are represented by an irregular interaction graph~\cite{li2024radiogat,chen2023graph,perdomo2025wirelessnet}. Spatial and spectral elements become nodes, while edges encode selected physical or network relations. The graph, rather than electromagnetic space itself, is non-Euclidean; an incorrect edge rule can therefore introduce bias even though message passing accommodates heterogeneous entities and non-uniform sampling.

RadioGAT \cite{li2024radiogat} addresses extreme spatial-spectral sparsity through a model-driven graph construction mechanism. It formulates a radio depth map that encodes obstruction proportions, distance fading, and frequency fading into a unified depth feature:
\begin{align}
D^{(k)}_{b,n,m} = T_{b,n,m}\big(C - \alpha\log_{10}d_{b,n,m} - \beta\log_{10}(f_k)\big),
\label{eq:radio_depth}
\end{align}
Here, $b$ is the block index, $n$ the grid index, $m$ the transmitter index, and $T_{b,n,m}$ is the non-building path proportion. Because logarithms require dimensionless arguments, the source-specific constant $C$ implicitly absorbs the reference distance and frequency together with other offsets. The units, normalization, and fitted $\alpha,\beta$ must therefore accompany this feature; it is not a universal path-loss law. The cited node-level masked objective supports semi-supervised training from incomplete targets~\cite{li2024radiogat}.

GNN-MDAR \cite{wen2024reconstruction} addresses graph structural noise from propagation mechanism drifts across domains. It introduces a graph structure learner guided by a variational information bottleneck objective that maximizes predictive utility while minimizing mutual information with domain-specific graph noise. The compression parameter $\beta$ is gradually increased during training to first extract propagation-relevant features before discarding domain-specific structure \cite{wen2024reconstruction}.
In plain terms, the information-bottleneck objective keeps graph features that predict the target and suppresses features that mainly identify a particular training domain. Its mutual-information term is a statistical regularizer, not an electromagnetic law.

WirelessNet \cite{perdomo2025wirelessnet} demonstrates that homogeneous graphs are inadequate for network-level modeling. It introduces a heterogeneous message passing framework with separate node types for user equipment and base stations, and distinct edge types for communication and interference signals.

\subsection{GAN-Based Methods}
\label{subsec:gan_forward}

GANs reformulate RM construction as conditional generation~\cite{zhang2023rme,chen2024act,dare2024development}. Adversarial training can encourage sharper spatial structure than a pointwise regression loss in the cited settings, but boundary fidelity, mode coverage, and condition consistency remain empirical properties. Table~\ref{tab:gan_methods} therefore compares the mechanisms and reported evidence of the principal variants rather than assigning an intrinsic sharpness advantage.


A conditional GAN presents the same condition $u$ to both generator and discriminator. One standard saturating objective is
\begin{align}
&\min_{G} \max_{D}\; \mathbb{E}_{x,u \sim p_{\mathrm{data}}}[\log D(x \mid u)]\notag \\
&\quad+ \mathbb{E}_{u \sim p_{\mathrm{data}},\,z \sim p_{z}}[\log(1-D(G(z,u)\mid u))].
\label{eq:cgan_objective}
\end{align}
The second discriminator term must retain $u$; otherwise the discriminator can judge realism without checking consistency with the environment or measurements. Practical generators often use the non-saturating alternative together with reconstruction or physics terms, so Eq.~\eqref{eq:cgan_objective} is a baseline game rather than a complete RM objective.

\subsubsection{Paired Conditional Generation}
\label{subsubsec:paired_gan}

RME-GAN uses a two-phase paired conditional strategy~\cite{zhang2023rme}. Its first phase constructs a global propagation template $z$ and compares four-direction spatial gradients. Writing the minimized cosine distance explicitly avoids confusing cosine similarity with a loss:
\begin{align}
\mathcal L_{\mathrm{Gradient}}=\sum_i\left(1-
\frac{\langle\nabla_s\tilde y(i),\nabla_s z(i)\rangle+\epsilon}
{\sqrt{\|\nabla_s\tilde y(i)\|_2^2+\epsilon}\,
 \sqrt{\|\nabla_s z(i)\|_2^2+\epsilon}}\right).
\label{eq:gradient_loss}
\end{align}
Here, $\nabla_s$ stacks finite differences in the four grid directions and $\epsilon>0$ has squared-gradient units. This stabilized cosine-distance form assigns zero distance to two flat neighborhoods, penalizes a flat/nonflat mismatch, and is minimized when nonzero gradients align. The published expression is written as a raw cosine-similarity sum; if that term is inserted with a positive coefficient into a minimized objective, its sign must be reversed or the implementation otherwise documented. The replacement above makes the intended minimized quantity explicit without claiming that it is the authors' exact code. The second phase adds geometry-aware sampling and local image losses~\cite{zhang2023rme}. The template biases global structure but does not make the generated field physically valid or restrict the discriminator to only local deviations.

\begin{table*}[!t]
\centering
\captionsetup{font={small}, skip=10pt}
\caption{Summary of diffusion-based methods for RM construction. Heterogeneous numerical highlights are omitted so that the table compares mechanisms and evidence types rather than unmatched scores.}
\vspace{-6pt}
\resizebox{0.98\linewidth}{!}{
\begin{tabular}{@{}l|l|l|l|l@{}}
\toprule
\textbf{Method} & \textbf{Target Problem} & \textbf{Core Methodology} & \textbf{Conditioning} & \textbf{Key Highlights} \\
\midrule
\multicolumn{5}{@{}l}{\textbf{Foundation \& Real-World Extreme Scenarios}} \\
\midrule
RadioDiff \cite{wang2024radiodiff} & Dynamic RM & Decoupled DM; adaptive FFT in U-Net & Tx loc., obstacles & Isolates drift/noise; sharp edges \\
RM-Gen \cite{luo2024rm} & Indoor PL & Mask-agnostic inpainting & Sparse RSS with variable masks & No mask-specific retraining \\
RRDDM \cite{zhang2025rrddm} & Vehicle trajectory RM & Dual-head U-Net; residual denoising & Trajectory RSS, PL prior & Structural residual + noise injection \\
RadioTrace \cite{yang2025radiotrace} & Unknown Tx, restricted sampling & Straight-through estimator & Sparse RSS, building map & Training-free; joint Tx estimation \\
\midrule
\multicolumn{5}{@{}l}{\textbf{Dimensional \& Network Expansion}} \\
\midrule
CGDM \cite{jin2025channel} & Ultra-grained channel fingerprint & Super-resolution; ELBO & Coarse-grained CF & Super-resolution from coarse fingerprints \\
BS-1-to-N \cite{dai2025bs} & Cross-BS CKM & Rotary PE; inter-CKM attention & Source BS CKM & Cell-free network inference \\
RadioDiff-3D \cite{11083758} & 3D volumetric RM & 3D conv.; autoregressive height & 3D map, Tx & PL + ToA + DoA output \\
BeamCKMDiff \cite{zhao2026beamckmdiff} & Beam-aware CKM & DiT; adaLN & Beam $\mathbf{w}$, env. & Zero-shot continuous beam gen. \\
\midrule
\multicolumn{5}{@{}l}{\textbf{Real-Time Acceleration Paradigms}} \\
\midrule
RadioDiff-Turbo \cite{11152929} & Latency bottleneck & SDE bound + pruning & Same as RadioDiff & Few-step distilled sampling \\
RadioDiff-Flux \cite{11282987} & Sequential scenes & Midpoint reuse; KL bound & Sequential env. matrices & Diffusion-trajectory state reuse \\
RadioFlow \cite{jia2025radioflow} & Real-time RM & CNF; flow matching (ODE) & Static/dynamic context & Single-evaluation flow matching \\
\bottomrule
\end{tabular}
}
\label{tab:diffusion_methods}
\end{table*}

FPTC-GANs \cite{wang2025two} address temporal inconsistency between stable environments and real-time vehicular interference. An RMP-GAN predicts a base map from stable parameters. An RMC-GAN then applies real-time corrections guided by sparse measurements.

RadioGen3D \cite{chen2026radiogen3d} extends adversarial RM generation to 3D low-altitude scenarios. It trains a 3D U-Net with a 3D PatchGAN discriminator on Radio3DMix, enabling one-shot volumetric inference from transmitter and/or sparse-measurement inputs together with environment maps.

\subsubsection{Unpaired and Unconditional Generation}
\label{subsubsec:unpaired_gan}

CycleGAN~\cite{zhu2017unpaired} eliminates pixel-to-pixel 
correspondence using cycle-consistency and identity losses. 
Roger~et~al.~\cite{roger2023deep} adapt this framework 
to V2X radio map reconstruction:
\begin{align}
\mathcal{L}_{cyc}(G,F) = \mathbb{E}_x[\|F(G(x))-x\|_1] + \mathbb{E}_y[\|G(F(y))-y\|_1].
\label{eq:cycle_loss}
\end{align}
RecuGAN \cite{sarkar2024recugan} further explores unconditional generation using InfoGAN principles with a Wasserstein objective and gradient penalty.

\subsubsection{Bridging to Physics-Aware Paradigms}
\label{subsubsec:bridge_physics}

TiRE-GAN \cite{zhou2025tire} bridges data-driven generation and physics-informed methods. It uses a radio depth map as a physics-embedded input channel. The generator objective includes a task-incentivized regularization term from a pre-trained outage detection network:
\begin{align}
L = L_G + \alpha L_{MSE} + \beta L_R(z, \hat{z}),
\label{eq:tire_loss}
\end{align}
where $z$ and $\hat z$ are downstream-task network outputs. This feedback encourages the generated map to preserve features used by the selected outage detector~\cite{zhou2025tire}; it neither forces a unique boundary nor constitutes an electromagnetic constraint. It foreshadows the deeper physics-integration strategies surveyed in Section~\ref{sec:physics_methods}.

\subsection{Diffusion-Based Methods}
\label{subsec:diffusion_forward}

Diffusion models establish a principled framework for capturing complex propagation distributions \cite{wang2024radiodiff, 11278649, 11282987}. Table~\ref{tab:diffusion_methods} provides a summary organized by application tier.

\subsubsection{Foundational Methods and Sparse Scenarios}
\label{subsubsec:diffusion_foundation}

RadioDiff formulates dynamic RM construction as a conditional generative task~\cite{wang2024radiodiff}. Its decoupled process separates data attenuation from noise addition and predicts corresponding reverse terms. The cited study reports improved stability under its implementation, while an adaptive fast Fourier transform module targets high-frequency obstacle features; neither property is an architecture-independent guarantee.

RM-Gen \cite{luo2024rm} models indoor path loss prediction as mask-agnostic image inpainting. RRDDM \cite{zhang2025rrddm} handles sparse vehicle-trajectory measurements through a dual-head U-Net. It injects structural residuals and Gaussian noise simultaneously:
\begin{align}
I_l = I_{l-1} + \alpha_l I^{res} + \beta_l \epsilon_{l-1}.
\label{eq:rrddm}
\end{align}
RadioTrace jointly estimates an RM and continuous transmitter coordinates in the source's unknown-location setting by using a straight-through estimator and an observation-consistency loss~\cite{yang2025radiotrace}:
\begin{align}
\mathcal{L}_{\mathrm{obs}} = \|\mathbf P_{\Omega}\odot(\hat{\mathbf{x}}_0^t-\mathbf O_{\Omega})\|_F^2.
\label{eq:radiotrace_loss}
\end{align}
Here, $\mathbf P_{\Omega}$ is a binary observation mask and $\mathbf O_{\Omega}$ stores the calibrated observations on that mask. This notation separates the observation mask from the RM symbol $\mathbf M$. The loss constrains measured pixels but does not by itself identify a unique transmitter or guarantee accuracy in unobserved regions.

\subsubsection{Dimensional and Architectural Expansion}
\label{subsubsec:diffusion_expansion}

CGDM \cite{jin2025channel} models ultra-grained channel fingerprint construction as image super-resolution, maximizing the evidence lower bound (ELBO) conditioned on coarse-grained data. BS-1-to-N \cite{dai2025bs} employs rotary position embedding and inter-CKM attention for cell-free network inference. RadioDiff-3D \cite{11083758} transitions to fully 3D tensor generation using autoregressive height-wise generation. BeamCKMDiff \cite{zhao2026beamckmdiff} shifts to a diffusion Transformer backbone. Continuous beam vectors $\mathbf{w}$ are injected via adaptive layer normalization for zero-shot generalization to unseen beam directions.
The ELBO is a tractable training objective that rewards agreement with the observations while regularizing a latent distribution; it is not an additional radio-map quality metric.

\begin{table*}[!t]
\centering
\captionsetup{font={small}, skip=10pt}
\caption{Scenario-oriented guide for source-agnostic RM reconstruction. Observation regimes are qualitative and source-specific; no universal sampling-rate threshold is implied.}
\vspace{-6pt}
\resizebox{0.98\linewidth}{!}{
\begin{tabular}{@{}l|c|c|c|l|l@{}}
\toprule
\textbf{Scenario} & \textbf{Observation Regime} & \textbf{Tx Location} & \textbf{Env. Prior} & \textbf{Recommended Paradigm} & \textbf{Representative Methods} \\
\midrule
Dense grid, known Tx & Relatively dense regular mask & Known & Building map & CNN masked completion & Masked AE \cite{teganya2021deep}, RobUNet \cite{shao2024robunet} \\
Sparse random, known Tx & Sparse random mask & Known & Building map & ViT / CNN+GAN & RadioFormer \cite{fang2025radioformer}, DeepREM \cite{chaves2023deeprem} \\
Extremely sparse point set & Few irregular observations & Known & Building map & ViT (equivariant) / Diffusion & STORM \cite{viet2025spatial}, \cite{luo2025denoising} \\
Trajectory-based sparse & Sparse trajectory & Known & Building map & Diffusion (residual) & RRDDM \cite{zhang2025rrddm} \\
Unknown Tx, sparse obs. & Sparse points; unknown source & \textbf{Unknown} & Building map & Diffusion (joint Tx estimation) & RadioTrace \cite{yang2025radiotrace} \\
No ground samples & No local field samples & Known & Satellite imagery & GAN (satellite-driven) & SC-GAN \cite{pan2025sc} \\
Multi-band tensor & Sparse observations per band & Known & Partial bands & CNN cross-band / Tensor & CF-CGN \cite{xie2025cf}, ROASMP \cite{zhou2023accurate} \\
3D volumetric sparse & Sparse volumetric samples & Known & 3D model & Latent diffusion / large generative model & \cite{11083758,liu2026radiolam} \\
Quantized wide-area observations & Sparse, noisy, low-bit samples & Latent/not explicit & Learned map prior & Gradient-free posterior diffusion & GSC~\cite{gu2025gsc} \\
\bottomrule
\end{tabular}
}
\label{tab:inverse_decision}
\end{table*}

\subsubsection{Real-Time Acceleration}
\label{subsubsec:diffusion_accel}

Iterative denoising causes high inference latency, which is a bottleneck for real-time applications \cite{11152929, 11282987}. Three acceleration paradigms form a trade-off chain.

RadioDiff-Turbo applies structure-group pruning and modifies the solver interval~\cite{11152929}. Its source reports a reduction from 1000 to 10 network evaluations and approximately 60~ms inference without retraining. These are network- and hardware-specific values; an end-to-end deployment must additionally time conditioning, transfers, and verification.

RadioDiff-Flux reuses intermediate states for scenes that share static structure~\cite{11282987}. Under the source's normalized forward corruption with $t\in(0,1]$, let $p_t=\mathcal N((1-t)z_i,t\mathbf I)$ and $q_t=\mathcal N((1-t)z_j,t\mathbf I)$. Their equal-covariance Gaussian divergence is
\begin{align}
D_{\mathrm{KL}}(p_t \| q_t) = \frac{1}{2}\frac{(1-t)^2}{t}\|z_i - z_j\|_2^2.
\label{eq:kl_bound}
\end{align}
Here, $z_i$ and $z_j$ are variational-autoencoder latents. The factor $(1-t)^2/t$ tends to zero as $t\to1$, so the specified noisy marginals become close when $\|z_i-z_j\|$ is bounded; it diverges as $t\to0$. This calculation alone does not prove that a particular state contains only static topology or that it is reusable. RadioDiff-Flux validates midpoint reuse empirically for its scene pairs and reports 3.5--58$\times$ acceleration with less than 0.15\% degradation under its metrics~\cite{11282987}. The result depends on the midpoint, scene similarity, cache cost, hardware, and timing boundary.

RadioFlow uses flow matching to learn an ordinary-differential-equation vector field $\mathrm dx/\mathrm dt=v_\theta(t,x)$~\cite{jia2025radioflow}. The cited implementation adopts a declared coupling and probability path between noise and data; a straight interpolation is not automatically the globally optimal transport path. Deterministic integration removes diffusion's per-step random injection but does not remove model or discretization error. The source reports single-step generation, 63~ms latency, and an 8.3$\times$ parameter reduction under its setup.

Outside the diffusion family, RadioVAR uses coarse-to-fine autoregressive token prediction~\cite{zhang2026radiovar}. The source calls this sampling-free relative to iterative diffusion; autoregressive decoding still performs sequential token prediction and may be deterministic or stochastic depending on the decoding rule. Its latency must therefore be reported with token count, decoding strategy, and hardware.

RMPrior explores a complementary acceleration axis by starting reverse diffusion from a noised propagation prior at an intermediate time rather than from pure Gaussian noise~\cite{guo2026rmprior}. The backbone and noise schedule remain unchanged, so the modification is inference-only. Its deployment benefit depends on the sum of prior-generation and truncated-denoising costs, and therefore requires the prior to be cheap or cacheable.

\subsubsection{Probabilistic Output and Diffusion Priors}
\label{subsubsec:diffusion_uq}

A diffusion model can produce multiple samples from its learned conditional distribution $p_\theta(\mathbf M\mid\mathbf E,\mathbf S)$~\cite{wang2024radiodiff,ha2025bayesian}. With independent random seeds giving conditionally independent samples $\{\gamma_j\}_{j=1}^{J}$, a Monte Carlo estimator of $\phi=\mathbb E_{p_\theta}[F(\gamma)]$ is
\begin{align}
\hat{\phi}_J = \frac{1}{J}\sum_{j=1}^J F(\gamma_j),
\label{eq:mc_estimate}
\end{align}
The estimator targets functionals under the learned distribution, not necessarily the true conditional law. A declared $F$ can represent capacity, outage, or bit-error calculations when all additional link parameters are supplied. A single deterministic point output cannot estimate this distributional expectation without an added ensemble or probabilistic model~\cite{levie2021radiounet,lee2023pmnet}.

Sample dispersion reflects variability represented by the learned conditional model; it should not automatically be labeled aleatoric uncertainty, and it does not quantify parameter uncertainty~\cite{ha2025bayesian}. If calibration is verified on scene-disjoint data, Monte Carlo estimates can support expected-utility or risk-aware decisions; robustness relative to a point estimate must still be measured~\cite{luo2025denoising}. Pretrained diffusion models can also act as learned priors for inverse reconstruction, but the prior regularizes rather than resolves ill-posedness: observation consistency and identifiability remain necessary~\cite{wang2025radiodiff-inv,gu2025gsc}.

\section{Learning-Based Methods for Source-Agnostic Radio Mapping}
\label{sec:inverse_methods}

Source-agnostic RM reconstruction includes latent or unavailable transmitter parameters and is generally non-injective under sparse observations~\cite{chaves2023deeprem,teganya2021deep}. Lower sampling usually enlarges the ambiguity, but sampling percentage alone does not define a transition from well-posed to ill-posed: placement, noise, scene, modality, prior, and operator rank also matter~\cite{ha2025bayesian,fang2025radioformer}. Table~\ref{tab:inverse_decision} therefore uses qualitative observation regimes rather than a universal threshold.

\subsection{CNN-Based Methods}
\label{subsec:cnn_inverse}

CNNs provide the foundational framework for source-agnostic reconstruction, treating the task as sparse-to-dense spatial completion \cite{teganya2021deep, chaves2023deeprem}. Table~\ref{tab:cnn_inverse} summarizes representative methods. The foundational masked reconstruction objective isolates the loss to observed indices $\Omega_t$:
\begin{align}
\min_w \frac{1}{T}\sum_{t=1}^{T}\big\|\mathbf{P}_{\Omega_t}\big(\tilde{\Psi}_t - p_w(\breve{\Psi}_t)\big)\big\|_F^2.
\label{eq:cnn_baseline}
\end{align}
This observed-index objective is the simple baseline analyzed by the cited completion-autoencoder source~\cite{teganya2021deep}. By itself, it does not directly penalize errors at unavailable locations and can learn an identity mapping on observed entries. The cited method therefore supplies the observation mask to distinguish missing values and, depending on the training regime, uses complete synthetic targets or disjoint input/output subsets of measured indices. A reproducible implementation must state which indices enter the input, loss, and evaluation; these sets cannot all be denoted by the same mask without clarification.

\begin{table*}[!t]
\centering
\captionsetup{font={small}, skip=10pt}
\caption{Summary of CNN-based methods for source-agnostic RM reconstruction. Observation regimes are described qualitatively because source-reported sampling percentages use different grids, masks, and scene splits.}
\vspace{-6pt}
\resizebox{0.98\linewidth}{!}{
\begin{tabular}{@{}l|l|c|l|l|l@{}}
\toprule
\textbf{Method} & \textbf{Core Architecture} & \textbf{Observation Regime} & \textbf{Input Type} & \textbf{Key Innovation} & \textbf{Loss} \\
\midrule
\multicolumn{6}{@{}l}{Masked Completion} \\
\midrule
Masked AE \cite{teganya2021deep} & Conv. autoencoder & Variable regular masks & Sparse grid & Binary mask for unmeasured voids & Masked Frobenius \\
DeepREM \cite{chaves2023deeprem} & U-Net + Pix2Pix cGAN & Sparse random mask & Sparse random & Adversarial loss for distribution consistency & MSE + adversarial \\
RobUNet \cite{shao2024robunet} & ResNet + dual attention & Sparse grids in unseen layouts & Sparse grid (unseen layouts) & Channel \& pixel attention for boundaries & MSE \\
\midrule
\multicolumn{6}{@{}l}{Vision Task Reformulation} \\
\midrule
SRResNet \cite{wang2025deep} & Super-resolution ResNet & Coarse low-resolution grid & Low-res grid & RM as super-resolution problem & Full-grid MSE \\
Template-Perturb. \cite{zhang2024radiomap} & Exemplar inpainting & Large voids & Sparse + radio depth & Fill sequence follows Tx$\to$Rx path & Physics-prioritized \\
CF-CGN \cite{xie2025cf} & Cycle-consistent GAN & Cross-band & Band $i \to j$ & Variable-weight cycle-consistency & Cycle loss \\
\midrule
\multicolumn{6}{@{}l}{High-Dimensional Tensor Extension} \\
\midrule
CAED \cite{ivanov2024deep} & 3D conv. autoencoder & Variable volumetric masks & 3D voxel & Decoder dilation + structural pruning & 3D Frobenius \\
Multitask SR \cite{wang2023super} & Multi-channel SR & Variable & Multi-param. tensor & Homoscedastic uncertainty weighting & $L_{MTL}$ \\
ROASMP \cite{zhou2023accurate} & CP decomp. + ADMM-RLS & Streaming 4D & Space$\times$freq$\times$time & Online robust tensor completion & $\ell_1$-reg. CP \\
\bottomrule
\end{tabular}
}
\label{tab:cnn_inverse}
\end{table*}

\subsubsection{Masked Completion Paradigm}
\label{subsubsec:masked_completion}

The earliest approaches use fully convolutional autoencoders with binary observation masks \cite{teganya2021deep}. The mask distinguishes true low-power regions from unmeasured voids. Skip connections and dual-path routing prevent irreversible mixing of signal and environment channels during early convolutions \cite{vankayala2021radio}. DeepREM reports operation at a 5\% sampling setting with adversarial loss for distribution consistency; this percentage is a study-specific operating point rather than a universal sparsity limit~\cite{chaves2023deeprem}. RobUNet \cite{shao2024robunet} advances generalization to unseen building layouts through dual channel and pixel attention.

\subsubsection{Vision Task Reformulation}
\label{subsubsec:vision_reformulation}

SRResNet \cite{wang2025deep} redefines RM reconstruction as super-resolution, recovering continuous maps from one-sixteenth sample density. The template-perturbation method \cite{zhang2024radiomap} bridges exemplar-based inpainting with radio depth maps, aligning the filling sequence with transmitter-to-receiver signal paths. For cross-band extrapolation, CF-CGN \cite{xie2025cf} employs variable-weight cycle-consistent generative networks:
\begin{align}
L_{cyc} = \mathbb{E}\|\Psi_{i,j}(\Psi_{j,i}(I_j)) - I_j\|_F^2 + \mathbb{E}\|\Psi_{j,i}(\Psi_{i,j}(I_i)) - I_i\|_F^2,
\label{eq:cycle_loss_inverse}
\end{align}
This penalty encourages approximate cycle consistency on the training distributions; it does not prove that either learned mapping is bijective or physically valid. Its value for cross-band precoding must be established on held-out bands with the relevant material, array, and channel targets.

\subsubsection{High-Dimensional Tensor Extension}
\label{subsubsec:tensor_extension}

CAED \cite{ivanov2024deep} extends the baseline to a 3D Frobenius norm for voxel environments with decoder dilation and structural pruning. A multitask super-resolution framework \cite{wang2023super} recovers path loss, delay spread, angular spread, and LoS status simultaneously. Homoscedastic uncertainty provides dynamic weight adaptation:
\begin{align}
L_{MTL}(W, \sigma) = \sum_m \frac{L_m}{2\sigma_m^2} + \sum_m \log(\sigma_m).
\label{eq:mtl_loss}
\end{align}
The probabilistic derivation requires $\sigma_m>0$ for every task. Implementations should therefore optimize an unconstrained log-scale and map it to a positive $\sigma_m$; otherwise the logarithm and inverse variance are undefined. These weights reflect learned task scales rather than calibrated predictive uncertainty.

ROASMP \cite{zhou2023accurate} formulates RM prediction as online robust 4D tensor completion spanning space, frequency, and time. It uses temporal pre-filling with an attention-based forgetting factor, followed by CP decomposition with $\ell_1$-norm regularization to reject outliers. Online optimization via ADMM with recursive least squares enables dynamic tracking of low-rank subspaces across streaming measurements \cite{zhou2023accurate}. For practical deployment, cross-AP inference paradigms \cite{dai2025generating} derive target RMs from neighboring AP maps, while MoENet \cite{jaiswal2025leveraging} fuses transfer learning with uncertainty-aware mixture of experts for data-scarce scenarios. NAS-REC \cite{he2025radio} further targets UAV communications by combining neural architecture search with deep regularization completion for outage map reconstruction without pre-training.
CP decomposition represents a tensor as a sum of rank-one outer products, whereas ADMM alternates between simpler constrained subproblems. These operations impose an algebraic prior and do not by themselves guarantee electromagnetic validity.

\subsection{ViT-Based Methods}
\label{subsec:vit_inverse}

Set-oriented Transformer variants offer a structural advantage for inverse problems because they can process sparse observations as coordinate-aware tokens with masks for variable cardinality~\cite{fang2025radioformer,hehn2023transformer}. A grid CNN instead rasterizes observations and requires a validity mask; zero filling without a mask would confound missing samples with physical zeros~\cite{liu2025paying}. The advantage therefore follows from the input interface, not from the Transformer label alone.

The Transformer structure restoration framework uses alternating self-attention and axial attention~\cite{chen2024radio}. For $n$ tokens arranged on an approximately square grid, axial attention reduces an all-pairs $\mathcal O(n^2)$ interaction to $\mathcal O(n^{3/2})$; the expression changes for nonsquare grids or sparse point sets. A high-receptive-field feature loss addresses structural smoothing using a pretrained extractor:
\begin{align}
L_{\mathrm{hrf}}=\mathbb E\!\left[\|\phi_{\mathrm{hrf}}(R)-\phi_{\mathrm{hrf}}(\tilde R)\|_2^2\right],
\label{eq:hrf_loss}
\end{align}
Here, $\phi_{hrf}$ is the structural feature extractor and the norm accounts for its vector or tensor output. The total objective $L_{\mathrm{final}}=\lambda_{L2}L_2+\lambda_{hrf}L_{hrf}$ encourages feature similarity; it does not ensure absolute power accuracy or electromagnetic consistency.

RadioFormer decouples environment tokens from sparse-observation tokens and couples them with cross-attention~\cite{fang2025radioformer}. Under the standard row-token convention, let $F_O\in\mathbb R^{N_O\times d_O}$ and $F_B\in\mathbb R^{N_B\times d_B}$; then
\begin{align}
Q = F_O W_Q,\quad K = F_B W_K,\quad V = F_B W_V,
\label{eq:cca_qkv} \\
\operatorname{CA}(Q,K,V)=\operatorname{Softmax}\!\left(\frac{QK^\top}{\sqrt d}\right)V.
\label{eq:cca}
\end{align}
with $W_Q\in\mathbb R^{d_O\times d}$ and $W_K,W_V\in\mathbb R^{d_B\times d}$. Thus $QK^\top\in\mathbb R^{N_O\times N_B}$ scores each observation token against building tokens. Writing $W_QF_O$ together with $QK^\top$ would instead be dimensionally inconsistent under this convention. Attention expresses a learned association with geometry, not a physical propagation operator.

STORM \cite{viet2025spatial} defines the task as continuous point-set function regression, eliminating grid resolution bottlenecks. It processes variable-sized sample sets and embeds physical equivariance for translation and rotation. A weighted direction vector determined by the strongest signal measurements establishes a canonical pose:
\begin{align}
v(r) = \sum_{n=1}^{N} \exp(\tilde{\gamma}_n)(r_n - r),
\label{eq:storm_direction}
\end{align}
where $\tilde{\gamma}_n$ is the normalized RSS at the $n$-th location. Physical equivariance is hard-coded by aligning coordinates with this dominant propagation direction:
\begin{align}
x_n = [\tilde{\gamma}_n,\; (U(r_n - r))^\top]^\top,
\label{eq:storm_feature}
\end{align}
Here, $U$ aligns a nonzero $v(r)$ with a canonical axis. Query-relative coordinates provide translation equivariance, and the cited canonicalization targets rotation equivariance~\cite{viet2025spatial}. The claim requires consistent transformation of the environment, transmitter, receiver coordinates, and output, and it becomes ambiguous when $v(r)=0$ or when symmetric configurations admit multiple canonical orientations. The construction can reduce augmentation for the modeled symmetries but does not eliminate the need to test these degeneracies. STORM also uses causal attention for joint estimation and next-best-point selection.

\subsection{GAN-Based Methods}
\label{subsec:gan_inverse}

Inverse GAN methods have shifted from point-by-point regression to cell-level global generation \cite{zheng2023cell, chen2024act}. The conditional GAN maps an environment $x$ to an RSRP distribution $y$. Physical priors from empirical models are integrated through residual estimation: $y = S_{\text{true}} - \hat{S}_{\text{empirical}} \odot \text{Mask}$. This restricts the generator to learning high-frequency residual shadows \cite{zheng2023cell}.

ACT-GAN \cite{chen2024act} extracts multi-scale contexts via aggregated contextual transformation blocks with varying dilation rates. Convolutional block attention modules enforce spatial attention on dense building edges. A composite loss combines MSE, perceptual, style, and adversarial terms \cite{chen2024act}.

SC-GAN \cite{pan2025sc} bypasses ground-level measurements by using satellite imagery. The input is a dual-channel tensor of a building map and transmitter map. No intermediate RSS measurements are required. Commercial satellite imagery provides 0.3--1 m resolution, meeting typical RM grid requirements. However, satellite imagery lacks building-height information, electromagnetic material properties, and up-to-date scene information. These limitations confine SC-GAN to macroscopic coverage estimation where building footprint geometry dominates \cite{pan2025sc}.

\begin{table*}[!t]
\centering
\captionsetup{font={small}, skip=10pt}
\caption{Capability comparison for source-agnostic RM reconstruction. The table compares observation interfaces and supported tasks, not source-reported accuracy or sampling thresholds.}
\vspace{-6pt}
\footnotesize
\setlength{\tabcolsep}{3pt}
\begin{tabularx}{0.99\linewidth}{@{}>{\raggedright\arraybackslash}p{2.7cm}|>{\raggedright\arraybackslash}X|>{\raggedright\arraybackslash}X|>{\raggedright\arraybackslash}X|>{\raggedright\arraybackslash}X@{}}
\toprule
\textbf{Capability} & \textbf{CNN Completion} & \textbf{ViT} & \textbf{GAN} & \textbf{Diffusion} \\
\midrule
Observation interface & Masked regular grid & Irregular points or tokens & Paired or unpaired image conditions & Masks, measurements, or external priors \\
Unknown-Tx handling & Method-dependent & Partial in cited methods~\cite{hehn2023transformer} & Possible if latent source is modeled & RadioTrace~\cite{yang2025radiotrace} \\
Multiple conditional samples & Requires probabilistic head/ensemble & Requires probabilistic head/ensemble & Possible through a stochastic latent; calibrate & Supported by repeated sampling; calibrate~\cite{ha2025bayesian} \\
Cross-band/tensor reconstruction & Latent PnP denoising~\cite{xu2025radio} & Architecture-dependent & CF-CGN~\cite{xie2025cf} & No representative isolated in this table \\
Latency structure & One network pass; implementation-dependent & One pass; token-pair cost & One generator pass after training & Step-dependent; accelerated variants~\cite{11152929,jia2025radioflow} \\
3D volumetric support & CAED~\cite{ivanov2024deep} & Architecture-dependent & RadioGen3D~\cite{chen2026radiogen3d} & RadioDiff-3D~\cite{11083758} \\
Zero-shot (no wireless training) & \ding{55} & \ding{55} & \ding{55} & \checkmark \cite{wang2025radiodiff-inv} \\
\bottomrule
\end{tabularx}
\label{tab:inverse_comparison}
\end{table*}

\subsection{Diffusion-Based Methods}
\label{subsec:diffusion_inverse}

A deterministic model normally returns one point estimate~\cite{levie2021radiounet,lee2023pmnet}. Under an ideal squared-error objective, the population optimum is the conditional mean, which is insufficient when a downstream application needs a nonlinear functional of conditional uncertainty~\cite{ha2025bayesian,luo2025denoising}. Let $P$ denote uncertain received power in linear scale. The function $C(P)=B\log_2(1+\alpha P)$ is concave for $P\geq0$ and $\alpha>0$. Jensen's inequality therefore gives $C(\mathbb E[P])\geq\mathbb E[C(P)]$; evaluating capacity at a posterior-mean estimate can overstate expected capacity. Conditional generative models can approximate $\mathbb E[F(\gamma)]$ from repeated samples, but only under their learned distribution and with all variables required by $F$ supplied. Equation~\eqref{eq:mc_estimate} states the estimator and its calibration boundary.

\subsubsection{Bayesian Inverse Estimation and Learned Priors}
\label{subsubsec:bayesian_inverse}

RadioDiff-Inverse \cite{wang2025radiodiff-inv} frames reconstruction as Bayesian maximum a posteriori estimation $\hat{x}_{MAP} = \arg\max_x\, p(x)p(y|x)$. A pre-trained diffusion model provides the prior $p(x)$. The reverse generation is corrected via likelihood-weighted Monte Carlo sampling:
\begin{align}
&p^{\mathrm{wgt}}_\theta(x^i_{t-1}\mid x_t,y_{t-1})\notag\\
&\quad\propto \exp\!\left(-\frac{\|x^i_{t-1}-\mu_\theta\|^2}{2\sigma_t^2}-\frac{\|y_{t-1}-Ax^i_{t-1}\|^2}{2c_t^2\sigma_t^2}\right).
\label{eq:bayesian_posterior}
\end{align}
The proportionality sign is required because the displayed exponential omits the normalizing constant of the weighted transition. Treating it as an equality would not in general define a probability density. The measurement operator $A$, noise scale $c_t\sigma_t$, and normalization or resampling procedure must be specified before this weight is interpreted as a posterior transition.

A notable property is training-free operation. It transfers a diffusion prior from natural images to the RF domain without wireless-specific fine-tuning \cite{wang2025radiodiff-inv}. Both domains share piecewise-smooth structures with sharp boundaries. Edge priors from natural images regularize shadow boundary reconstruction. However, three conditions degrade this transfer. First, RMs exhibit distance-dependent power decay with no natural image counterpart. Second, multipath fading patterns are physically meaningful but a natural image prior smooths them away. Third, the RM dynamic range exceeds 100 dB, far beyond 8-bit images. Natural image priors are effective for macro-scale shadowing but should not be relied upon for fine-grained multipath recovery \cite{ha2025bayesian}. Domain-specific priors trained on wireless data are preferable for such cases. CKMDiff provides a 2026 example of this domain-specific route by learning a CKM prior and coupling it with an explicit inverse-problem observation model~\cite{fu2026ckmdiff}. Its formulation reinforces a general rule: the prior should model plausible fields, while the likelihood or measurement-consistency term should preserve the information actually observed.

\subsubsection{Architectural Adaptations for Blind Completion}
\label{subsubsec:blind_completion}

When environmental priors are absent, the conditional decoupled diffusion model \cite{luo2025denoising} splits the forward process into image attenuation and noise enhancement. High-dimensional pixels are mapped to a condensed latent space via a variational autoencoder. Multi-scale conditions are extracted using a Swin-B Transformer. A dual-branch U-Net simultaneously predicts drift $\mathbf{c}_\theta$ and noise $\boldsymbol{\epsilon}_\theta$:
\begin{align}
\min_\theta \mathbb{E}_{t, \mathbf{x}_0, \boldsymbol{\epsilon}, \mathbf{y}} \!\left[ \|\mathbf{c} - \mathbf{c}_\theta\|^2 + \|\boldsymbol{\epsilon} - \boldsymbol{\epsilon}_\theta\|^2 \right].
\label{eq:decoupled_loss}
\end{align}

The latent-domain plug-and-play framework of Xu et al.~\cite{xu2025radio} addresses spatio-spectral RM estimation. It models $\mathcal X=\sum_{r=1}^{R}S_r\circ c_r$, where $S_r$ is a spatial loss field and $c_r$ is the corresponding power-spectral-density vector, and inserts a denoiser into an ADMM update on the latent spatial factors. The factorization, nonnegativity assumptions, denoiser behavior, observation operator, and stated convergence conditions bound its recoverability claims; it is not a diffusion method and should not be attributed to DULRTC-RME.

RadioLAM \cite{liu2026radiolam} targets fine-grained 3D RM construction at a specified receiver height under ultra-low sampling. It combines propagation-model augmentation, a diffusion-based large artificial intelligence model with mixture-of-experts routing, and propagation-guided selection among candidate maps.

\begin{figure*}[!t]
    \centering
    \captionsetup{font={small}, skip=10pt}
    \includegraphics[width=0.85\linewidth]{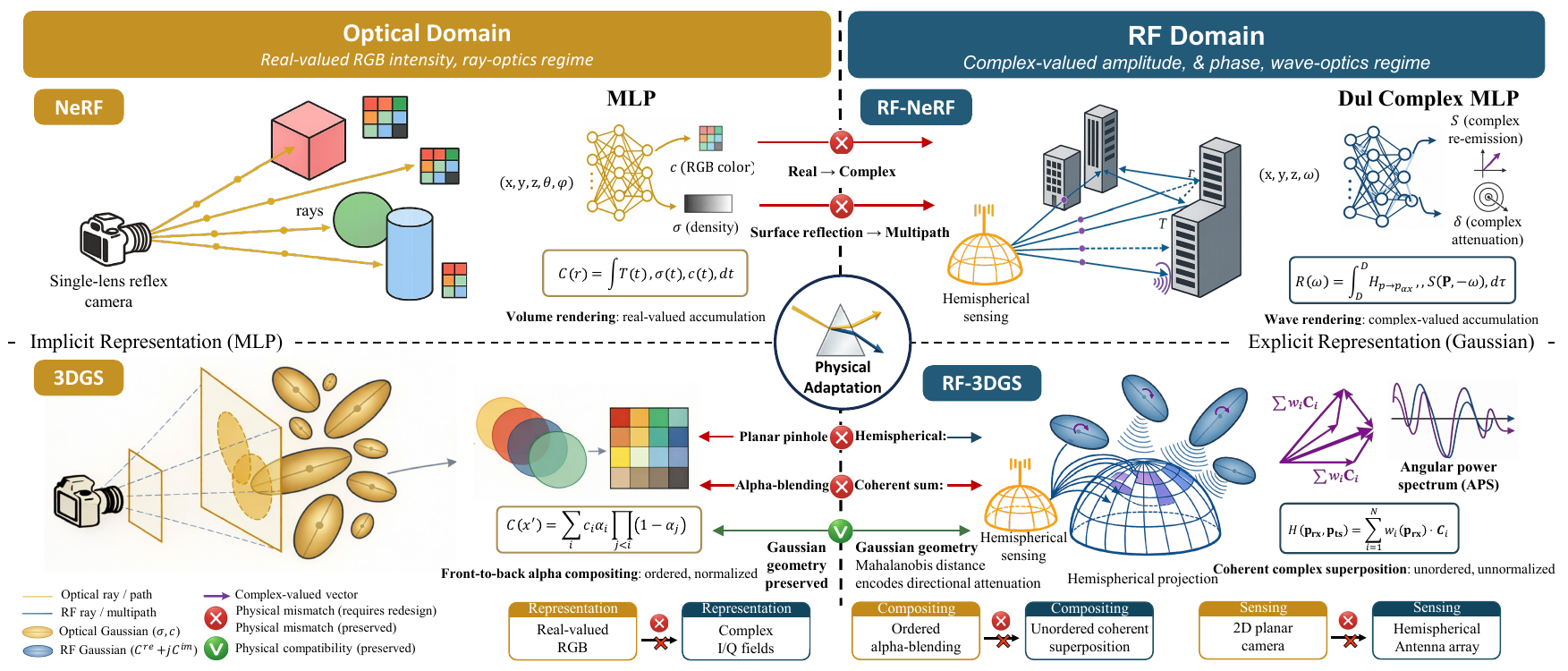}
    \caption{Architectural mapping from optical neural rendering to cited RF adaptations. The top row contrasts optical real-valued volume rendering with complex-field RF rendering. The bottom row shows that inverse-source RF formulations can replace front-to-back optical alpha compositing by complex superposition; other RF-3DGS renderers may retain ordered attenuation or projection operations. Red crosses and green checks indicate the modeling mismatches and compatibilities asserted by the illustrated formulations, not universal electromagnetic validity.}
    \label{fig:optical_to_rf_mapping}
\end{figure*}

\subsubsection{Handling Extreme Observational Degradations}
\label{subsubsec:extreme_degradation}

For low-bit measurements $\mathbf y=\mathcal Q(\mathbf H\mathbf x_0+\mathbf n)$, a hard step quantizer has zero pathwise derivative almost everywhere~\cite{gu2025gsc}. Direct backpropagation through $\mathcal Q$ therefore vanishes, although surrogate gradients or a differentiable quantized likelihood remain alternatives. Under the cited method's Gaussian approximation $\mathbf x_0\mid\mathbf x_t\sim\mathcal N(\hat{\mathbf x}_{0\mid t},\gamma_t^2\mathbf I)$, i.i.d. noise variance $\sigma_y^2$, diagonal sampling mask $\mathbf H$, and bin intervals $(l_i,u_i]$, the truncated-Gaussian moments give
\begin{align}
&\hat{\mathbf x}_{0\mid t,\mathbf y}=\hat{\mathbf x}_{0\mid t}+\frac{\gamma_t^2}{s_t}\mathbf H\boldsymbol\Delta,\qquad s_t=\sqrt{\gamma_t^2+\sigma_y^2}.
\label{eq:quantized_posterior}
\end{align}
Here, $\boldsymbol\Delta$ contains the bin-conditioned standardized Gaussian moment corrections given by the cited source~\cite{gu2025gsc}. The expression is closed form only under these prior, noise, masking, and scalar-quantization assumptions; it is not a model-free posterior update.

\subsubsection{Physics-Informed Sensing and Localization}
\label{subsubsec:physics_sensing}

RadioDiff-Loc addresses NLoS source localization from sparse RSS without known transmit power by combining relative-power normalization with multi-channel conditioning~\cite{wang2025radiodiff-loc}. Its sampling rule is motivated by a simplified far-field knife-edge boundary model. Under that model, the source reports a Fisher-information contribution $J_{jj}\propto 1/s_j^2$ near an edge coordinate $s_j$ and a separate Gaussian-prior mutual-information criterion. These analyses motivate sampling near high-diffraction vertices; they do not prove that every building vertex is universally optimal under multipath, material uncertainty, or a different localization model. The cited study reports sub-one-percent sampling for its stated geometry, sensor rule, and localization target, so that operating point should not be transferred to a different map resolution or reconstruction task without remeasurement.

\subsection{Summary and Practitioner Guidance}
\label{subsec:inverse_summary}

Table~\ref{tab:inverse_comparison} supports a capability-based selection rather than a sampling-rate ranking. CNN completion is the lowest-complexity baseline for masked regular grids~\cite{teganya2021deep,chaves2023deeprem}. ViT architectures are attractive when observations arrive as irregular point sets or when long-range interactions dominate~\cite{viet2025spatial,fang2025radioformer}. GAN methods provide fast conditional generation and can use paired, unpaired, or satellite-image supervision~\cite{pan2025sc,roger2023deep}. Diffusion models represent non-unique solutions through conditional sampling, but their iterative inference adds latency~\cite{ha2025bayesian,luo2025denoising}. Source-reported operating points remain useful evidence within each study, but they do not define universal boundaries between these families.

A unifying insight is that probabilistic generation becomes justified when the stated observations and physical metadata admit materially different maps~\cite{ha2025bayesian}. No fixed sampling percentage establishes this transition because identifiability also depends on sensor placement, scene geometry, frequency, output resolution, and noise. Once non-uniqueness is demonstrated within a controlled setting, the model must draw upon a prior and its reconstruction quality is bounded by the relevance of that prior~\cite{wang2025iradiodiff,luo2025denoising}.

\section{Optics-Inspired Electromagnetic Mapping Methods}
\label{sec:optics_methods}


Many methods in Sections~\ref{sec:forward_methods} and~\ref{sec:inverse_methods} return values on an image or voxel grid. This section reviews coordinate- and primitive-based alternatives inspired by neural rendering~\cite{zhao2023nerf2,krijestorac2023deep,zhang2026rf,wen2025wrf}. RF NeRF and 3DGS methods combine continuous or explicit scene representations with source-specific volumetric, attenuation, or superposition rules; the representation is not automatically physics-grounded merely because it is continuous. Fig.~\ref{fig:optical_to_rf_mapping} illustrates the mapping used by cited formulations, and Table~\ref{tab:when_to_use} states the regimes in which their query interface may be preferable to a grid network.

\begin{table*}[!t]
\centering
\captionsetup{font={small}, skip=10pt}
\caption{When to use optics-inspired methods versus traditional neural networks. Entries are qualitative capabilities under the stated regime, not universal thresholds or a controlled accuracy ranking. \rev{Table~\ref{tab:within_study_resources} gives the consolidated resource evidence.}}
\vspace{-6pt}
\scriptsize
\setlength{\tabcolsep}{3pt}
\begin{tabularx}{0.99\linewidth}{@{}>{\raggedright\arraybackslash}p{3.2cm}|>{\raggedright\arraybackslash}p{2.6cm}|>{\raggedright\arraybackslash}p{2.6cm}|>{\raggedright\arraybackslash}p{2.6cm}|>{\raggedright\arraybackslash}X@{}}
\toprule
\textbf{Deployment Condition} & \textbf{Traditional NN} & \textbf{NeRF} & \textbf{3DGS} & \textbf{Rationale} \\
\midrule
Fixed environment; variable Tx locations & Supported when Tx is an input & Source-specific; often refit & Source-specific; often refit & Audit what changes without scene optimization \\
Fixed scene; continuous queries prioritized & Coordinate output possible & Native coordinate/ray queries & Native primitive queries & Fidelity still depends on measurements and renderer \\
Strict query-latency budget & One network pass & $RQ$ field evaluations & $G_{\mathrm{vis}}$ projection/sorting & Compare under matched output, precision, hardware, and timing boundary \\
Angular power spectrum required & Supported if trained as target & Supported by directional output & Supported by directional basis & APS support follows the output interface, not the family name \\
Complex-valued CSI required & Supported by complex/two-channel output & Requires an RF complex renderer & Requires complex coefficients and renderer & No unmodified optical formulation supplies RF phase \\
Generalization to unseen environments & Possible for represented conditions & GRaF reports a study-specific test & Requires shared encoder/initializer evidence & Per-scene fitting alone does not demonstrate transfer \\
City-scale scene with frequent updates & Candidate if represented at training & Refit/query cost can dominate & Primitive update/storage can dominate & Measure construction and query stages separately \\
Severely sparse scene measurements & Requires an explicit prior & Requires an explicit prior & Requires an explicit prior & Representation alone does not resolve missing information \\
Dynamic Tx and Rx (6D CKM) & Possible with explicit joint conditioning & Requires source/receiver conditioning & BiWGS provides study-specific evidence & Bidirectionality and reciprocity assumptions must be validated \\
\bottomrule
\multicolumn{5}{l}{\footnotesize Capabilities require the stated input/output interface; GRaF and BiWGS evidence remains source- and protocol-specific.} \\
\end{tabularx}
\label{tab:when_to_use}
\end{table*}

\rev{The reported spectrum comparisons provide experimental quality evidence alongside the analytic cost drivers in Table~\ref{tab:arch_roadmap}. One study reports median SSIM values of 0.56, 0.78, and 0.82 for DCGAN, NeRF$^2$, and WRF-GS~\cite{wen2025wrf}. In the geometric-algebra study, GAI-GS reaches 0.91 compared with 0.84 for NeRF-APT and 0.82 for WRF-GS~\cite{shen2026gaigs}. Its quality gain accompanies distinct fitting, inference, and rendering trade-offs. A grid-regression CNN and an RF field require matched scenes, observations, output modality, resolution, and quality requirements before their resource figures support a direct choice.}

\subsection{Neural Radiance Fields for RM Construction}
\label{subsec:nerf_rm}

Migrating optical NeRF to the RF domain presents significant physical challenges \cite{zhao2023nerf2, krijestorac2023deep}. Unlike visible light, RF signals are complex-valued with phase governing constructive and destructive interference. The optical volume density $\sigma(\mathbf{r})$ is replaced by a complex attenuation $\delta = \Delta a\, e^{j\Delta\theta}$, the real-valued color output becomes a complex re-emitted signal, and the transmittance accumulation shifts from real-valued to complex-valued integration. Furthermore, RF hardware provides far lower spatial sampling than optical cameras.

\subsubsection{Foundational Voxelized Modeling}
\label{subsubsec:nerf_voxel}

NeRF$^2$ \cite{zhao2023nerf2} discretizes 3D space into micro-voxels based on the Huygens-Fresnel principle. Each voxel is a secondary source characterized by complex attenuation $\delta$ and re-emitted signal $S$. The received signal from direction $\omega$ is:
\begin{align}
R(\omega) = \int_{0}^{D} H_{P(r,\omega)\rightarrow P_{RX}}\, S(P(r,\omega),-\omega)\,dr,
\label{eq:nerf_rendering}
\end{align}
Here, $D$ is the scene integration limit. In the cited NeRF$^2$ formulation, cumulative attenuation is written $H(r,\omega)=\exp\!\left(\int_0^r\hat\delta(P(\tilde r,\omega))\,\mathrm d\tilde r\right)$, where $\hat\delta$ is a learned log-attenuation density~\cite{zhao2023nerf2}. A discrete implementation must include the ray-sample spacing, and a complex logarithmic representation requires a consistent phase convention; the exponential does not remove branch or discretization error. Dual complex-domain networks and positional encoding fit the measured scene, while turbo learning augments a downstream task with source-generated data~\rev{\cite{zhao2023nerf2}}. Neither step supplies unmeasured sub-wavelength information without an appropriate prior.

\subsubsection{Physics-Embedded Neural Channel Synthesis}
\label{subsubsec:nerf_physics}

\rev{NeWRF embeds physical propagation rules into neural channel synthesis~\cite{lu2024newrf}.} A key innovation decouples free-space propagation from object reflection. Reflection points are treated as virtual transmitters, relieving the network from learning distance-based attenuation. The neural channel synthesis equation is:
\begin{align}
H(f) = \sum_{r \in \mathcal{L}} \sum_{p \in \mathcal{P}_r} w_{rp}\, A_{rp}\, e^{j(\psi_{rp} - 2\pi f d_{rp} / c)},
\label{eq:channel_synthesis}
\end{align}
Here, $r$ indexes a path or ray, $p\in\mathcal P_r$ indexes its samples, $d_{rp}$ is the associated propagation distance, and $w_{rp}$, $A_{rp}$, and $\psi_{rp}$ are a rendering weight, amplitude term, and residual phase~\rev{\cite{lu2024newrf}}. Using $d_p$ without a ray index would imply that the same sample distance applies to every path. The expression is a study-specific synthesis rule rather than a general solution of Maxwell's equations.

A critical detail is phase prediction. The complex exponential $e^{j\psi}$ is differentiable, but a wrapped scalar phase target is discontinuous at a chosen $2\pi$ branch cut. Regressing in-phase and quadrature components, $I=A\cos\psi$ and $Q=A\sin\psi$, avoids assigning a large loss to complex values that are adjacent across that cut. Amplitude is $A=\sqrt{I^2+Q^2}$, and $\psi=\operatorname{atan2}(Q,I)$ preserves the quadrant when $(I,Q)\ne(0,0)$. Phase is physically undefined at zero amplitude, gradients can be ill-conditioned near the origin, and any reported wrapped phase still has a branch cut; losses should therefore operate on $I$ and $Q$ or a normalized complex phasor whenever possible~\rev{\cite{lu2024newrf,zhang2026rf}}.

\subsubsection{Generalizable Radio-Frequency Radiance Fields}
\label{subsubsec:graf}

The preceding frameworks rely on per-scene optimization. Generalizable radio-frequency radiance fields (GRaF) \cite{yang2025graf} address this limitation by treating the target spatial spectrum as an interpolation of adjacent transmitters. A multi-head Transformer represents voxels sampled along a ray as a token sequence. Cross-attention learns cumulative attenuation weights driven by spatial geometry. Specialized attention heads encode low-level physical knowledge \cite{yang2025graf}.

The GRaF study reports zero-shot cross-scene evaluation. Under its unseen-environment protocol, it reports a 26.9\% MSE reduction and a 10.2\% PSNR improvement over NeRF$^2$, and its generated spatial spectra reduce angle-of-arrival error by 61.6\% relative to the study's 50\%-real-data setting~\cite{yang2025graf}. These values depend on that split, baseline, and metric implementation and are not a matched comparison with the CNN and 3DGS studies in \rev{Tables~\ref{tab:within_study_resources} and~\ref{tab:when_to_use}}.

\subsection{3DGS-Based Methods}
\label{subsec:3dgs_rm}

In the cited RF study, 3DGS reduces post-fit query latency relative to its NeRF baseline by using explicit anisotropic Gaussian primitives and parallelizable projection/rasterization~\cite{zhang2026rf}. This is not a universal end-to-end speed ordering because fitting, primitive count, visibility, sorting, hardware, and query density differ. Migrating optical 3DGS to phase-sensitive RF tasks requires a declared complex-domain field representation and an angular antenna/query model rather than an unmodified optical color-and-camera interpretation.

\subsubsection{Complex-Domain Gaussian Splatting}
\label{subsubsec:wrfgs}

WRF-GS \cite{wen2025wrf} reinterprets 3D Gaussians as virtual transmitters. Optical color and opacity are replaced with signal strength and environmental attenuation. The multipath received signal is:
\begin{align}
y = \sum_{l=0}^{L-1} A\, \Delta A_l\, e^{j(\varphi + \Delta\varphi_l)},
\label{eq:wrfgs}
\end{align}
where $\Delta A_l$ and $\Delta\varphi_l$ are the amplitude and phase of the $l$-th path. A Mercator projection maps angular coordinates to a planar sensing domain~\cite{wen2025wrf}. Its linear scale grows as $\sec\theta_{el}$ and its area scale as $\sec^2\theta_{el}$; the latter already equals $4/3$ at $30^\circ$. Projection distortion should therefore be measured rather than declared negligible, and high-elevation UAV or satellite links require compensation or a spherical parameterization.

Optical alpha-blending is replaced by EM splatting following cascaded wave attenuation. WRF-GS+ introduces deformable Gaussians to decouple static large-scale fading from dynamic small-scale fading \cite{wen2025neural}.

\rev{The same study reports median spectrum SSIM of 0.82 for WRF-GS and 0.90 for WRF-GS+, with single-spectrum synthesis increasing from 5 to 8~ms on an RTX 3090~\cite{wen2025neural}. This is a measured quality--latency trade-off for the combined representation and deformation network. Training-set sweeps from 200 to 4,800 measurements support data-efficiency comparisons within that scene, while transfer to a new environment requires separate evaluation.}

\subsubsection{MLP-Free Fourier-Legendre Synthesis}
\label{subsubsec:gsrf}

GSRF \cite{yang2025gsrf} treats each 3D Gaussian as a secondary RF source via the Huygens-Fresnel principle. It replaces spherical harmonics with a Fourier-Legendre expansion (FLE):
\begin{align}
\psi_k(\alpha,\beta) = \sum_{l=0}^{L}\sum_{m=-l}^{l} c_{ml}^{(k)}\, e^{im\alpha}\, P_l^m(\cos\beta),
\label{eq:fle}
\end{align}
which decouples azimuth and elevation with complex coefficients $c_{ml}^{(k)}$ encoding amplitude and phase. Combined with orthographic splatting and a frequency-domain consistency loss, GSRF operates entirely without MLPs \cite{yang2025gsrf}.

\rev{Its full-training-set RFID ablation reports PSNR of 22.64~dB for the complete model, 20.51~dB with spherical harmonics replacing FLE, 20.89~dB without phase, and 21.30~dB without the Fourier loss~\cite{yang2025gsrf}. The basis comparison uses degree three and 16 coefficients per basis. These ablations quantify distinct representation and loss contributions. The Fourier loss adds FFT evaluation during training. Queries use the same rendering operators, with cost still dependent on the fitted Gaussian population. Isolated incremental runtime and memory are unreported, so the complete-model profiles in Table~\ref{tab:within_study_resources} characterize a different measurement scope.}


\subsubsection{Complex Superposition via Inverse Source Formulation}
\label{subsubsec:ngrf}

nGRF \cite{umer2025neural} formulates the problem as an electromagnetic inverse source problem governed by the Helmholtz equation. It models distance- and direction-dependent attenuation via the Mahalanobis distance of anisotropic Gaussian bases. Crucially, nGRF identifies that optical alpha-blending is physically inadequate for RF \cite{umer2025neural}. Electromagnetic waves undergo coherent superposition rather than strict occlusion. The aggregate signal is an unordered complex summation:
\begin{align}
H(p_{\mathrm{rx}},p_{\mathrm{tx}})=\sum_{i=1}^{N}w_i(p_{\mathrm{rx}})\,C_i(p_{\mathrm{tx}}),
\label{eq:ngrf}
\end{align}
The explicit $p_{\mathrm{tx}}$ dependence is carried by the complex source coefficient $C_i(p_{\mathrm{tx}})=C_i^{\mathrm{re}}(p_{\mathrm{tx}})+jC_i^{\mathrm{im}}(p_{\mathrm{tx}})$, while $w_i(p_{\mathrm{rx}})$ describes receiver-side spatial influence in this simplified notation. Constructive and destructive interference emerge from the coefficient phases. Unlike optical alpha blending, these spatial weights are not constrained to a front-to-back convex combination. The cited nGRF formulation reports a 220$\times$ inference speedup over its NeRF baselines~\cite{umer2025neural}; primitive count, query set, hardware, precision, and timing boundary prevent transferring that factor to another study.

\subsubsection{6D Channel Knowledge Maps via Bidirectional Splatting}
\label{subsubsec:biwgs}

BiWGS \cite{zhou20256d} extends to 6D channel knowledge maps (CKMs) for dynamic transmitter and receiver movement. It introduces bidirectional spherical harmonics (BSH) to fit complex scattering coefficients from multiple virtual projection planes.

BiWGS encodes reciprocal symmetry through parameter sharing across bidirectional spherical-harmonic coefficient pairs~\cite{zhou20256d}. This construction enforces $h_{A\to B}=h_{B\to A}$ within the model when the propagation medium and terminal calibration satisfy reciprocity. The constraint must be disabled or modified for nonreciprocal components, active relays, or direction-dependent hardware responses.

BiWGS applies a logarithmic-domain loss to reduce domination by strong paths~\cite{zhou20256d}. A 100~dB power span corresponds to a $10^{10}$ power ratio and a $10^5$ field-amplitude ratio. For a squared dB residual, differentiation with respect to predicted linear power introduces an inverse-power factor together with the current log residual. This reweighting can emphasize weak paths, but its exact gradient range depends on the prediction error and the positive power floor; it is not universally $\mathcal{O}(\log R)$. The floor and output domain must therefore be reported whenever a log-domain loss is compared with a linear-domain objective.

\subsubsection{\rev{Geometric Algebra-Informed Splatting}}
\label{subsubsec:gaigs}
\rev{GAI-GS combines Gaussian primitives with geometric-algebra attention to learn ray--object interaction features~\cite{shen2026gaigs}. A tokenizer aggregates high-opacity Gaussian anchors with transmitter context, and the resulting scene token conditions attenuation and signal branches. This is a learned structural prior rather than an exact electromagnetic solver. The source evaluates measured indoor RSSI and RFID spectra. The comparisons in Table~\ref{tab:within_study_resources} and the accompanying discussion characterize the joint architecture under the reported indoor evaluation conditions.}

\subsection{Discussion}
\label{subsec:optics_discussion}

NeRF and 3DGS are attractive when the requested output is a continuous complex field or directional spectrum within a fixed, well-observed scene~\cite{zhao2023nerf2, zhang2026rf, yang2025gsrf}. Their representation can encode field variation at arbitrary query coordinates, whereas a grid network is bounded by its output discretization. This fidelity is conditional on adequate scene measurements and a valid RF rendering rule; it is not an unconditional accuracy advantage over a CNN.

The principal bottleneck is scene-specific construction~\rev{\cite{lu2024newrf,umer2025neural}}. A change in geometry, material, transmitter, or output definition may require refitting primitives or network parameters. Reported construction and query times vary with the number of rays, samples, Gaussians, output channels, and hardware, so cross-paper wall-clock values are not used as a common ranking here. A conditioned grid network is preferable when the deployment requires rapid reuse across many scenes or transmitter settings.

\rev{GSRF provides a concrete sparse-data operating point. With 220 RFID training instances, it reports median MSE reductions of 56.4\% against NeRF$^2$ and 19.3\% against WRF-GS. On an RTX 3080 Ti, the respective GSRF, NeRF$^2$, and WRF-GS fitting times are 0.27, 5.01, and 1.61~h, and query times are 4.18, 352.73, and 7.58~ms~\cite{yang2025gsrf}. At 0.8 measurements/ft$^3$, GSRF achieves MSE comparable to that of NeRF$^2$ at 7.8 measurements/ft$^3$. This supports sparse fitting within the measured scene. GSRF explicitly requires scene data and updates or refitting after environmental changes.}

Hybrid and amortized representations aim to reduce scene-specific fitting cost. A shared encoder can initialize scene parameters, and a local update can refine primitives affected by dynamic objects~\cite{wen2025neural,yang2025graf}. Amortization does not itself establish real-time operation: feature extraction, initialization, refinement, rendering, and physical verification must all be included in the measured update latency. Optics-inspired methods are candidates for fixed-scene continuous-query digital twins, whereas conditioned feed-forward networks are candidates when cross-scene reuse or frequent updates dominate. Table~\ref{tab:when_to_use} and the consolidated resources in Table~\ref{tab:within_study_resources} distinguish these regimes without asserting an unconditional winner.

\section{Physics-Informed Methods for RM Construction}
\label{sec:physics_methods}

The methods in Sections~\ref{sec:forward_methods}--\ref{sec:optics_methods} are predominantly data-driven~\cite{levie2021radiounet,wang2024radiodiff,zhao2023nerf2}. They minimize empirical loss and may have no explicit mechanism for checking electromagnetic consistency. Representative training data can make a purely data-driven model effective within its evaluated distribution, but an out-of-distribution output can violate an applicable physical constraint~\cite{karniadakis2021physics,jiang2024physics}. Physics-informed RM construction introduces electromagnetic knowledge through input features, penalty terms, or network structure~\cite{feng2025physics,chen2026radiodun,11278649}. The resulting constraint is only as valid as its field modality, source model, materials, boundaries, and discretization.

The three integration levels act at distinct pipeline stages. Data-level integration converts geometry or electromagnetic models into input tensors and can support cross-frequency transfer when the features and training data cover the target material and band. Loss-level integration penalizes PDE or boundary residuals but cannot ensure physical consistency under a mismatched modality, grid, source, or material model. Architecture-level integration maps a physical algorithm or symmetry into the network topology; this stronger coupling can increase activation memory and can propagate an incorrect physical assumption through every layer. Table~\ref{tab:physics_modules} therefore records whether a module is backbone-agnostic, its incremental cost, and the regime in which its evidence applies. Recent electromagnetic inverse-scattering studies illustrate two further choices: a physics-driven network can preserve a conventional solver update, whereas a physics-informed neural operator learns a mesh-independent mapping while penalizing governing-equation residuals~\cite{du2026scattering,dong2026operator}. These studies recover material or source parameters rather than an RM directly, so their operator designs transfer only when an RM construction model jointly infers the environment and field.

\subsection{A Taxonomy of Embeddable Physical Knowledge}
\label{subsec:physics_taxonomy}

Physical information enters a learning pipeline with different assumptions and costs~\cite{karniadakis2021physics}. We separate three integration locations---input data, loss, and architecture---without treating them as a guaranteed ordering of fidelity.

\subsubsection{Geometric and Environmental Knowledge}
\label{subsubsec:geometric_knowledge}

Data-level geometry includes binary or height-encoded building maps, LoS/NLoS labels, fractional-LoS ratios, and edge or corner candidates~\cite{levie2021radiounet,lu2025sip2net}. Its cost depends on scene size, ray--grid tests, dimensionality, and update rate. Raster features can be concatenated to grid backbones, while graph, point, mesh, or continuous models require a compatible encoder. Geometry identifies candidate wave--obstacle interactions but does not specify the electromagnetic response~\cite{chen2024diffraction}.

\begin{table*}[!t]
\centering
\captionsetup{font={small}, skip=10pt}
\caption{Physics-informed modules by integration level. Backbone compatibility follows each integration method (\checkmark: agnostic; \ding{55}: dependent; Partial: conditional). \rev{Costs are analytic unless measured values are given. Evidence distinguishes paired ablations, bundled comparisons, and qualitative integration examples. NR denotes an unreported numerical cost.}}
\vspace{-6pt}
\scriptsize
\setlength{\tabcolsep}{2.5pt}
\begin{tabularx}{0.99\linewidth}{@{}>{\raggedright\arraybackslash}p{2.15cm}|>{\raggedright\arraybackslash}p{3.6cm}|>{\raggedright\arraybackslash}X|>{\raggedright\arraybackslash}p{2.7cm}|>{\raggedright\arraybackslash}X@{}}
\toprule
\textbf{Module} & \textbf{Physical prior and integration} & \textbf{Incremental overhead} & \rev{\textbf{Candidate regime}} & \rev{\textbf{Evidence and scope}} \\
\midrule
\multicolumn{5}{@{}l}{\textbf{Panel A: Data-level modules}} \\
\midrule
FSPL + antenna gain channel & Friis equation + 3D gain slice\newline Concat as input channel \checkmark & $\mathcal{O}(MHW)$; one tensor per setting & Frequency or antenna varies; geometry known & \rev{SIP2Net \cite{lu2025sip2net}: physical input channels. Integration example} \\
ITU material reflectance/\allowbreak transmittance & Fresnel equations + ITU-R P.2040\newline Concat as input channel \checkmark & $\mathcal{O}(FCHW)$ preprocessing/storage & Cross-band or cross-material transfer & \rev{SIP2Net \cite{lu2025sip2net}: material inputs. Integration example} \\
Fractional LoS map & 3D Bresenham ray-obstacle intersection\newline Concat as input channel \checkmark & $\mathcal{O}(MHWL)$ ray-grid tests. \rev{Measured feature cost NR} & Known Tx; blockage dominates & \rev{Agile \cite{yapar2024overview}: normalized RMSE $0.0514\to0.0461$, fixed MSE loss. Paired input comparison} \\
Radio depth map & Distance fading + obstruction ratio\newline Concat as input channel \checkmark & $\mathcal{O}(MHW)$ plus one channel & Sparse labels with reliable geometry & TiRE-GAN \cite{zhou2025tire}; RadioGAT \cite{li2024radiogat} \\
$k_{eff}^2$ amplitude-curvature mask & Helmholtz-derived amplitude curvature\newline Spatial mask for diffusion \checkmark & $\mathcal{O}(HW)$ finite differences. \rev{Model resources in Table~\ref{tab:within_study_resources}} & Sharp fading boundaries; dense prior available & \rev{RadioDiff-$k^2$ \cite{11278649}: SRM NMSE reduced by 43\% versus RadioDiff. Guidance-pipeline comparison} \\
\midrule
\multicolumn{5}{@{}l}{\textbf{Panel B: Loss-level modules}} \\
\midrule
LoS path loss anchor & Free-space propagation law\newline Additive penalty term \checkmark & $\mathcal{O}(MHW)$ per batch & Known sources and calibrated LoS samples & \rev{\cite{liu2025pinn}: LoS-consistency penalty. Integration example} \\
VIE/MoM residual & Volume integral equation + Green's func.\newline Physical loss $\mathcal{L}_{PHY}$ Partial$^*$ & Dense operator can require $\mathcal{O}((HW)^2)$ memory & Small static scenes with known materials & PEFNet \cite{jiang2024physics}: VIE-constrained prediction \\
Helmholtz PDE residual & $\nabla^2 u + k^2 u = -f$ discretized\newline $\mathcal{L}_{PINN}$ penalty Partial$^*$ & $\mathcal{O}(HW)$ stencil per update. \rev{Measured incremental cost NR} & Sparse data with valid source/boundary model & \rev{PhyRMDM \cite{jia2025rmdm}: RMSE $0.0182\to0.0125$ with PINN loss. Source Table 4} \\
Log-domain loss transform & Dynamic range compression\newline $\mathcal L=\|\ell(\hat P)-\ell(P)\|^2$, $\ell(P)=10\log_{10}((P+P_{\mathrm{floor}})/P_0)$ \checkmark & Small elementwise cost; positive floor and domain conversion required & Large power dynamic range & \cite{zhou20256d}: log-domain optimization \\
\rev{Fourier spectrum loss} & \rev{Spectrum structure}\newline \rev{Fourier-domain matching} \rev{\checkmark} & \rev{$\mathcal{O}(HW\log(HW))$ during training. Extra time/memory NR. Training-only loss evaluation} & \rev{Fine spectrum structure} & \rev{GSRF \cite{yang2025gsrf}: PSNR $21.30\to22.64$~dB. Paired loss ablation} \\
\midrule
\multicolumn{5}{@{}l}{\textbf{Panel C: Architecture-level modules}} \\
\midrule
ADMM unrolling & Iterative optimization structure\newline Layer-by-layer mapping \ding{55} & $\mathcal{O}(KHW)$ stored activations & Data-poor inverse tensor completion & DULRTC-RME \cite{wang2025dulrtc}: unrolled inverse estimator \\
Geometric-optics cross-attention & Transmission boundaries as prompts\newline Spatial cross-attention \checkmark$^\dagger$ & Attention cost grows with token pairs. \rev{Model profile in Table~\ref{tab:within_study_resources}} & Diffraction and transmission boundaries matter & \rev{iRadioDiff \cite{wang2025iradiodiff}: ALG RMSE $9.619\to6.357$~dB with all physics cues. Bundled ablation} \\
Reciprocity parameter sharing & EM reciprocity $h_{A\to B} = h_{B\to A}$\newline Hard weight symmetry \ding{55} & Small query overhead; reduced parameter freedom and calibration checks & Reciprocal bidirectional channels & BiWGS \cite{zhou20256d}: reciprocal channel constraint \\
\bottomrule
\multicolumn{5}{@{}p{0.99\linewidth}@{}}{\scriptsize $M$: transmitters; $F$: frequencies; $C$: feature channels; $L$: ray samples; $K$: unrolled blocks. $^*$Operator precomputation required; $^\dagger$cross-attention backbone required.} \\
\end{tabularx}
\label{tab:physics_modules}
\end{table*}

\subsubsection{Electromagnetic Propagation Knowledge}
\label{subsubsec:em_knowledge}

The second category derives from Maxwell's equations and their asymptotic approximations \cite{jiang2024physics, kouyoumjian1974uniform}. This includes FSPL from the Friis equation, material-dependent reflection and transmission coefficients from Fresnel equations using ITU-R P.2040, diffraction losses from the uniform theory of diffraction (UTD), and projected antenna radiation patterns \cite{chen2024diffraction, jaensch2025radio}. These features encode quantitative electromagnetic responses. However, they require frequency-dependent parameters and may need per-pixel computation, incurring substantial preprocessing cost.

\subsubsection{Wave Equation and PDE-Level Knowledge}
\label{subsubsec:pde_knowledge}

The deepest category derives from governing partial differential equations (PDEs), principally the Helmholtz equation $\nabla^2 u + k^2 u = -f$ \cite{11278649, jia2025rmdm}. Embedding this knowledge requires evaluating differential operators on the predicted field. This serves as either preprocessing features or loss-level constraints. \rev{Its physical meaning and computational cost depend on the field modality, discretization, and evaluation frequency~\cite{jiang2024physics,bayliss1983iterative}. A training-only stencil has a different deployment cost from an additional physical network.}

\subsection{Data-Level Integration: Electromagnetic Feature Engineering}
\label{subsec:data_level}

Data-level integration transforms electromagnetic laws into physics-informed feature tensors concatenated with the network input \cite{chen2024diffraction, lu2025sip2net}. This approach is backbone-agnostic. The resulting tensors can be paired with any architecture without modifying the loss or topology. Since input design choices are discussed in Section~\ref{subsec:input_representation}, this subsection focuses on the specific engineering techniques and their computational trade-offs.

\subsubsection{Material and Antenna Properties}
\label{subsubsec:material_features}

Multi-band operations require explicit electromagnetic feature fusion \cite{chen2024diffraction, jaensch2025radio}. Rather than feeding raw permittivity values, researchers compute physical reflectance and transmittance from ITU standards. Radio depth maps that physically anchor attenuation to building distributions are generated via logarithmic distance path loss and inverse distance weighting \cite{zhou2025tire, li2024radiogat}. The 3D antenna gain is projected onto the 2D propagation plane using trigonometric decomposition:
\begin{align}
\begin{aligned}
\mathrm{FSPL}(D,f)&=20\log_{10}\!\left(\frac{4\pi Df}{c}\right),\\
G_{2D}&=G(\Phi_{2D},\Theta_{2D}).
\end{aligned}
\label{eq:fspl_gain}
\end{align}
Separating the two features avoids a link-budget sign ambiguity: antenna gains add in a received-power equation but are subtracted from an effective path-loss quantity. The free-space expression assumes far-field unobstructed propagation with $D$ and $f$ in units consistent with $c$. The planar gain slice is a source-specific approximation for limited elevation variation~\cite{jaensch2025radio}; active 3D beam steering may require a spherical or volumetric representation. Within represented frequency, material, geometry, and antenna regimes, these features provide an explicit transfer mechanism, but a held-out band still requires direct validation~\cite{chen2024diffraction}.

\subsubsection{Geometric Occlusion Features}
\label{subsubsec:occlusion_features}

\rev{Agile uses a fractional-LoS map generated by Bresenham's line algorithm as an additional occlusion feature~\cite{krijestorac2023agile,yapar2024overview}. On the challenge test set, adding this feature with the MSE objective retained changes normalized RMSE from 0.0514 to 0.0461, a 10.3\% reduction~\cite{yapar2024overview}. The paired comparison isolates the input addition from the separate KL-loss variant. Incremental feature-generation time and memory are unreported in that comparison.} Specialized terrain simulators can instead extract cumulative knife-edge diffraction loss and three-dimensional elevation angles for mountainous environments~\cite{suto2023propagation,chen2025gprt}.

\subsubsection{PDE-Derived Feature Maps}
\label{subsubsec:pde_features}

The RadioDiff-$k^2$ study reports numerical difficulty when a Helmholtz residual is applied directly to its non-physical latent representation and instead computes an effective-wavenumber proxy during preprocessing~\cite{11278649}:
\begin{align}
k_{\mathrm{eff}}^2(\mathbf{x}) \triangleq -\frac{\nabla^2 A(\mathbf{x})}{A(\mathbf{x}) + \epsilon},
\label{eq:keff}
\end{align}
Here, $A$ must be a nonnegative linear-amplitude proxy, not an unconverted dB map, and spatial derivatives must include the physical grid spacing. Then $k_{eff}^2$ has units of inverse length squared, while $\epsilon$ has the same units as $A$. Near an amplitude null, the selected $\epsilon$ controls both magnitude and sign stability. Under the cited construction, negative values rank strong amplitude curvature near fading dips or shadow boundaries and are transformed into diffusion masks~\cite{11278649}. The source reports a 45.5\% NMSE reduction in its conditioning ablation; this value does not establish a universal gain for every grid, carrier, modality, or backbone.

\subsubsection{Topological Filtering for Dynamic Environments}
\label{subsubsec:tda_filter}

Dynamic environments can introduce transient multipath peaks into a spatial feature map~\cite{gao2025time,zhao2025temporal}. Topological persistence measures how long a connected component or feature survives as a filtration threshold changes. A high-persistence peak can be ranked as a source candidate and a low-persistence peak as an artifact candidate, but persistence alone does not establish either physical cause. Standard boundary-matrix reduction has quadratic storage and cubic worst-case time in the number of simplices; sparse, structured-grid, and problem-specific implementations can be substantially cheaper. The filtration, complex size, implementation, memory, and wall-clock preprocessing time must therefore be reported separately from network inference.

\subsubsection{Reflection on Preprocessing Costs}
\label{subsubsec:preprocessing_costs}

Explicit physical feature engineering can improve generalization but transfers cost from the network to the input pipeline~\rev{\cite{chen2024diffraction,yapar2024overview}}. For $M$ transmitters, an $H\times W$ grid, and at most $L$ grid-intersection tests per ray, a line-of-sight feature extractor requires $\mathcal{O}(MHWL)$ tests. Storing $C_p$ precomputed feature maps with $s$-byte precision requires $sHWC_p$ bytes per two-dimensional scene and $sHWDC_p$ bytes for a $D$-layer volume. Four float32 feature maps at $1024\times1024$ resolution therefore occupy 16~MiB per scene before batching, model activations, optimizer states, or cached frequency variants are included. One float32 channel on a $256^3$ voxel grid already occupies 64~MiB. These deterministic storage calculations expose why apparently lightweight feature concatenation becomes a memory bottleneck in volumetric or multi-frequency training.

Operational latency must be reported as $t_{\mathrm{total}}=t_{\mathrm{feature}}+t_{\mathrm{network}}+t_{\mathrm{verification}}$. Caching static features can remove $t_{\mathrm{feature}}$ for repeated queries, but dynamic geometry, moving transmitters, or frequency-dependent materials force recomputation. A reproducible cost statement must identify the grid size, feature count, precision, batch size, processor or accelerator, implementation language, parallelization strategy, and whether cached features were used. Without these conditions, a wall-clock number cannot be transferred to another deployment.

\rev{A target-scene adaptation example further connects labels, error, and computational work. Chen and Chen report that, with 10\% target data, initialization from a pretrained model yields an NMAE of 0.106 in 50 epochs, compared with 0.125 in 100 epochs from random initialization~\cite{chen2024diffraction}. The epoch counts cover target adaptation. Source pretraining remains an additional cost. This operating point supports adaptation with target observations. Zero-shot transfer requires a separate held-out-environment test.}

\subsection{Loss-Level Regularization: Embedding Physical Constraints}
\label{subsec:loss_level}

Loss-level integration translates electromagnetic principles into penalty terms~\cite{jiang2024physics,jia2025rmdm}. It biases predictions toward the selected constraints in unmeasured regions, but a finite penalty does not prove that the output satisfies the underlying physical law~\cite{karniadakis2021physics}.

\subsubsection{Multipath Consistency}
\label{subsubsec:multipath_consistency}

Under sparse sampling, a calibrated line-of-sight (LoS) propagation term can regularize a model when its far-field assumptions apply~\cite{liu2025pinn}. The free-space path-loss anchor is
\begin{align}
\mathrm{FSPL}(d,f_c)=20\log_{10}\!\left(\frac{4\pi d f_c}{c}\right).
\label{eq:fspl_anchor}
\end{align}
Equation~\eqref{eq:fspl_anchor} assumes far-field free-space propagation with $d$ in meters and $f_c$ in hertz; kilometers and megahertz require the corresponding additive constant. Antenna gains, polarization mismatch, feeder loss, and near-field effects are separate link-budget terms. An absolute anchor also requires a known transmitter and calibrated power convention. A source-agnostic model must estimate these quantities jointly or use a relative consistency term~\cite{hehn2023transformer}. Likewise, a one-sided NLoS-delay penalty should use a calibrated margin because timing resolution and model mismatch preclude a universal strict inequality~\cite{liu2025pinn}.

\subsubsection{Numerical Solver Embedding}
\label{subsubsec:solver_embedding}

PEFNet \cite{jiang2024physics} directly embeds the volume integral equation (VIE) and its method of moments (MoM) discretization:
\begin{align}
\mathcal{L}_{PHY} = \frac{1}{N}\|(I + W\chi)E'^{tot} - E^{inc}\|,
\label{eq:vie_loss}
\end{align}
where $I$ is the identity matrix, $W$ the Green's function coefficient matrix, and $\chi$ the permittivity contrast matrix. The data loss becomes a residual compensation mechanism:
\begin{align}
\mathcal{L}_{DAT} = \frac{1}{N}\|pl' - pl^{dat}\|.
\label{eq:data_loss}
\end{align}
In the cited RadioMapSeer experiment, PEFNet reports $R^2>0.99$ under its cross-scene split and approximately 1.1~s inference, compared with the source's classical-solver timing~\cite{jiang2024physics}. These values are study-specific because the hardware, solver tolerance, scene size, output modality, and timing boundary are not a shared benchmark.

\begin{figure*}[!t]
    \centering
    \captionsetup{font={small}, skip=10pt}
    \includegraphics[width=0.85\linewidth]{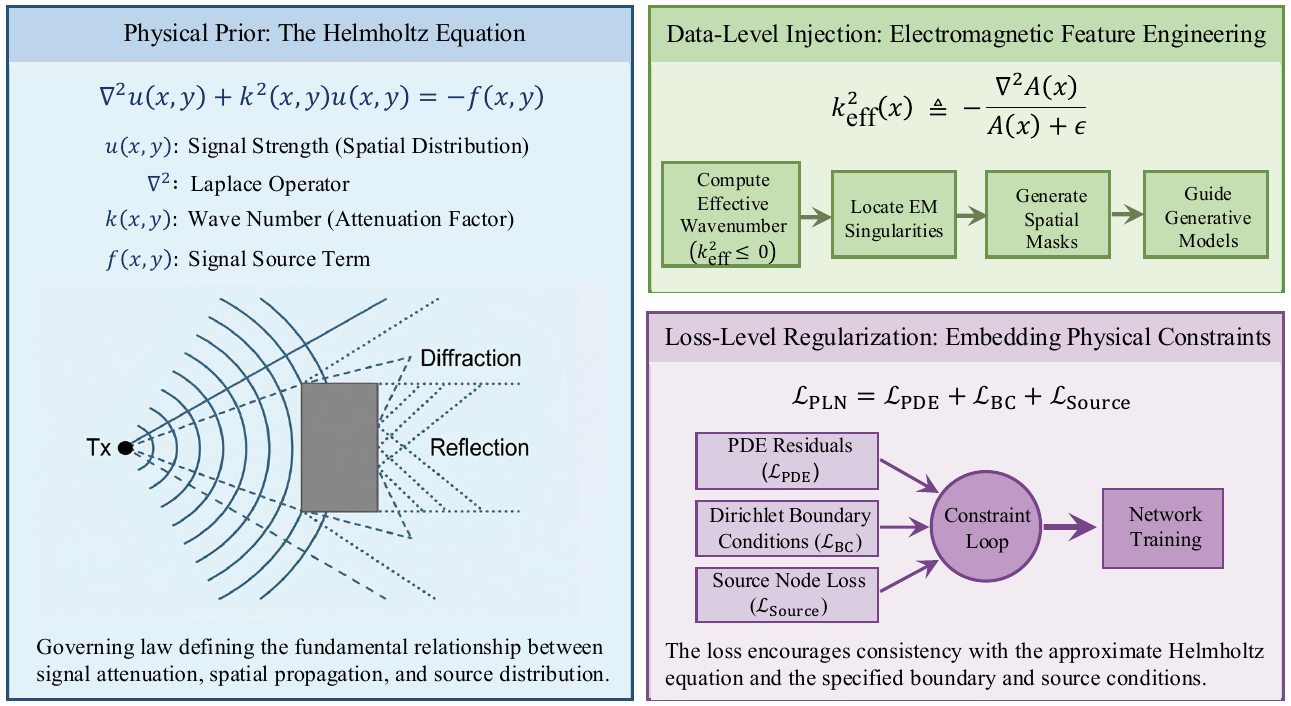}
    \caption{Illustration of a Helmholtz-informed RM training pipeline. The left panel shows the wave-equation proxy used by the cited method; its validity depends on the field modality, discretization, sources, materials, and boundary data. The top-right branch injects the source's effective-wavenumber proxy ($k_{\text{eff}}^2$) to rank strong amplitude curvature. The word ``singularity'' inside the original schematic denotes this empirical mask and not a proven mathematical or electromagnetic singularity. The bottom-right branch combines a PDE residual, the declared boundary model, and a source loss in $\mathcal{L}_{\text{PLN}}$. \rev{Finite loss penalties encourage consistency with the approximate Helmholtz equation and the specified boundary and source conditions.}}
    \label{fig:physics_paradigm}
\end{figure*}

\subsubsection{Gradient Pathologies and Dynamic Balancing}
\label{subsubsec:gradient_pathology}

Combining physical and data losses can produce poorly balanced gradients~\cite{karniadakis2021physics,jiang2024physics}. The two residuals may use different units and numerical scales, for example a field-equation residual and a path-loss error in dB. A quoted difference of several orders of magnitude is implementation-specific rather than universal. Each study should report normalized loss definitions and gradient norms during training; otherwise either the data or physics term can dominate for purely numerical reasons.

Three balancing strategies appear in the surveyed RM studies. Dynamic adjustment uses relative training rates to alter component weights~\cite{wang2023super}. Homoscedastic task weighting minimizes terms of the form $\mathcal L_m/(2\sigma_m^2)+\log\sigma_m$ with $\sigma_m>0$; implementations should optimize an unconstrained log-scale and map it to a positive value. The learned $\sigma_m$ balances task scales but is not, by itself, calibrated predictive uncertainty. Adaptive annealing gradually increases $\lambda_{\mathrm{PHY}}$ from zero over a warm-up interval~\cite{jia2025rmdm}. These heuristics can reduce gradient imbalance, but their benefit and final constraint satisfaction must be checked from per-term losses and gradient norms rather than assumed from the weighting rule.

\subsubsection{Helmholtz Equation Embedding}
\label{subsubsec:helmholtz}

PhyRMDM \cite{jia2025rmdm} operates a dual-network structure where the initial stage applies physical constraints via the 2D Helmholtz equation:
\begin{align}
\nabla^2 u(x,y) + k^2(x,y)u(x,y) = -f(x,y),
\label{eq:phyrmdm_helmholtz}
\end{align}
where $u(x,y)$ is treated as a field-amplitude proxy derived from the RM, $k(x,y)$ the wavenumber, and $f(x,y)$ the source term. The continuous Laplacian is discretized using central differences \cite{bayliss1983iterative}. As illustrated in Fig.~\ref{fig:physics_paradigm}, the physics-informed loss $\mathcal{L}_{PLN}$ is decomposed into three complementary components that jointly form a constraint loop during network training:
\begin{align}
\mathcal{L}_{PLN} = \mathcal{L}_{PDE} + \mathcal{L}_{BC} + \mathcal{L}_{Source}.
\label{eq:pinn_loss}
\end{align}
The three loss terms have complementary roles. $\mathcal{L}_{PDE}$ penalizes a discrete residual of~\eqref{eq:phyrmdm_helmholtz}, $\mathcal{L}_{BC}$ penalizes the boundary model used by the cited implementation, and $\mathcal{L}_{Source}$ anchors a known source~\rev{\cite{jia2025rmdm,karniadakis2021physics}}. They regularize the prediction only with respect to those declared approximations. A second-order central difference has truncation error $\mathcal{O}(\Delta x^2)$ for a sufficiently resolved smooth field~\cite{erlangga2004class}; this asymptotic statement does not justify applying a carrier-resolved Helmholtz residual on an RM grid much coarser than the wavelength. At 28~GHz, $\lambda\approx10.7$~mm, so a 1~m cell spans about 93 wavelengths and cannot resolve the carrier field. A physically interpretable loss on such a grid requires a validated envelope or effective-wavenumber model, or a separate wavelength-resolving mesh coupled to the RM grid. Merely increasing the finite-difference order does not recover sub-grid phase information. \rev{In the cited loss ablation, adding the PINN term while retaining the data and regularization terms changes normalized RMSE from 0.0182 to 0.0125, a 31.3\% reduction~\cite{jia2025rmdm}. The isolated training-time increment is unreported.}

\subsection{Architecture-Level Mapping: Structural Isomorphism}
\label{subsec:arch_level}

Architecture-level integration maps a selected physical update rule, operator, or symmetry into network topology~\cite{chen2026radiodun,wang2025dulrtc}. It therefore makes the strongest structural commitment among the three integration levels, but it does not by itself establish Maxwell consistency outside the assumptions encoded by that structure.

\subsubsection{Algorithm Unrolling}
\label{subsubsec:unrolling}

Algorithm unrolling maps classical iterative optimizers onto deep cascaded networks layer by layer \cite{wang2025dulrtc, chen2026radiodun}. DULRTC-RME \cite{wang2025dulrtc} formulates RM estimation as low-rank tensor completion:
\begin{align}
\min_{\mathcal{X},\mathcal{E}} \sum_{i=1}^{3} \alpha_i \|\mathcal{X}_{(i)}\|_* + \lambda\|\mathcal{E}\|_1 + f(\mathcal{X}) + g(\mathcal{E}),
\label{eq:dulrtc}
\end{align}
Here, $\alpha_1,\alpha_2,\alpha_3$ are learnable nuclear-norm weights optimized through backpropagation. The cited DULRTC-RME study reports 25.81~dB PSNR at its 10\% sampling operating point, with CNNs replacing selected proximal operators in an unrolled ADMM structure~\cite{wang2025dulrtc}. RadioDUN maps an alternating sparse-recovery algorithm into a physics-inspired architecture and reports reconstruction from nine measurements under its stated map size, sampling rule, and evaluation protocol~\cite{chen2026radiodun}. Neither operating point is a universal sampling threshold.

A practical constraint is activation memory. With $K$ unrolled blocks, batch size $B$, $C$ stored channels, and an $H\times W$ map, the retained forward activations alone scale as $\mathcal{O}(BKCHW)$. At $s$-byte precision, their lower-bound storage is $sBKCHW$ bytes before gradients, optimizer states, temporary operators, and CNN features are counted. This linear growth in unrolling depth can make a theoretically interpretable architecture less deployable than a shallow data-driven baseline.

\subsubsection{Differentiable Physical Components}
\label{subsubsec:diff_physics}

Hard-thresholded occlusion logic can yield zero or undefined gradients at visibility changes. A cited differentiable-propagation formulation replaces the threshold with the soft indicator $I=1-\tanh(\mathrm{POL})$, where $\mathrm{POL}=\sum O_L$ is cumulative intersection depth~\cite{bakirtzis2024solving}. This approximation permits gradient-based environment reconstruction but introduces a softness parameter and model mismatch that must be validated. Likewise, multi-head attention with positional encoding has been used to approximate complex diffraction behavior that a recursive model such as Vogler's would otherwise compute~\cite{chen2025learning}; it is a learned surrogate, not an exact replacement for every diffraction regime.

\subsubsection{Physics-Informed Generative Architectures}
\label{subsubsec:physics_gen}

iRadioDiff extracts transmission boundaries and signal-discontinuity priors from geometrical optics and injects them as prompts through spatial cross-attention~\cite{wang2025iradiodiff}. Its source-reported ablation changes RMSE from 6.36 to 9.62~dB when those priors are removed, a 51\% increase under that study's data and protocol. This supports the value of the prompts within the reported experiment but is not a cross-dataset effect size.

DeepRT constructs a wireless-environment knowledge pool from ray-tracing principles and maps reflection and diffraction priors into a large-model latent space through relation-aware modules~\cite{li2025deeprt}. The cited setup reports 4~ms network inference and an approximately $3\times10^5$ speedup relative to its own ray-tracing baseline. The comparison does not include every deployment-dependent feature-generation or verification cost, and it should not be transferred to different hardware or scene complexity without remeasurement. The connection between language-model sequence processing and electromagnetic propagation remains an open question; the reported gain may arise from representational capacity and structured prior injection rather than an intrinsic linguistic--electromagnetic alignment.


\subsubsection{Physical Hallucinations: Definition and Detection}
\label{subsubsec:hallucinations}

A critical limitation of architecture-level methods in non-physical latent spaces is the inability to guarantee compliance with Maxwell's equations \cite{karniadakis2021physics, jiang2024physics}. This introduces physical hallucinations, defined as follows.

\begin{quote}
A physical hallucination is a generated field that matches the requested data format but violates a macroscopic electromagnetic constraint under the stated source, material, and boundary model. The violation must be tied to a testable constraint, rather than to visual appearance alone.
\end{quote}

For a predicted complex field $\hat u$ on interior samples $\Omega_I$, the normalized Helmholtz residual and a boundary residual are
\begin{align}
&r_{\mathrm{PDE}}=\frac{\|\nabla^2\hat u+k^2\hat u+f\|_{2,\Omega_I}^2}{\|k^2\hat u\|_{2,\Omega_I}^2+\|f\|_{2,\Omega_I}^2+\epsilon_{\mathrm{PDE}}},
\label{eq:hallucination_pde}\\
&r_{\mathrm{BC}}=w_D\frac{\|\hat u-g_D\|_{2,\Gamma_D}^2}{\|g_D\|_{2,\Gamma_D}^2+\epsilon_D}+w_N\frac{\|\partial_n\hat u-g_N\|_{2,\Gamma_N}^2}{\|g_N\|_{2,\Gamma_N}^2+\epsilon_N},
\label{eq:hallucination_bc}
\end{align}
Here, $f$ is the source term and $g_D$ and $g_N$ are imposed Dirichlet and Neumann data. Nonnegative weights $w_D+w_N=1$ average only the boundary types that are actually imposed; an unavailable or empty component receives zero weight. The stabilizers $\epsilon_{\mathrm{PDE}}$, $\epsilon_D$, and $\epsilon_N$ are separately chosen in the squared units of their respective denominators. Field values and normal derivatives are normalized separately because they have different units. Finite differences are meaningful only on a grid that resolves the modeled field, while automatic differentiation can evaluate derivatives of a continuous representation but cannot repair an invalid field model~\cite{11278649,jia2025rmdm}.

Two complementary tests examine reciprocity and energy. Reciprocity requires a linear reciprocal medium and calibrated interchange of terminal reference planes, antennas, and polarizations. The energy test assumes time-harmonic steady state, a control volume $V$ containing every represented source, and no unmodeled incident power through its boundary. Under these conditions, define
\begin{align}
&r_{\mathrm{rec}}=\frac{|\hat h_{A\rightarrow B}-\hat h_{B\rightarrow A}|^2}{|\hat h_{A\rightarrow B}|^2+|\hat h_{B\rightarrow A}|^2+\epsilon_{\mathrm{rec}}},
\label{eq:hallucination_reciprocity}\\
&r_{\mathrm{eng}}=\frac{\left[\int_{\partial V}\langle\hat{\mathbf S}\rangle\!\cdot\mathbf n\,\mathrm dA-P_{\mathrm{in}}\right]_+}{P_{\mathrm{in}}+\epsilon_P},
\label{eq:hallucination_energy}
\end{align}
Here, $\epsilon_{\mathrm{rec}}$ has squared-channel units. The time-averaged Poynting vector $\langle\hat{\mathbf S}\rangle=\tfrac12\operatorname{Re}(\hat{\mathbf E}\times\hat{\mathbf H}^{*})$ requires compatible complex electric- and magnetic-field estimates, $P_{\mathrm{in}}$ is accepted source power, and $\epsilon_P$ has power units. A power-only RM cannot supply this Poynting test. The positive-part operator flags unexplained outward excess but does not test excess absorption or phase error. Reciprocity must be disabled for nonreciprocal media, active components, or uncalibrated direction-dependent hardware~\cite{zhou20256d}.

Power-only maps require tests that use quantities present in their output. Let $\Omega_m$ contain calibrated RSS observations $y_i$ in dB with standard deviations $\sigma_i$, and let $\hat m_i$ be the predicted RSS. When the accepted transmit power $p_{\mathrm{tx}}$ and a conservative location-dependent gain bound $g_i^{\max}$ are known, define
\begin{align}
&r_{\mathrm{obs}}=\frac{1}{|\Omega_m|}\sum_{i\in\Omega_m}\frac{(\hat m_i-y_i)^2}{\sigma_i^2+\epsilon_{\mathrm{obs}}},
\label{eq:hallucination_observation}\\
&r_{\mathrm{bud}}=\frac{1}{|\Omega|}\sum_{i\in\Omega}\frac{[\hat p_i-p_{\mathrm{tx}}g_i^{\max}]_+^2}{(p_{\mathrm{tx}}g_i^{\max})^2+\epsilon_{\mathrm{bud}}},
\label{eq:hallucination_budget}
\end{align}
Here, $\hat p_i$ is predicted linear received power. $\epsilon_{\mathrm{obs}}$ has squared-dB units, whereas $\epsilon_{\mathrm{bud}}$ has squared-power units; neither is interchangeable with the PDE or reciprocity stabilizer. The first residual tests measurement agreement after calibration uncertainty, and the second flags predictions above a declared source-and-propagation budget. The budget applies only when antenna gains, near-field effects, active repeaters, and all relevant sources are included in $g_i^{\max}$. Monotonic LoS-envelope or reciprocal path-loss checks require verified links and calibrated fading margins; multipath prevents either from serving as a universal rule.

A map is flagged when any applicable residual exceeds a threshold $\tau_j$. Each $\tau_j$ should be calibrated as a high quantile of the same residual on held-out physically valid simulations or measurements, because discretization, measurement noise, and material mismatch prevent a universal zero threshold. A complete detector reports all applicable residuals rather than only a binary label. Detector performance should additionally be reported through the area under the receiver-operating-characteristic curve (AUROC), the true-positive rate at a declared false-positive rate, and downstream decision error, rather than through only a residual mean. Complex-field outputs can support $(r_{\mathrm{PDE}},r_{\mathrm{BC}},r_{\mathrm{rec}},r_{\mathrm{eng}})$, whereas power-only outputs can support $(r_{\mathrm{obs}},r_{\mathrm{bud}})$ when their required metadata are available. This modality-specific rule prevents an incomplete output representation from being judged by an unavailable field quantity.

\subsection{Summary and Practitioner Guidance}
\label{subsec:physics_summary}

The three levels represent complementary paradigms with distinct costs and capabilities. Table~\ref{tab:physics_modules} provides a structured reference for practitioners. \rev{PhyRMDM's paired loss ablation reduces source-reported RMSE by 31.3\% when the PINN term is added while retaining the data and regularization terms~\cite{jia2025rmdm}. iRadioDiff's 33.9\% ALG RMSE reduction concerns the complete physical-cue bundle~\cite{wang2025iradiodiff}. GSRF separately quantifies phase and Fourier-loss contributions in Sec.~\ref{subsubsec:gsrf}, while the WRF-GS+ comparison in Sec.~\ref{subsubsec:wrfgs} pairs its accuracy gain with measured query cost. Table~\ref{tab:within_study_resources} supplies model-level resource measurements. The stencil and attention expressions in Table~\ref{tab:physics_modules} describe individual operators. A physical addition is justified by its matched accuracy benefit and the complete cost of the deployed pipeline.}

Physics awareness is not automatically preferable. A purely data-driven model is often the stronger engineering choice when representative labeled data densely cover the deployment distribution, physical metadata are missing or unreliable, the governing discretization is mismatched to the target resolution, or an edge device cannot afford feature generation and residual evaluation. An incorrect material map or boundary condition can introduce systematic bias that additional training cannot remove. Physics-informed and data-driven variants should therefore be compared under the same backbone, parameter budget, data split, and latency accounting. The physical variant is justified only when its gain persists under plausible perturbations of the assumed source, material, and geometry parameters.

\subsubsection{Scenario-Specific Recommendations}
\label{subsubsec:physics_recommendations}

\rev{Start with a feed-forward grid model when representative labels cover the deployment environments. With scarce labels, evaluate physical features using matched learning curves and reliable geometry or material data~\cite{chen2024diffraction,yapar2024overview}. Loss-level constraints require a valid output modality, source, boundary, and discretization model~\cite{jia2025rmdm,jiang2024physics}. For unseen environments, distinguish zero-shot prediction, limited target-data adaptation, and scene refitting~\cite{wang2026radiodifffs}.}

\rev{Select the lowest-cost candidate meeting the held-out accuracy requirement under the same observations and adaptation budget. Frequent updates require end-to-end timing of feature construction, inference, and verification. For a fixed scene, the cost of neural-field fitting can be amortized over many queries, while geometry changes require accounting for updates or refitting. Interpret the resource values in Table~\ref{tab:within_study_resources} at the stated workloads and sampling budgets.}

\subsubsection{Open Challenges}
\label{subsubsec:physics_challenges}

Three challenges span the integration levels. First, online physical-feature computation must be accelerated or cached without hiding its cost in the latency report~\cite{chen2024diffraction,11278649}. Second, data and physical losses require explicit normalization and balance checks because their gradients can have different units and scales~\cite{karniadakis2021physics,ivanov2024deep}; no automatic rule is universally necessary or sufficient. Third, the residual detector in Section~\ref{subsubsec:hallucinations} requires standardized calibration sets, modality-specific thresholds, and measured false-positive rates under imperfect material and boundary models. Lightweight verification layers must preserve these diagnostics without becoming more expensive than the RM construction model itself.

\section{Open Issues and Future Directions}
\label{sec:open_issues}


Learning-based RM construction has achieved significant milestones \cite{zeng2024tutorial, 10640063}. However, the transition toward ubiquitous 6G environmental awareness remains constrained by several open challenges. This section identifies the most pressing issues, drawing upon the architectural insights, physical embedding strategies, dataset limitations, and deployment risks discussed throughout this tutorial. The challenges are organized along four complementary axes: complex propagation scenarios, advanced inversion and optimization, trustworthy data and model operation, and foundation models with unified frameworks.

\subsection{Modeling Complex and Emerging Propagation Scenarios}
\label{subsec:emerging_scenarios}

\subsubsection{Beam-Aware RM Construction}
\label{subsubsec:beam_aware}

Current research predominantly addresses spatial field prediction for omnidirectional or fixed-beam antennas \cite{levie2021radiounet, wang2024radiodiff}. Beam-aware modeling, where the RM is conditioned on a continuous beamforming vector $\mathbf{w} \in \mathbb{C}^{N_t}$, remains largely unexplored beyond BeamCKMDiff \cite{zhao2026beamckmdiff, chen2026channel}. Two bottlenecks impede progress. First, beam-specific propagation datasets are scarce. BeamCKM \cite{wang2025beamckm} is one of the few public benchmarks. Second, integrating continuous beam parameters increases the conditioning dimensionality. Architectures must decouple spatial geometry from high-dimensional beam sweeping \cite{zeng2024tutorial}. The adaLN-based conditioning in Section~\ref{subsubsec:condition_embedding} provides a template. Extending it to joint spatial-beam-frequency conditioning at scale remains open. RadioDiff-v2 shifts the target from a scalar power field to an angular power spectrum and evaluates the resulting map through multi-beam selection and localization, thereby making the downstream beam objective explicit~\cite{wang2026radiodiffv2,huang2026map2aps}.

\subsubsection{Non-Terrestrial and Confined Environments}
\label{subsubsec:nonterrestrial}

Satellite-to-ground links often have a dominant LoS component together with atmospheric, ionospheric, blockage, beam, and mobility effects whose importance depends on band and elevation~\cite{al2022survey,hao2021satellite,davarian2002earth}. A model may need atmospheric or remote-sensing variables only when the target band and attenuation mechanism justify them~\cite{al2022next}. Maritime surfaces introduce time-varying reflection and scattering geometry~\cite{peng2025machine}, while tunnels and enclosed metallic structures can support strong multipath~\cite{cheng2019space}. Curved routes and variable effective ranges do not make fixed-size tensors mathematically incompatible, but padding or resampling can waste resolution and obscure distance; variable-length sequences, multiscale grids, or graphs are alternatives whose benefit must be measured~\cite{perdomo2025wirelessnet}. The beam-aware satellite formulation of Wang et al. treats the active set of multiple satellites and beam-dependent fields as latent variables and uses a parametric propagation model rather than a terrestrial image prior~\cite{wang2026beamaware}. Its evaluation should separate satellite identity, beam footprint, elevation, and orbital or atmospheric state so that a fixed constellation cannot be memorized.

\subsection{Frontiers in Advanced Inversion and Optimization}
\label{subsec:inversion_frontiers}

\subsubsection{\rev{Active Sampling and Reconstruction}}
\label{subsubsec:active_sampling}

\rev{Optimizing measurement locations can reduce the data required for accurate RM reconstruction. RMSampler uses spatial covariance among diffusion-generated candidate RMs to select sampling locations by the expected relative reductions in global reconstruction error and emitter-localization error~\cite{cheng2026rmsampler}. Each acquired received-power measurement updates the estimates guiding the next sampling decision. Future work should extend this joint sampling and reconstruction approach to mobile sensing with travel, energy, and access constraints~\cite{li2026uavadaptive3d}, while calibrating uncertainty in unseen environments. Comparisons should use a common reconstructor and measurement budget to isolate the benefit of location selection, and include the computation needed to select sampling locations and the associated travel costs.}

\subsubsection{Multi-Source Radiation Inversion}
\label{subsubsec:multisource}

Most source-agnostic RM studies surveyed here evaluate a single radiation source~\cite{viet2025spatial,ha2025bayesian}. With multiple unknown emitters, overlapping spatial or spectral signatures enlarge the set of configurations consistent with sparse observations~\cite{lim2023interference,zhao2025imnet}. Its growth rate depends on the source model, discretization, priors, and measurement operator, so no blanket exponential law is assumed. The surveyed learning methods also do not provide a general identifiability theorem for this setting. Conditional diffusion samples can explore several plausible reconstructions~\cite{ha2025bayesian,luo2025denoising}, but sample diversity is not proof of posterior correctness and requires calibration. Open tasks include stating sufficient identifiability conditions, designing source-aware sampling, and evaluating recovery as the number and separation of emitters vary.

\subsubsection{Blind Parameter Inversion Under Uncalibrated Observations}
\label{subsubsec:blind_inversion}

Real-world sensor data are corrupted by unknown noise floors, position drift, and device-specific antenna variations \cite{sallouha2024rem, becvar2024machine, cisse2024fine}. Developing unsupervised blind inversion frameworks that maintain fidelity despite uncalibrated observations is essential for autonomous network operation. The physics-informed loss strategies of Section~\ref{subsec:loss_level} provide a starting point. However, their current formulations assume calibrated measurements \cite{jiang2024physics, jia2025rmdm}. Extending them to jointly estimate noise model parameters alongside the RM remains open. Wireless Digital Twin Calibration offers a complementary 2026 direction by refining DFT-domain channel information against measured evidence~\cite{luo2026calibration}. For RM deployment, an analogous calibration layer should distinguish a global simulator mismatch from a local map residual, otherwise the learned model may absorb hardware or solver bias as spatial propagation structure.

\subsubsection{Differentiable Inversion for Network Optimization}
\label{subsubsec:diff_inversion}

A trained neural network acts as a differentiable surrogate model \cite{levie2021radiounet, jiang2024learnable}. Given a target coverage, gradients can be backpropagated to optimize base station parameters. However, as discussed in Section~\ref{subsubsec:receptive_field}, this faces severe optimization pathology. Transmitter locations encoded as binary masks or Gaussian heatmaps produce fragmented gradients under naive descent \cite{wang2024radiodiff}. Overcoming this requires novel differentiable rendering techniques, hybrid discrete-continuous solvers, or reinforcement learning formulations that bypass gradient pathology \cite{wang2022green, okawa2024optimal}.

\subsection{Security, Integrity, and Privacy}
\label{subsec:security_privacy}

An RM pipeline is exposed to threats at both its measurement and environment interfaces. Measurement poisoning or spoofing can alter a small subset of RSS, CSI, or beam reports so that a coverage hole is hidden or a false hotspot is created; federated collection additionally permits adaptive client poisoning~\cite{li2022multitentacle}. Signal-level spoofing studies in positioning systems further show why authentication and cross-sensor consistency belong at the data interface, although their attack models are not identical to RM poisoning~\cite{psiaki2016gnss}. A reproducible threat model should state the compromised fraction, temporal persistence, perturbation budget in dB or complex amplitude, and whether the attacker knows the model. The evaluation should report clean RMSE/NMSE, attacked RMSE/NMSE, attack success rate for a specified coverage decision, and the increase in the applicable physical residuals from Section~\ref{subsubsec:hallucinations}.

Environment tensors require a separate, physically realizable threat model. Perturbations should correspond to permitted edits of geometry, occupancy, material permittivity or conductivity, antenna orientation, or source parameters, and should be rendered through the same propagation solver used for the reference map. Reporting only an arbitrary pixel norm can overstate or understate risk. Useful axes are the perturbation magnitude in meters or material units, the fraction of changed voxels, the resulting robust RMSE, and the utility retained by the downstream coverage or beam-selection task. Certified or randomized defenses should be compared at matched preprocessing, training, and inference budgets.

Contributed traces also reveal sensitive information, including a user's route, home or workplace, device identity, and participation in a measurement campaign. Before release, coordinates should be coarsened or protected by a documented privacy mechanism, and raw identifiers should be separated from the learning record. A privacy audit should test trajectory or location reconstruction, membership inference---whether an attacker can determine that a particular trace or user was included in the training set---and device re-identification, while measuring the corresponding RM utility loss and calibration change~\cite{shokri2011location,shokri2017membership}. Here, a privacy parameter is the formal leakage budget of the selected mechanism---for example, $\epsilon$ together with $\delta$ when applicable for differential privacy---rather than a generic privacy score~\cite{geyer2017differential}. The release protocol should specify the threat model, privacy parameter, aggregation granularity, retention period, and whether an attacker can combine the RM with public maps.

\subsection{Toward Foundation Models and Unified Frameworks}
\label{subsec:foundation_models}

\subsubsection{Cross-Domain Pre-Training and Transfer}
\label{subsubsec:cross_domain}

The success of DINOv2-ViT \cite{mkrtchyan2025vision} and RadioDiff-Inverse \cite{wang2025radiodiff-inv} in transferring visual priors to the wireless domain suggests cross-domain pre-training is viable \cite{jaiswal2025data, jaiswal2025leveraging}. However, natural images differ from RMs in distance-dependent decay, dynamic range, and multipath patterns \cite{wang2025iradiodiff}. Developing domain-specific foundation models pre-trained on large-scale wireless data is a more principled path \cite{jiang2025unirm}. The key question is whether a single model can span indoor offices, urban canyons, and maritime surfaces without catastrophic forgetting.

\subsubsection{Actionable Evaluation Protocols}
\label{subsubsec:evaluation_protocols}

Benchmark standardization should begin with scene-disjoint train, validation, and test splits. Tiles from the same building, street segment, or simulator seed must not cross a split, and the released files should include fixed observation masks, coordinate frames, dB/linear conventions, clipping ranges, and unit tests for leakage. Every method should be evaluated in three regimes: interpolation within seen scenes, zero-shot transfer to unseen scenes or frequencies, and adaptation with a declared number of target labels. The report should include RMSE or NMSE in the declared domain, SSIM or PSNR only with its dynamic range, peak memory, accelerator-hours, preprocessing time, inference latency, checkpoint, and evaluation code.

Uncertainty claims should be tested rather than inferred from visual diversity. For a nominal $(1-\alpha)$ predictive interval, report empirical coverage, average interval width, negative log-likelihood when a density is available, and coverage stratified by distance, blockage, and observation density. Calibration must be measured on a scene-disjoint set and after any target-scene adaptation. Continual-learning studies should stream scenes or frequencies in chronological order, evaluate both the new and all previously learned scenes after each update, and report average accuracy, forgetting $F=\max_t e_{i,t}-e_{i,T}$, update time, and retained memory.

Cross-frequency transfer requires a held-out frequency band rather than a random sample from every band. The zero-shot result, few-shot adaptation curve, and performance after full retraining should be reported together, with frequency, material, and antenna metadata exposed to the model. Foundation-model evaluations should additionally compare frozen zero-shot inference, a linear probe (a trained linear prediction head over frozen representations), and full fine-tuning under matched data and compute budgets. Pre-training corpora, scene overlap checks, deduplication rules, data-license constraints, and possible simulator or test-scene leakage must be disclosed; otherwise a large score can reflect memorization rather than transferable propagation knowledge.

\subsubsection{Amortized Inference for Real-Time Optics-Inspired Methods}
\label{subsubsec:amortized}

Per-scene optimization in NeRF and 3DGS limits rapid reuse in dynamic environments because scene or transmitter changes can require refitting~\cite{zhao2023nerf2,zhang2026rf,umer2025neural}. Amortized inference instead trains a feed-forward encoder $\mathcal{E}_\phi$ to predict scene-representation parameters from sparse measurements. Its intended benefit is to replace iterative per-scene fitting with one encoder pass; a minutes-to-milliseconds claim would still require matched end-to-end timing that includes measurement encoding, representation construction, and verification. For a 3DGS-based method, the encoder could map $\mathcal{M}=\{(\mathbf r_i,\mathbf s_i)\}_{i=1}^{N}$ to the Gaussian parameters:
\begin{align}
\{\boldsymbol{\mu}_k,\boldsymbol{\Sigma}_k,C_k^{\mathrm{re}},C_k^{\mathrm{im}}\}_{k=1}^{K}=\mathcal{E}_\phi(\mathcal{M}),
\label{eq:amortized}
\end{align}
Here, $\boldsymbol\mu_k$ is a position, $\boldsymbol\Sigma_k$ must be positive semidefinite, and $C_k^{re}+jC_k^{im}$ is a complex coefficient. A valid decoder should enforce covariance positivity, for example through a scale--rotation or Cholesky parameterization, and should be permutation-aware because Gaussian order has no physical meaning. Fixed $K$ also imposes a capacity assumption; variable scene complexity requires birth--death mechanisms, masking, or set prediction. GRaF provides a related amortized idea for RF radiance fields~\cite{yang2025graf}, but the surveyed material did not provide a validated direct RF-3DGS parameter predictor with matched end-to-end timing.

\subsubsection{Closing the Physical Hallucination Gap}
\label{subsubsec:closing_hallucination}

The residual tests in Section~\ref{subsubsec:hallucinations} provide a post-hoc detector for complex-field and power-only outputs when the required physical metadata are available~\cite{karniadakis2021physics,jiang2024physics}. Standardized calibration sets, false-positive characterization, low-latency implementation, and generation-time correction nevertheless remain absent from the literature. In particular, operational detectors that are standardized and validated across power-only studies remain absent. Three research directions follow. First, lightweight differentiable verification layers should evaluate the applicable PDE, energy, observation, or source-budget residual during inference and report detection error under model mismatch~\cite{11278649,jia2025rmdm}. Second, constrained generation can project diffusion outputs toward the feasible electromagnetic set at each denoising step, but the projection must respect the information content of the predicted modality~\cite{luo2025denoising,ha2025bayesian}. Third, standardized hallucination benchmarks should pair physically valid maps with controlled violations and publish calibrated thresholds, false-positive rates, and downstream decision errors~\cite{boban2023white}.

\section{Concluding Remarks}
\label{sec:conclusion}

This tutorial has surveyed the data foundations, neural architectures, and physics-informed strategies for learning-based radio map construction. We have organized the design space by the forward-inverse problem dichotomy and introduced a three-level physics-integration framework spanning data, loss, and architecture. The resulting guidance connects model choice to identifiability, generalization variables, physical fidelity, and computation rather than to architecture names alone. We have also shown that physical residuals, uncertainty calibration, and scene-disjoint evaluation are necessary to distinguish a numerically accurate map from a reliable deployment artifact. Data integrity and privacy must be treated as first-class constraints: poisoning, physically realizable environment perturbations, and leakage from contributed traces can invalidate an otherwise accurate model. The actionable protocols in Section~\ref{subsubsec:evaluation_protocols} provide a common basis for reporting zero-shot transfer, adaptation, continual learning, calibration, resource use, and foundation-model data leakage. Future radio-map systems will therefore combine differentiable propagation and amortized inference with auditable security controls and modality-appropriate physical verification. Sustained collaboration among wireless communications, computational electromagnetics, and machine learning will be essential for turning these methods into trustworthy components of network planning and operation.

\bibliography{ref}
\bibliographystyle{IEEEtran}
\ifCLASSOPTIONcaptionsoff
  \newpage
\fi
\end{document}